\pdfoutput=1

\documentclass[iop]{emulateapj}
\usepackage{graphicx}
\usepackage{epstopdf}
\usepackage{lineno}
\usepackage{footnote}
\usepackage{epsfig}
\usepackage{graphics}
\usepackage{amssymb}
\usepackage{amsmath}
\usepackage{mathtools}
\usepackage{enumerate}
\usepackage{pifont}

\usepackage[colorlinks=true,linkcolor=blue,citecolor=blue,urlcolor=blue]{hyperref}

\shorttitle{Blazar candidates behind the MCs}
\shortauthors{\.{Z}ywucka et al.}

\usepackage[utf8]{inputenc}
\def\dg{\hbox{$^\circ$}}
\def\arcmin{\hbox{$^\prime$}}
\def\arcsec{\hbox{$^{\prime\prime}$}}
\def\utw{\smash{\rlap{\lower5pt\hbox{$\sim$}}}}
\def\udtw{\smash{\rlap{\lower6pt\hbox{$\approx$}}}}

\def\fm{\hbox{$.\!\!^{\rm m}$}}

\def\fdg{\hbox{$.\!\!^\circ$}}

\def\farcs{\hbox{$.\!\!^{\prime\prime}$}}

\begin{document}

\title{Identification of blazar candidates behind Small and Large Magellanic Clouds}

\author{Natalia \.{Z}ywucka\altaffilmark{1}, Arti Goyal\altaffilmark{1}, Marek Jamrozy\altaffilmark{1}, {\L}ukasz Stawarz\altaffilmark{1}, Micha{\l} Ostrowski\altaffilmark{1},\\ Szymon Koz{\l}owski\altaffilmark{2}, and Andrzej Udalski\altaffilmark{2}}
\altaffiltext{1}{Astronomical Observatory of the Jagiellonian University, ul. Orla 171, 30-244 Krak\'ow, Poland}
\altaffiltext{2}{Astronomical Observatory of the University of Warsaw, Al. Ujazdowskie 4, 00-478 Warszawa, Poland} 
\email{n.zywucka@oa.uj.edu.pl}

\begin{abstract}

We report the identification of blazar candidates behind the Magellanic Clouds. The objects were selected from the Magellanic Quasars Survey (MQS), which targeted the entire Large Magellanic Cloud (LMC) and 70\% of the Small Magellanic Cloud (SMC). Among the 758 MQS quasars and 898 of unidentified (featureless spectra) objects, we identified a sample of 44 blazar candidates, including 27 flat spectrum radio quasars and 17 BL Lacertae objects, respectively. All the blazar candidates from our sample were identified with respect to their radio, optical, and mid-infrared properties. The newly selected blazar candidates possess the long-term, multi-colour photometric data from the Optical Gravitational Lensing Experiment, multi-colour mid-infrared observations, and archival radio data for one frequency at least. In addition, for nine of them the radio polarization data are available. With such data, these objects can be used to study the physics behind the blazar variability detected in the optical and mid-infrared bands, as a tool to investigate magnetic field geometry of the LMC and SMC, and as an exemplary sample of point like sources most likely detectable in $\gamma$-ray range with the newly emerging Cherenkov Telescope Array.
\end{abstract}

\keywords{galaxies: active --- galaxies: jets --- BL Lacertae objects: general --- Magellanic Clouds --- quasars: general --- radio continuum: galaxies}

\section{Introduction}

Highly polarized \citep[the degree of optical linear polarization PD$_{\rm{o}}$ $>$ 3\%,][]{ange80} flat-spectrum radio quasars (FSRQs) and BL Lacertae objects (BL Lacs) constitute a class of active galactic nuclei (AGN) called ``blazars'', whose total radiative energy output is dominated by the Doppler-boosted, non-thermal emission from relativistic jets launched by accreting supermassive black holes from the centers of massive elliptical galaxies \citep[e.g.,][]{ange80,bege84,urry95}. Blazars of the FSRQ type exhibit in addition prominent emission lines in their optical spectra due to the thermal plasma contribution; such lines are weak or even absent in the BL Lac subclass. The broad-band spectral energy distribution (SED) of blazar sources is characterized by two prominent peaks where, in the framework of the \textit{leptonic} scenario, the radio-to-optical/X-ray segment is believed to originate from the synchrotron radiation of electron-position pairs accelerated up to TeV energies, while the high frequency X-ray-to-$\gamma$-ray segment is due to the Inverse Compton scattering of various ambient photon fields (produced both internally and externally to the outflow) by the jet electrons \citep{ghis98}. Alternatively, the hadronic scenario explains the broad-band contiuum emission of blazars by assuming that protons, accelerated to ultra high energies ($\geq1$ EeV), produce $\gamma$-rays via either the direct synchrotron emission or a meson decay and the synchrotron emission of secondaries from proton-photon interactions \citep[e.g.,][]{bott13}. 

Based on the position of the synchrotron peak frequency, $\nu_{\rm{peak}}$ in the $\nu$F$_{\nu}$ plane, where F$_{\nu}$ is the observed energy flux spectral density, BL Lacs can be further divided into low frequency peaked BL Lacs (LBLs; $\nu_{\rm{peak}}$ $\lesssim$ 10$^{14}$ Hz), intermediate frequency peaked BL Lacs (IBLs; 10$^{14}$ $\lesssim$ $\nu_{\rm{peak}}$ $\lesssim$ 10$^{15}$ Hz), and high frequency peaked BL Lacs \citep[HBLs; $\nu_{\rm{peak}}$ $\gtrsim$ 10$^{15}$ Hz;][]{abdo10a}. FSRQs are characterized by $\nu_{\rm{peak}}$ $<$ 10$^{14}$ Hz but larger peak intensity ratio between the Inverse Compton and synchrotron emission components (aka the ``Compton dominance'') when compared with the LBLs. The so called \textit{blazar sequence} formed by the decreasing $\nu_{\rm{peak}}$ with the increasing Compton dominance, was proposed to result from stronger and stronger radiative cooling suffered by the emitting electrons due to the increasing energy density of the seed photon population for the Inverse Compton scattering \citep[e.g.,][]{ghis98,foss98,ghis08}.

Blazars are the ideal candidates to test the extreme physics of relativistic jets, due to their large observed luminosities $\lesssim$10$^{48}$ ergs s$^{-1}$ \citep{ghis14} and high-amplitude intensity changes on various timescales from years down to minutes \citep{cui04,ahar07,acke16b}, as well as to examine their evolution and large scale structure formation \citep{acke17}. Blazars show the most extreme behavior in the radio-optical polarization variability on different timescales, ranging from years to minutes, among all the AGN. The polarization at radio frequencies is weaker than in the optical bands with the characteristic radio linear polarization degree PD$_{\rm{r,1.4}}$ $>$ 1\% at 1.4 GHz \citep{iler97}. \citet{fan08} showed that the polarization is linked to the amount of relativistic beaming, that the averaged polarization increases with the radio frequency for FSRQs, and that, statistically, radio-selected BL Lacs are more polarized than FSRQs. Polarimetric observations can be used as a tool to study the location of the emission regions in blazar jets \citep[e.g.,][]{jors07,blin09,zhan16}, to constrain the structure of the jet magnetic field \citep[e.g.,][]{alle03,mars14}, and to probe the interstellar medium of foreground galaxies via the Faraday rotation measure method \citep[RM; e.g.,][] {gaen05, mao12}.

In the present study, we report the identification of blazar candidates behind the LMC and SMC which are selected from the catalogue of the Magellanic Quasar Survey \citep[MQS;][]{kozl13} and a ``featureless spectra'' (FS) objects list (Section~\ref{InSamSel}). In Section~\ref{dataset} we shortly describe the optical, mid-IR, and radio data sets analyzed, while Section~\ref{iden_proc} contains details of the selection and identification procedures. The main results of the studies are discussed in the final Section~\ref{discussion}.

\section{Input sample selection}\label{InSamSel}

The AGN identification based on the optical colour selection is extremely challenging in dense stellar fields such as the Galactic plane or LMC and SMC, due to large densities of stars ($\sim$10$^{6}$ deg$^{-2}$) in the interstellar medium as compared to quasars ($\sim$25 deg $^{-2}$) from the background, for a limiting magnitude I $<$ 20. With the main goal of studying the optical variability of quasar light curves based on the Optical Gravitational lensing Experiment \textit{(OGLE)}-III phase \citep{udal08a, udal08b}, the MQS survey was designed to increase the number of the identified AGN behind both Magellanic Clouds (MCs). This survey covers 42 deg$^{2}$ in the sky and completed 100\% and 70\% of its planned LMC and SMC areas, respectively. Below, we briefly recall the step-by-step AGN selection procedure that led to the MQS input sample \citep[see,][]{kozl09, kozl10, kozl11, kozl12, kozl13}.

\subsection{The MQS catalogue} \label{mqs}

The MQS quasar candidates were identified in the four-step selection procedure which involves: (i) mid-infrared (mid-IR) properties, (ii) optical variability, (iii) ROSAT X-ray sources’ counterparts, and (vi) optical spectroscopy study:
\begin{enumerate}[(i)]
\item \citet{kozl09} identified 4,699 and 657 quasar candidates in the LMC and SMC, respectively. These quasar candidates were selected based on the mid-IR and optical colour-colour selection by cross-matching the mid-IR data from the Spitzer SAGE Survey data \citep{meix06} and the S3MC Survey \citep{bola07}, and the OGLE-III optical survey. As a result, mid-IR colour-colour, mid-IR colour-magnitude, and mid-IR-to-optical colour planes were created in order to separate AGN from stars, young stellar objects (YSOs), planetary nebulae (PNe), and other galactic and extragalactic sources. The selected sample was further divided into the QSO and YSO groups and subgroups. All of them may contain quasars, as well as other sources, but only the QSO-Aa group is considered to include the highest number of AGN, while the other groups are mostly populated by contaminating sources \citep{kozl09}.
\item The OGLE-III light curves were analyzed using the damped random walk (DRW) model \citep{kell09, kozl10} and the structure function (SF) method \citep{kozl16}. The DRW model fits the light curve as a stochastic process with exponential covariance matrix characterized by the two parameters: a damping time scale $\tau$ and an amplitude of the driving Gaussian noise $\sigma$. \citet{kozl10} showed that quasars occupy a well defined locus in the $\sigma$-$\tau$ space, where over 1,000 quasar candidates were selected for the MQS input sample.
\item The LMC and SMC OGLE variable source positions were cross-matched with the ROSAT catalogues \citep{habe99, habe00} to find any X-ray counterparts at the level of 3$\sigma$. As a result, a small sample of 205 X-ray detected candidates was selected \citep{kozl12}.
\item The final confirmation whether an object is a quasar or not was based on spectroscopic observations of 3,014 sources, including 2,248 behind the LMC and 766 behind the SMC, selected according to at least \emph{one} of the (i)--(iii) methods. The spectroscopic data were obtained using the 3.9 m Anglo-Australian Telescope (AAT) and the AAOmega spectrograph. Only the sources with at least two visible emission lines in each spectrum, except for the objects at the redshifts between 0.7 and 1.2 (where only Mg II line is visible), were used for the AGN identification. In such a manner, \citet{kozl13} confirmed spectroscopically 565 quasars in the LMC and 193 quasars in the SMC, totalling 758 sources in the MQS catalogue.
\end{enumerate}
Prior to the MQS survey, only 66 quasars were known behind the MCs. Owing to all the aforementioned selection criteria, 758 quasars were selected from the OGLE data, including 713 newly identified quasars: 565 (within which 527 are new) and 193 (186), behind the LMC and SMC, respectively. For all spectroscopically observed MQS objects there exist long-term, densely sampled optical light curves from the OGLE survey.

\subsection{The FS list} \label{fs}
The spectroscopic observations also resulted in 898 objects (669 in the LMC and 229 in the SMC fields) without any distinguishing features in their spectra, either intrisically or due to a low S/N statistics for the line detection. These objects could be associated with either BL Lac type blazars, or OB stars, YSOs, PNe, and other types of Galactic or extragalactic objects.

\begin{deluxetable*}{cccccc}[th!]
\tabletypesize{\footnotesize}
\tablecolumns{6}
\tablewidth{0pt}
\tablecaption{Summary of radio catalogues and surveys used in this study. \label{RadioSurvs}}
\tablehead{
\colhead{Survey} & \colhead{Telescope} & \colhead{Frequency} & \colhead{Area} & \colhead{beam size} & \colhead{rms} \\ 
& & [GHz] &  & [\arcsec] & [mJy/beam] \\ 
\tiny{(1)} & \tiny{(2)} & \tiny{(3)} & \tiny{(4)} & \tiny{(5)} &  \tiny{(6)}
}
\startdata
SUMSS$^{\rm{1}}$ & MOST & 0.843 & $\delta$ $\leq$ $-$30 & 43 $\times$ 43 csc$\mid\delta\mid$ & $1 \div 2$ \\     
AT20G$^{\rm{2}}$ & ATCA & 5, 8, 20 & $\delta$ $\leq$ $-$15 & 9.9, 5.5, 2.4 & $0.1 \div 0.5$ \\ 
PMN$^{\rm{3}}$ & Parkes 64 m & 4.85 & $\delta$ $\leq$ 10 & 252 & $5 \div 7$ \\
ATPMN$^{\rm{4}}$ & ATCA & 4.8, 6.8 & $-$87 $<$ $\delta$ $<$ $-$38 & 1.8, 1.2 & $15 \div 20$ \\ 
SMC multi$^{\rm{5}}$ & ATCA and Parkes & $1.42 \div 8.64$ & 100\% of SMC & $98 \div 15$ & $1.8 \div 0.4$  \\
LMC dual$^{\rm{6}}$ & ATCA & 4.8, 8.6 & 100\% of LMC & 33, 20 & 0.1 \\
SMC dual$^{\rm{7}}$ & ATCA & 4.8, 8.6 & 100\% of SMC & 35, 22 & 0.15 
\enddata
\tablecomments{$^{\rm{1}}$\citealp{bock99};~$^{\rm{2}}$\citealp{murp10};~$^{\rm{3}}$\citealp{grif93}; $^{\rm{4}}$\citealp{mcco12};~$^{\rm{5}}$\citealp{fili02};~$^{\rm{6}}$ \citealp{dick05};~$^{\rm{7}}$\citealp{dick10} \\Columns: (1) name of sky survey; (2) telescope used; (3) central frequency of observations; (4) sky coverage; (5) typical resolution (beam size) achieved; (6) typical noise level.} 
\end{deluxetable*}

\section{Multiwavelength observations} \label{dataset}

\subsection{Mid-IR data}\label{MidIRdata}

\subsubsection{Wide Infrared Survey Explorer} \label{wise}
 
The Wide-field Infrared Survey Explorer (WISE) is a satellite launched in December 2009 and designed to map the entire sky and to monitor individual astrophysical objects in mid-IR range. This 40\,cm diameter telescope is equipped with IR detectors collecting data in four bands centered at 3.4 (W1), 4.6 (W2), 12 (W3), and 22 (W4) $\mu$m with the angular resolutions of 6\farcs1, 6\farcs4, 6\farcs5, and 12\farcs0, respectively \citep{wrig10}.

\subsubsection{SAGE LMC and SMC IRAC Source Catalogues} \label{irac}

Both the LMC and SMC regions were observed using the four-band Infrared Array Camera (IRAC) and the three-band Multiband Imaging Photometer for SIRTE (MIPS) instruments, on-board the Spitzer Space Telescope \citep{wern04}, as a part of the program Surveying the Agents of a Galaxy’s Evolution \citep[SAGE\footnote{\url{http://sage.stsci.edu/}}][]{meix06,gord11}. The angular resolution is 1\farcs7, 1\farcs7, 1\farcs9, and 2\farcs0 in the IRAC bands, centered at 3.6, 4.5, 5.8, and 8.0 $\mu$m, respectively, while it is 6\farcs0, 18\farcs0, and 40\farcs0 in the MIPS bands centered at 24, 70, and 160 $\mu$m, respectively. In our study, we have used the simultaneous IRAC data only, in order to minimize uncertainties of the selection procedure.
 
\subsection{Radio continuum surveys} \label{radioData}

Self-absorbed radio cores and the high degree of optical polarization are the defining characteritics of blazars \citep[e.g.,][]{blan79}. Therefore, in order to ascertain the blazar nature of sources from the MQS and the FS lists, we searched for radio counterparts of the optically detected sources in various radio sky surveys and catalogues. Description of the sky surveys and catalogues are given below and their main characteristics are listed in Table~\ref{RadioSurvs}.

\subsubsection{The SUMSS catalogue} \label{sumss}

The SUMSS sky survey \citep{bock99} was performed with the Molonglo Observatory Synthesis Telescope (MOST) and covered 2.47 sr of the southern sky with $\delta$ $\leq$ $-$30\dg\, at the frequency of 0.843 GHz. The observations were completed in 2007 and are publicly available \footnote{\url{http://www.physics.usyd.edu.au/sifa/Main/SUMSS}}. The SUMSS catalogue consists of 210,412 radio sources with the peak brightness limit of 6 mJy/beam at $\delta$ $\leq$ $-$50\dg\, and 10 mJy/beam at $\delta$ $>$ $-$50\dg\,. The synthesized beam of the maps is 43\arcsec $\times$ 43\arcsec cosec$|\delta|$. The position accuracy of the radio sources in the catalogue are within 1$-$2\arcsec\, for the objects with the peak fluxes $\geq$ 20\,mJy/beam, while for the remaining sources with lower radio fluxes, the position accuracy is not worse than 10\arcsec.

\subsubsection{The AT20G catalogue} \label{atca}

The AT20G sky survey \citep{murp10} was conducted between 2004 and 2008 with the Australia Telescope Compact Array (ATCA) and covered 6.12 sr of the southern sky ($\delta$ $<$ 0\dg\,). The AT20G catalogue contains 5,890 radio sources with the flux-density limit of 40 mJy. Near-simultaneous measurements at 5 and 8 GHz south of $\delta$ $<$ $-$15\dg\, were also performed. The synthesized beam of this sky survey is 2\farcs4\,, 5\farcs5\,, and 9\farcs9\, at the frequencies of 20, 8, and 5 GHz, respectively. In addition, the total polarized intensity was measured and collected in the catalogue for 1,559 sources at one up to three frequencies.

\subsubsection{The PMN catalogue} \label{pmn}

The PMN sky survey \citep{grif93,wrig94} of the southern hemisphere was performed in 1990 with the Parkes 64 m telescope at 4.85 GHz and covered the sky region of 4.51 sr of $-$87\fdg5 $<$ $\delta$ $<$ +10\dg. The radio survey was conducted in four areas and divided into the following subparts: Southern with the declination of $-$87\fdg5 $<$ $\delta$ $<$ $-$37\dg\,, Zenith with $-$37\dg $<$ $\delta$ $<$ $-$29\dg\,, Tropical $-$29\dg $<$ $\delta$ $<$ $-$9\fdg5\,, and Equatorial with $-$9\fdg5 $<$ $\delta$ $<$ +10\dg. The resolution of the survey is 252\arcsec. The PMN catalogue consists of 50,814 radio sources including 23,277 sources of the Southern Survey. The flux density limit varies between 50 mJy/beam at the northern edge ($\delta$ $\approx$ $-$37\dg\,) and 20 mJy/beam at the southern edge of this region. 

\begin{figure*}[h!]
\begin{center}
\includegraphics[angle=0,scale=0.65]{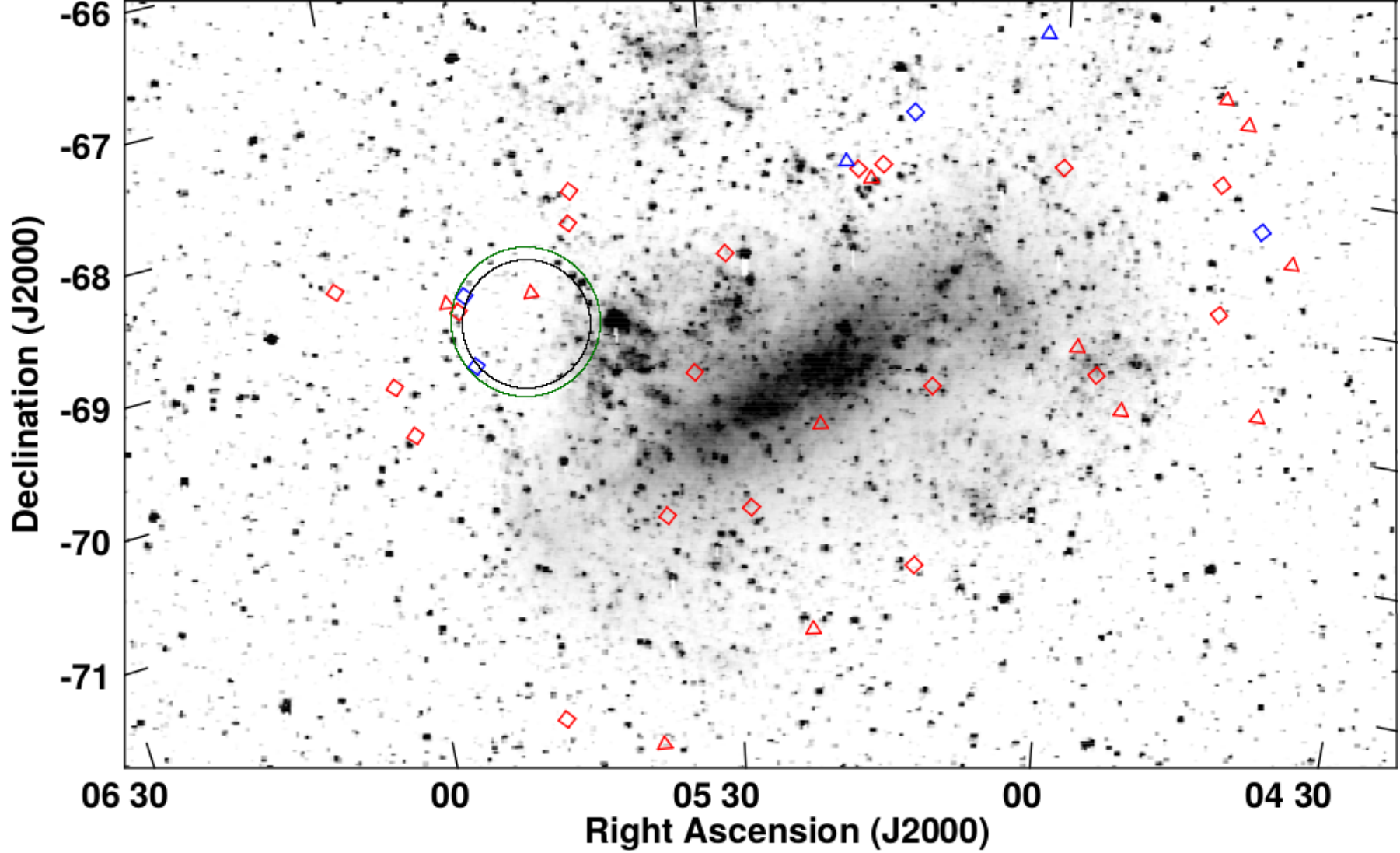}
\includegraphics[angle=0,scale=0.65]{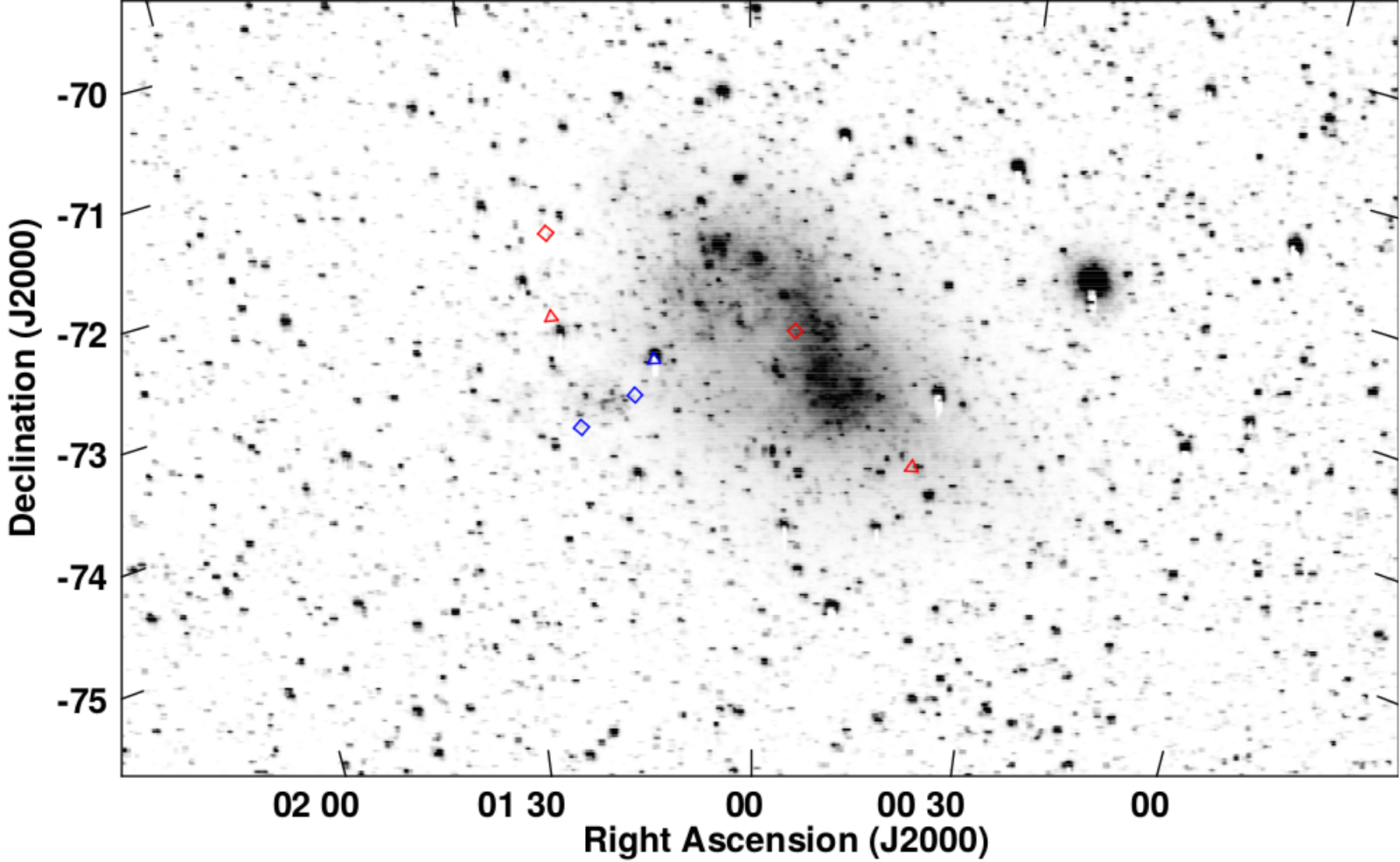}
\caption{Distribution of the selected blazar candidates behind the LMC (top panel) and SMC (bottom panel). The optical positions of the FSRQ candidates are shown with red diamond symbols, while red triangular symbols denote the BL Lac candidates. Objects with polarimetric measurements are additionally marked with blue symbols. Black and green circles show possible areas of the $\gamma$-ray flaring activity detected by the \textit{Fermi}-LAT during the 4th month of observations and, later, in April 2015 (see, Section~\ref{HE}). The optical images of both MC (grey scale) are taken from \citet{both88}.\label{clouds}} 
\end{center}
\end{figure*}

\subsubsection{The ATPMN catalogue} \label{atpmn}

The ATPMN follow-up catalogue \citep{mcco12} is a result of observations of PMN sources with the ATCA. The data were carried out at 4.8 and 8.6 GHz and covered the region of 2.25 sr ($-$87\dg $<$ $\delta$ $<$ $-$38\fdg5\,). The ATPMN catalogue consists of 9,040 sources with the flux density of 70 mJy/beam at $\delta$ $>$ $-$73\dg and 50 mJy/beam at $\delta$ $\leq$ $-$73\dg. The angular resolution of the ATPMN radio survey is high and varies between 1\farcs2\, and 1\farcs8.

\subsubsection{The ATCA/Parkes SMC survey} \label{filip}

The ATCA radio-contiunuum study of the SMC was performed by \citet{fili02} based on the observations by the ATCA and the Parkes 64 m telescopes. The catalogues contain 717 radio sources detected in the SMC at 1.42, 2.37, 4.8, and 8.64 GHz. The beam size varies from 98\arcsec\, at 1.42 GHz to 15\arcsec\, at 8.64 GHz, and the positional accuracy is less than 1\arcsec.

\subsubsection{The 4.8 and 8.6 GHz Survey of the MCs} \label{MCsSurv}

The 4.8 and 8.6 GHz Surveys of LMC \citep{dick05} and SMC \citep{dick10} were performed using the ATCA and the Parkes 64~m telescopes. The work was intended to study a diffuse emission in the Galaxy, supernova remnants and HII regions, as well as bright point sources. The simultaneous observations of the total intensity and the polarized flux density at 4.8 and 8.6 GHz were gathered to produce 6\degr $\times$ 6\degr maps with the resolutions of 33\arcsec\, and 20\arcsec\, at 4.8 and 8.6 GHz, respectively, for the LMC, and 4\fdg5 $\times$ 4\fdg5\, maps with the resolutions of 35\arcsec and 22\arcsec at 4.8 and 8.6 GHz, respectively, for the SMC. In our work, we have used the maps obtained with the ATCA array only, since the Parkes 64 m telescope is suitable for extended radio emission. Since no catalogue has resulted from these surveys, we have measured the total intensity and polarized fluxes directly from the maps provided by \citet{dick05}\footnote{\url{http://www.atnf.csiro.au/research/lmc\_ctm/index.html}} and \citet{dick10}\footnote{\url{http://www.atnf.csiro.au/research/smc\_ctm/index.html}}.

\section{Identification of blazar candidates} \label{iden_proc}

As stated earlier, the MQS catalogue contains 758 quasars. Statistically, around 10$-$15\% of them (i.e.,  75$-$110) should be ``radio-loud'', meaning the ratio of radio 5 GHz to optical B-band flux spectral densities $R \equiv$ F$_{5\rm{GHz}}$/F$_{B}$ $\geq$ 10 \citep{kell89,stoc92,ho02}. Among those radio-loud quasars, at least \emph{some} \citep[$\lesssim 10\%$; see, e.g.,][]{Ghisellini13} should be blazars of the FSRQ type. The FS list, on the other hand, may contain BL Lac type blazars, which should be radio-loud as well.

The blazar candidates from both lists are, therefore, selected using the radio and optical positioning, i.e., the cross-matching procedure. The identified sample of blazar candidates is further analyzed by investigating characteristic radio, mid-IR, and optical properties of the sources, such as spectral radio and mid-IR indices, radio-loudness parameter, and fractional linear polarization.

\subsection{Cross-matching of source positions} \label{crosm}

We searched for the radio counterparts by cross-matching the optical positions of the MQS and FS sources with the positions of radio sources listed in the radio catalogues with the standard spherical trigonometric distance procedure. In particular, we defined the radius $\Delta$r of the error circle as:
\begin{equation} 
\Delta r = \sqrt{\left(\frac{1}{3} \, r_{\rm{b_1}}\right)^2 + \left(\frac{1}{3}  \, r_{\rm{b_2}}\right)^2} 
\end{equation} 
where $r_{\rm{b_1}}$ is the resolution of the OGLE catalogue and $r_{\rm{b_2}}$ is the angular resolution of various radio catalogues considered. With such, the radio counterpart was considered to be coincident with the optical source if their spherical distance was found to be equal or less than $\Delta r$.

Out of the 758 AGN in the MQS catalogue, 17 were found to have measured radio fluxes at more than one radio frequency, 10 at one frequency only, and the remaining 731 quasars could not be matched with any radio conterparts. Similarly, out of the 898 sources in the FS list, seven were found to have measured radio fluxes at more than one frequency, 10 at one frequency, while 881 sources have no radio countreparts. All the sources identified in the cross-matching procedure along with their radio fluxes are reported in Table~\ref{blazarFluxes}. These are termed hereafter as ``blazar candidates'' and are further sumarized in Table~\ref{blazarlist}. Figure~\ref{clouds} presents the overlays of the optical images of the both MCs with the positions of blazar candidates. 

\subsection{Radio indices, radio-loudness, and redshifts} \label{params}

For the newly selected blazar candidates, we estimate the radio spectral indices $\alpha_r$, defined here as $F_{\nu} \propto \nu^{-\alpha_r}$. The radio spectral index value indicates whether a radio spectrum of a source is flat ($\alpha_r$ $<$ 0.5) or steep ($\alpha_r$ $>$ 0.5). Flat radio spectra are the defining characteristics of blazars, as mentioned previously. It is noteworthy here that blazars are extremely variable objects, hence simultaneous observations are needed to calculate the robust values of radio spectral indices. In this work, we are, however, forced to rely on the $\alpha_{r}$ values estimated based on non-simultaneous data. The radio indices were calculated for the selected objects, which possess at least two measurements at different frequencies available in the aforementioned data sets. The radio flux spectral densities at different frequencies and resulting $\alpha_r$ values are presented in Table~\ref{blazarFluxes} and Table~\ref{blazarlist}, respectively. 

\begin{figure}[!th]
\includegraphics[angle=0,scale=1.0]{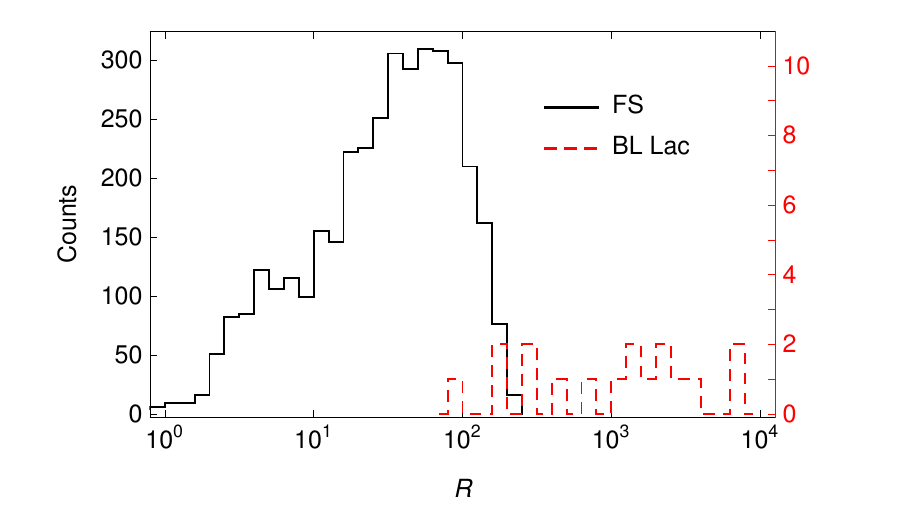}
\includegraphics[angle=0,scale=1.0]{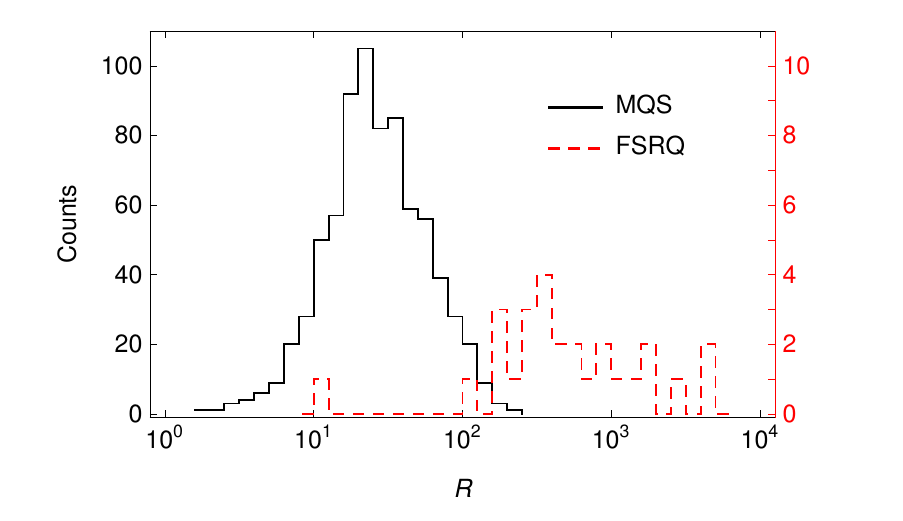}
\caption{Radio-loudness distributions of the FSRQ (upper panel) and the BL Lac (bottom panel) type blazar candidates, shown by red dashed lines. Black solid lines correspond to the 5\,GHz ATCA upper limits for the sources catalogued in the MQS or the FS lists.} \label{radioLoud} 
\end{figure}

We have also estimated the values of the radio-loudness parameter $R$ for our blazar candidates, as listed in Table~\ref{blazarlist}. Since the radio data at exact 5 GHz were available for four sources only, i.e., two FSRQ and two BL Lac blazar candidates, when estimating the $R$ values, we have extrapolated radio fluxes from lower frequencies assuming the radio spectral index as $\alpha_{r}$ = 0.5. The $R$ distributions for the blazar candidates (red dashed line) are shown in Figure~\ref{radioLoud}. For comparison, we have also included in the histograms the upper limits on $R$ (black solid line) estimated for the remaining objects from both lists based on the 5\,GHz ATCA flux upper limit.

\begin{deluxetable*}{lcccccccccccc}
\tabletypesize{\scriptsize}
\tablecolumns{13}
\tablewidth{0pt}
\tablecaption{The radio fluxes of FSRQ type blazar candidates. \label{blazarFluxes}}
\tablehead{
\multicolumn{1}{c}{Object} & \colhead{F$_{\textnormal{0.84}}$} & \colhead{F$_{\textnormal{1.42}}$} & \colhead{F$_{\textnormal{2.37}}$} & \colhead{F$_{\textnormal{4.8}}$} & \colhead{F$_{\textnormal{4.85}}$} & \colhead{F$_{\textnormal{5.0}}$} & \colhead{F$_{\textnormal{8.0}}$} & \colhead{F$_{\textnormal{8.6}}$} & \colhead{F$_{\textnormal{20.0}}$} \\ 
  & [mJy] & [mJy] & [mJy] & [mJy] & [mJy] & [mJy] & [mJy] & [mJy] & [mJy]\\
\multicolumn{1}{c}{\tiny{(1)}} & \tiny{(2)} & \tiny{(3)} & \tiny{(4)} & \tiny{(5)} & \tiny{(6)} & \tiny{(7)} & \tiny{(8)} & \tiny{(9)} & \tiny{(10)}
}
\startdata
\multicolumn{10}{c}{FSRQ blazar type candidates}\\\hline
J0054$-$7248$^{\rm{a}}$ & & & & & 36.0$\pm$6.0 & & & & \\     
J0114$-$7320 & 129.0$\pm$4.0 & 86.8$\pm$4.3 & 70.3$\pm$3.5 & 19.0$\pm$7.0 & & & & 15.0$\pm$10.0 & \\
 & & & & 18.5$\pm$1.1$^{\rm{e}}$ & & & & 17.4$\pm$1.5$^{\rm{e}}$ & \\
 & & & & 288.0$\pm$14.4 & & & & 22.0$\pm$1.1 & \\
J0120$-$7334 & 56.2$\pm$3.0 & 45.8$\pm$2.3 & 31.6$\pm$1.6 & 24.0$\pm$1.4$^{\rm{e}}$ & 29.0$\pm$6.0 & & & 15.3$\pm$1.2$^{\rm{e}}$ & \\
J0122$-$7152 & 6.8$\pm$1.0 & & & & & & & & \\ 
J0442$-$6818 & 9.2$\pm$1.0 & & & & 29.0$\pm$7.0 & 51.0$\pm$3.0 & 60.0$\pm$4.0 & & 74.0$\pm$4.0 \\
&&&&&&&&&&&&\\
J0445$-$6859 & 25.3$\pm$1.7 & & & & & & & & \\
J0446$-$6758 & 6.0$\pm$0.8 & & & & & & & & \\
J0455$-$6933 & 80.4$\pm$4.0 & & & 36.5$\pm$2.0$^{\rm{e}}$ & & 17.0$\pm$7.0 & & 27.0$\pm$1.8$^{\rm{e}}$ & \\
J0459$-$6756$^{\rm{a}}$ & & & & & 34.0$\pm$7.0 & & & & \\ 
J0510$-$6941 & 31.2$\pm$1.3 & & & 8.9$\pm$0.8$^{\rm{e}}$ & & & & 6.2$\pm$2.0$^{\rm{e}}$ &  \\
&&&&&&&&&&&&\\
J0512$-$7105 & 11.0$\pm$0.9 & & & 2.8$\pm$0.7$^{\rm{e}}$ & & & & &  \\
J0512$-$6732 & 59.2$\pm$2.2 & & & 45.7$\pm$2.4$^{\rm{e}}$ & & 47.0$\pm$2.0 & 56.0$\pm$3.0 & 48.5$\pm$2.6$^{\rm{e}}$ & 46.0$\pm$3.0 \\
J0515$-$6756$^{\rm{a}}$ & & & & & 27.0$\pm$7.0 & & & & \\
J0517$-$6759$^{\rm{a}}$ & & & & & 29.0$\pm$7.0 & & & & \\ 
J0527$-$7036 & 137.8$\pm$4.2 & & & 34.0$\pm$7.0 & & &  & 11.0$\pm$10.0 & \\
& & & & 39.8$\pm$2.1$^{\rm{e}}$ & & & & 24.7$\pm$1.5$^{\rm{e}}$ & \\  
&&&&&&&&&&&&\\
J0528$-$6836 & 31.1$\pm$1.5 & & & 31.1$\pm$1.7$^{\rm{e}}$ & 49.0$\pm$7.0 & & & 22.7$\pm$1.4$^{\rm{e}}$ & \\
J0532$-$6931 & 10.3$\pm$1.3 & & & 12.6$\pm$2.6$^{\rm{e}}$ & & & & & \\
J0535$-$7037$^{\rm{a}}$ & & & & & 35.0$\pm$7.0 & & & & \\ 
J0541$-$6800 & 54.5$\pm$1.9 & & & 45.7$\pm$2.4$^{\rm{e}}$ & 59.0$\pm$ 7.0 & & & 32.3$\pm$1.8$^{\rm{e}}$ &  \\
J0541$-$6815 & 45.8$\pm$1.6 & & & 48.8$\pm$2.5$^{\rm{e}}$ & & & & 38.8$\pm$2.1$^{\rm{e}}$ &  \\
&&&&&&&&&&&&\\
J0547$-$7207 & 24.6$\pm$1.9 & & & 8.0$\pm$0.9$^{\rm{e}}$ & & & & 7.1$\pm$1.8$^{\rm{e}}$ &\\
J0551$-$6916 & & & & 32.9$\pm$1.8$^{\rm{e}}$ & 35.0$\pm$7.0 & & & 14.8$\pm$1.2$^{\rm{e}}$ & \\
J0551$-$6843 & 158.5$\pm$6.4 & & & 32.6$\pm$0.8$^{\rm{e}}$ & & & & 14.0$\pm$1.8$^{\rm{e}}$ & \\
J0552$-$6850$^{\rm{a}}$ & & & & & 49.0$\pm$7.0 & & & &  \\
J0557$-$6944 & 49.0$\pm$1.7 & & & & 40.0$\pm$7.0 & & & &\\
&&&&&&&&&&&&\\
J0559$-$6920$^{\rm{a}}$ & & & & & 27.0$\pm$7.0 & & & &  \\
J0602$-$6830 & 52.0$\pm$1.7 & & & & 40.0$\pm$7.0 & & & &  \\ \hline
\multicolumn{10}{c}{BL Lac type blazar candidates}\\\hline
J0039$-$7356 & 8.5$\pm$1.1 & & 4.2$\pm$0.2 & & & & \\
J0111$-$7302 & 86.4$\pm$4.2 & 111.5$\pm$5.6 & 77.4$\pm$3.9 & 69.0$\pm$7.0 & 72.0$\pm$7.0 & 72.0$\pm$4.0 & 72.0$\pm$4.0 & 61.0$\pm$10.0 & 74.0$\pm$4.0 \\
 & & & & 62.9$\pm$3.2$^{\rm{e}}$ & & & & 53.6$\pm$2.8$^{\rm{e}}$ &  \\
J0123$-$7236 & 10.8$\pm$0.9 & & 4.6$\pm$0.2 & & & \\
J0439$-$6832 & 36.2$\pm$1.4 & & & & & & &\\
J0441$-$6945 & 59.4$\pm$3.0 & & & & & & &\\
&&&&&&&&&&&&\\
J0444$-$6729 & 34.3$\pm$1.3 & & & & & & &\\
J0446$-$6718 & 9.9$\pm$1.1 & & & & & & &\\
J0453$-$6949 & 6.2$\pm$0.9 & & & & & & & \\ 
J0457$-$6920$^{\rm{a}}$ & & & & & 30.0$\pm$7.0 & & &  \\
J0501$-$6653 & 51.7$\pm$1.8 & & & 17.7$\pm$1.1$^{\rm{e}}$ & & & & 9.7$\pm$1.0$^{\rm{e}}$ &\\
&&&&&&&&&&&&\\
J0516$-$6803$^{\rm{a}}$ & & & & 29.0$\pm$7.0 & & & &  \\  
J0518$-$6755$^{\rm{a}}$ & & & & 67.71$\pm$3.5$^{\rm{e}}$ & 95.0$\pm$8.0 & & & 33.40$\pm$2.0$^{\rm{e}}$ & \\
J0521$-$6959$^{\rm{a}}$ & & & & 52.6$\pm$2.7$^{\rm{e}}$ & 58.0$\pm$7.0 & & & 29.3$\pm$2.0$^{\rm{e}}$ & \\
J0522$-$7135$^{\rm{a}}$ & & & & & 89.0$\pm$8.0 & & & \\
J0538$-$7225 & 7.0$\pm$1.0 & & & & & & \\
&&&&&&&&&&&&\\
J0545$-$6846 & 176.3$\pm$7.4 & & & 41.7$\pm$2.2$^{\rm{e}}$ & 26.0$\pm$7.0 & & & 23.1$\pm$1.7$^{\rm{e}}$ &  \\
J0553$-$6845 & 22.2$\pm$1.1 & & & & & & &
\enddata
\tablecomments{$^{\rm{a}}$Dubious objects; Columns: (1) source designation, (2) - (10) radio flux density at 0.843, 1.42, 2.37, 4.8, 4.85, 5.0, 8.0, 8.6, and 20.0 GHz, respectively; $^{\rm{e}}$ radio flux measurements based on the 4.8 and 8.6 GHz radio surveys of the MCs.}
\end{deluxetable*}

\begin{deluxetable*}{lccccccc}
\tabletypesize{\footnotesize}
\tablecolumns{8}
\tablewidth{0pt}
\tablecaption{Newly identified FSRQ and BL Lac type blazar candidates. \label{blazarlist}}
\tablehead{
\multicolumn{1}{c}{Object} & \colhead{RA} & \colhead{DEC} & \colhead{R} & \colhead{z} & \colhead{$\alpha_{\textnormal{r}}$} & \colhead{$\alpha_{\textnormal{IR}}$} & \colhead{I [mag]} \\ 
\multicolumn{1}{c}{\tiny{(1)}} & \tiny{(2)} & \tiny{(3)} & \tiny{(4)} & \tiny{(5)} & \tiny{(6)} & \tiny{(7)} & \tiny{(8)}
}
\startdata
\multicolumn{7}{c}{FSRQ type blazar candidates}\\\hline
J0054$-$7248$^{\rm{a}}$ & 00 54 44.70  & -72 48 13.68 & 1730 & 0.505 &  & 1.89$\pm$0.50 & 20.73\\     
J0114$-$7320 & 01 14 05.57 & -73 20 06.50 & 246 & 0.937 & 0.58$\pm$0.31 & 1.33$\pm$0.13 & 18.26\\
J0120$-$7334 & 01 20 56.05 & -73 34 53.51 & 195 & 1.565 & 0.56$\pm$0.06 & 1.88$\pm$0.07 & 17.66\\
J0122$-$7152 & 01 22 58.49 & -71 52 07.00 & 267 & 0.939 & & 0.47$\pm$0.18 & 19.84\\ 
J0442$-$6818 & 04 42 45.19 & -68 18 38.99 & 371 & 0.964 & -0.57$\pm$0.15 & 1.20$\pm$0.09 & 18.22 \\
&&&&&&\\
J0445$-$6859 & 04 45 36.60 & -68 59 46.10 & 285 & 1.714 & & 1.73$\pm$0.02 & 18.87\\
J0446$-$6758 & 04 46 33.91 & -67 58 55.88 & 169 & 1.301 & & 1.72$\pm$0.15 & 19.96\\
J0455$-$6933 & 04 55 59.10 & -69 33 29.09 & 336 & 1.319 & 0.47$\pm$0.04 & 1.30$\pm$0.06 & 18.78\\
J0459$-$6756$^{\rm{a}}$ & 04 59 54.27 & -67 56 35.59 & 898 & 1.687 & & 1.45$\pm$0.14 & 19.29\\ 
J0510$-$6941 & 05 10 45.85 & -69 41 26.48 & 165 & 1.061 & 0.72$\pm$0.01 & 0.60$\pm$0.02 & 18.37\\
&&&&&&\\
J0512$-$7105 & 05 12 21.49 & -71 05 55.61 & 489 & 0.286 & 0.79$\pm$0.00 & 1.71$\pm$0.03 & 20.70\\
J0512$-$6732 & 05 12 22.48 & -67 32 20.00 & 557 & 2.557 & 0.08$\pm$0.04 & 1.99$\pm$0.09 & 18.37\\
J0515$-$6756$^{\rm{a}}$ & 05 15 03.49 & -67 56 53.02 & 4440 & 0.374 & & 1.83$\pm$0.05 & 20.20 \\
J0517$-$6759$^{\rm{a}}$ & 05 17 10.31 & -67 59 01.21 & 4450 & 0.427 & & 1.01$\pm$0.19 & 20.42\\ 
J0527$-$7036 & 05 27 49.08 & -70 36 41.69 & 769 & 0.774 & 0.73$\pm$0.02 & 1.13$\pm$0.05 & 18.39\\  
&&&&&&\\ 
J0528$-$6836 & 05 28 47.51 & -68 36 20.99 & 580 & 3.320 & 0.08$\pm$0.11 & 0.49$\pm$0.18 & 18.32\\
J0532$-$6931 & 05 32 12.24 & -69 31 30.90 & 12 & 1.353 & -0.12$\pm$0.00 & 1.39$\pm$0.04 & 16.54\\
J0535$-$7037$^{\rm{a}}$ & 05 35 47.71 & -70 37 50.70 & 2670 & 0.731 & & 1.49$\pm$0.12 & 20.34 \\ 
J0541$-$6800 & 05 41 34.97 & -68 00 40.61 & 494 & 0.934 & 0.18$\pm$0.08 & 0.93$\pm$0.06 & 19.64\\
J0541$-$6815 & 05 41 58.96 & -68 15 42.30 & 351 & 1.586 & 0.04$\pm$0.07 & 1.42$\pm$0.08 & 18.59\\
&&&&&&\\
J0547$-$7207 & 05 47 58.68 & -72 07 45.30 & 121 & 0.793 & 0.61$\pm$0.06 & 0.96$\pm$0.08 & 18.79\\
J0551$-$6916 & 05 51 33.21 & -69 16 33.82 & 304 & 2.226 & 1.38$\pm$0.06 & 1.71$\pm$0.07 & 17.93\\
J0551$-$6843 & 05 51 40.55 & -68 43 08.40 & 987 & 1.595 & 1.00$\pm$0.09 & 1.34$\pm$0.05 & 18.39\\
J0552$-$6850$^{\rm{a}}$ & 05 52 22.53 & -68 50 01.72 & 1970 & 1.740 & & 1.57$\pm$0.02 & 19.34\\
J0557$-$6944 & 05 57 59.59 & -69 44 12.19 & 1530 & 0.480 & 0.12$\pm$0.00 & 0.59$\pm$0.09 & 20.14\\
&&&&&&\\
J0559$-$6920$^{\rm{a}}$ & 05 59 01.63 & -69 20 09.38 & 1050 & 1.817 & & 1.66$\pm$0.21 & 19.54\\
J0602$-$6830 & 06 02 34.31 & -68 30 41.40 & 356 & 1.086 & 0.23$\pm$0.00 & 1.37$\pm$0.10 & 18.39\\\hline
\multicolumn{7}{c}{BL Lac type blazar candidates}\\\hline
J0039$-$7356 & 00 39 42.51 & -73 56 15.50 & 1570 & & 0.68$\pm$0.00 & 0.97$\pm$0.12 & 20.85 \\
J0111$-$7302 & 01 11 33.24 & -73 02 03.41 & 171 & & 0.14$\pm$0.05 & & 19.10\\
J0123$-$7236 & 01 23 17.47 & -72 36 05.40 & 199 & & 0.83$\pm$0.00 & 2.04$\pm$0.14 & 20.49\\
J0439$-$6832 & 04 39 50.17 & -68 32 22.60 & 1030 & & & 1.34$\pm$0.15 & 20.21\\
J0441$-$6945 & 04 41 04.99 & -69 45 40.72 & 3840 & & & 1.29$\pm$0.06 & 21.11\\
&&&&&&\\
J0444$-$6729 & 04 44 55.57 & -67 29 40.42 & 689 & & & 0.15$\pm$0.14 & 19.58\\
J0446$-$6718 & 04 46 55.72 & -67 18 38.48 & 478 & & & 1.33$\pm$0.08 & 20.53\\
J0453$-$6949 & 04 53 29.83 & -69 49 28.60 & 285 & & & 1.30$\pm$0.11 & 20.74\\ 
J0457$-$6920$^{\rm{a}}$ & 04 57 50.55 & -69 20 51.61 & 2040 & & & 1.65$\pm$0.44 & 19.93\\ 
J0501$-$6653 & 05 01 39.74 & -66 53 53.48 & 1490 & & 0.66$\pm$0.06 & -0.44$\pm$0.17 & 19.37\\
&&&&&&\\
J0516$-$6803$^{\rm{a}}$ & 05 16 07.14 & -68 03 37.30 & 1750 & & & 3.07$\pm$0.41 & 20.08\\  
J0518$-$6755$^{\rm{a}}$ & 05 18 09.01 & -67 55 23.48 & 89 & & 1.37$\pm$0.34 & 2.00$\pm$0.00 & 18.60\\
J0521$-$6959$^{\rm{a}}$ & 05 21 00.02 & -69 59 06.58 & 2010 & & 1.03$\pm$0.11 & 1.51$\pm$0.24 & 19.18\\
J0522$-$7135$^{\rm{a}}$ & 05 22 23.92 & -71 35 31.30 & 7020 & & & 1.19$\pm$0.15 & 20.08\\
J0538$-$7225 & 05 38 16.96 & -72 25 00.10 & 290 & & & 1.78$\pm$0.18 & 20.32\\
&&&&&&\\
J0545$-$6846 & 05 45 51.78 & -68 46 03.00 & 6900 & & 0.86$\pm$0.04 & 0.81$\pm$0.08 & 21.27\\
J0553$-$6845 & 05 53 15.71 & -68 45 27.20 & 3080 & & & 1.93$\pm$0.10 & 20.00 
\enddata 
\tablecomments{$^{\rm{a}}$Dubious objects; Columns: (1) source designation, (2) right ascension (RA) in J2000.0, (3) declination (DEC) in J2000.0, (4) the radio-loudness $R$ parameter, (5) redshift $z$, (6) radio spectral index $\alpha_{r}$, (7) mid-IR spectral index $\alpha_{IR}$, and (8) I band magnitude. The coordinates, $z$, and I band magnitude were taken from the MQS catalogue or FS list.}
\end{deluxetable*}

 As shown in the figure, our sample of blazar candidates consists of the sources characterized by the highest values of $R$, among both the MQS and FS lists. Note, however, that due to relatively high flux limits of the radio surveys considered (when compared to the optical flux limit of the OGLE project), many quasars from the MQS catalogue which have not been matched here with any radio counterpart, could still be radio-loud, as the upper limits for the $R$ parameter for the majority of such sources are rather high, namely $R<100$. Hence, our sample of blazar candidates does not contradict the expectation for $10\% - 15\%$ of the MQS quasars being radio-loud; instead, Figure~\ref{radioLoud} indicates only that our selection procedure picks up $\sim 3.6\%$ of the most radio-loud quasars from the MQS sample, roughly in agreement with what is expected for the beamed (blazar) population in the parent population of quasars.

Figure~\ref{redshift} shows the $z$ distribution of the MQS quasars (black solid line) and the FSRQ blazar candidates (red dashed line). Among $\sim$10 MQS quasars at $z>3$, we have selected one blazar candidate, i.e. J0528-6836 with $z=3.32$. 
\vspace{0.5cm}

\subsection{Comparison with Roma-BZCAT} \label{redshift}

We compared our sample with 1,425 BL Lacs and blazar candidates and 1,909 FSRQs and blazar candidates listed in \textit{the 5th edition of the Roma-BZCAT: Multi-frequency Catalogue of Blazars}\footnote{\url{http://www.asdc.asi.it/bzcat}} \citep[Roma-BZCAT,][]{mass15}. The following parameters from Roma-BZCAT were taken into account in this investigation: the $z$ distribution, radio flux densities at 1.4 GHz from the  Faint Images of the Radio Sky at Twenty-cm (FIRST) or 0.843 GHz from the SUMSS catalogue (if not available in the former), and optical apparent magnitudes in R filter from the tenth Sloan Digital Sky Survey data release (SDSS DR10). 

\begin{figure}[!b]
\includegraphics[angle=0,scale=1.0]{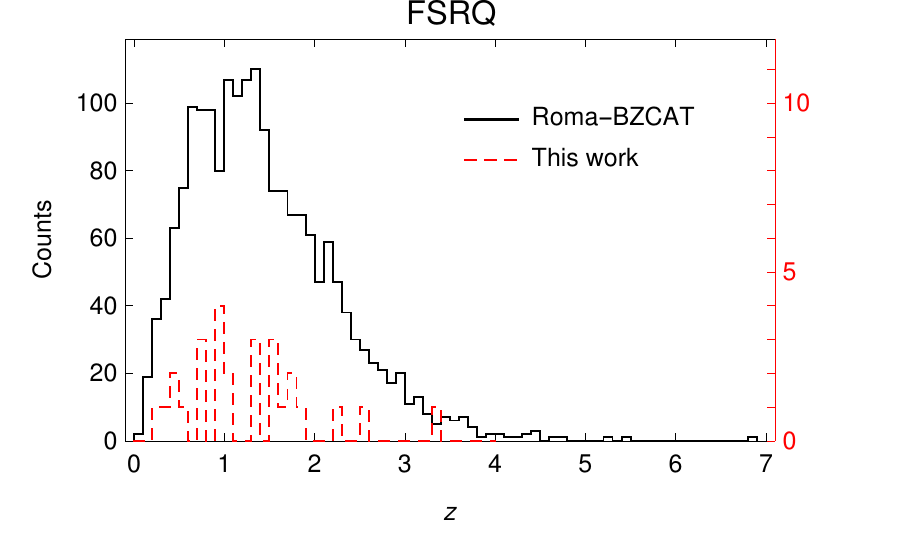}
\caption{Redshift distributions of the FSRQ blazar candidates (red dashed line) and the FSRQ blazars from Roma-BZCAT (black solid line).} \label{redshiftComparison} 
\end{figure}

Figure~\ref{redshiftComparison} shows the $z$ distributions of the OGLE FSRQ candidates (red dashed line) and the Roma-BZCAT FSRQs (black solid line). Note that 29 FSRQs, which are listed in the Roma-BZCAT catalogue with ''?'' symbol are not included in this study due to the uncertain $z$ values. 

\begin{figure}[!th]
\includegraphics[angle=0,scale=1.0]{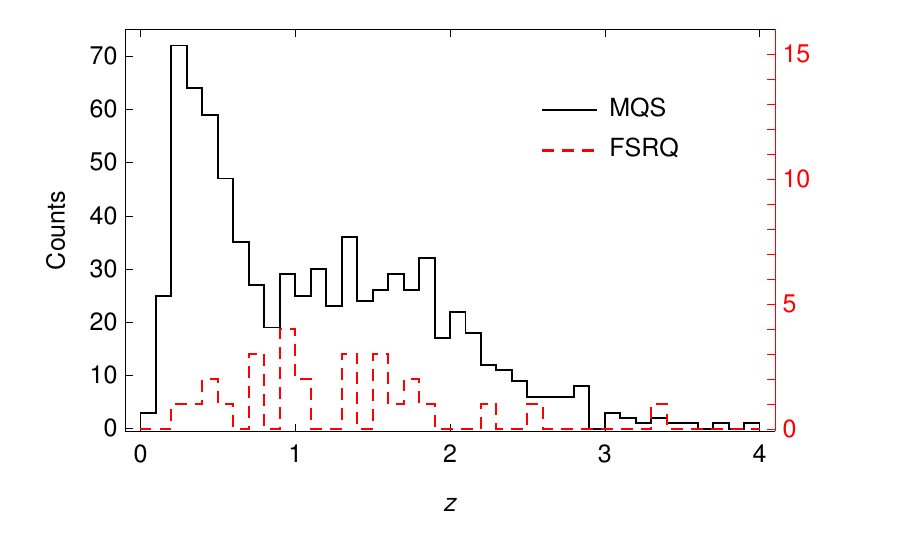}
\caption{Redshift distributions of the MQS quasars. Black solid line shows the distribution of the quasars lacking radio counterparts, while red dashed line is for the FSRQ blazar candidates.} \label{redshift}
\end{figure}

\begin{figure}[!t]
\includegraphics[angle=0,scale=1.0]{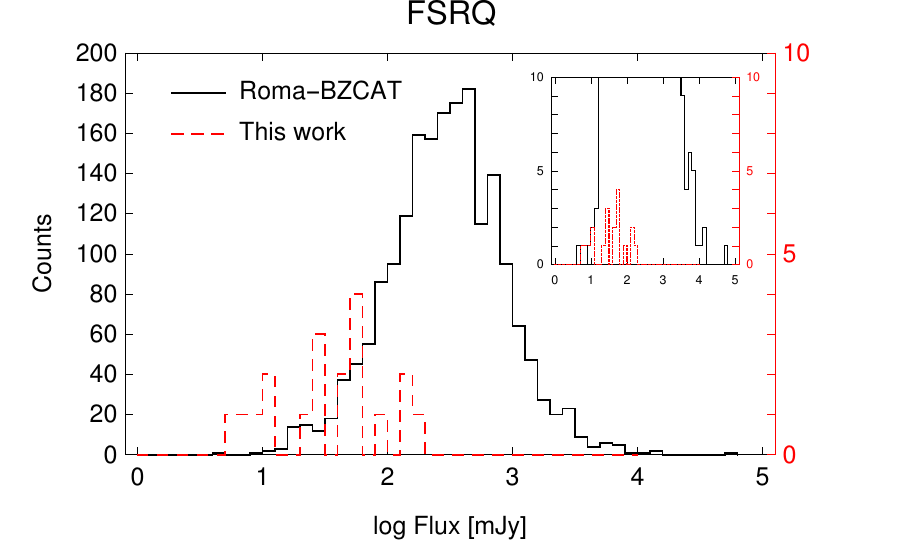}
\includegraphics[angle=0,scale=1.0]{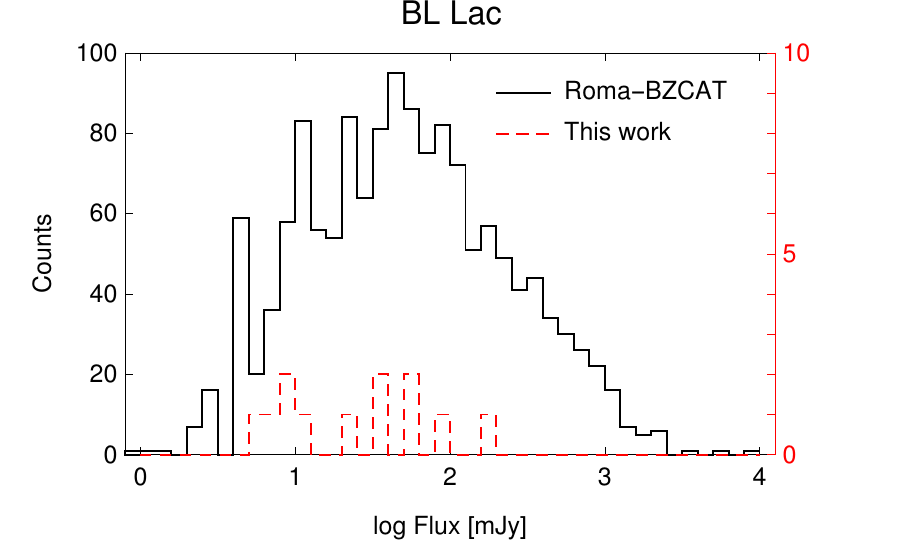}
\caption{Distributions of 0.84 GHz flux of the FSRQ (upper panel) and BL Lac (bottom panel) blazars from Roma-BZCAT and blazar candidates identified in this work. } \label{fluxComparison}
\end{figure}

\begin{figure}[!th]
\includegraphics[angle=0,scale=1.0]{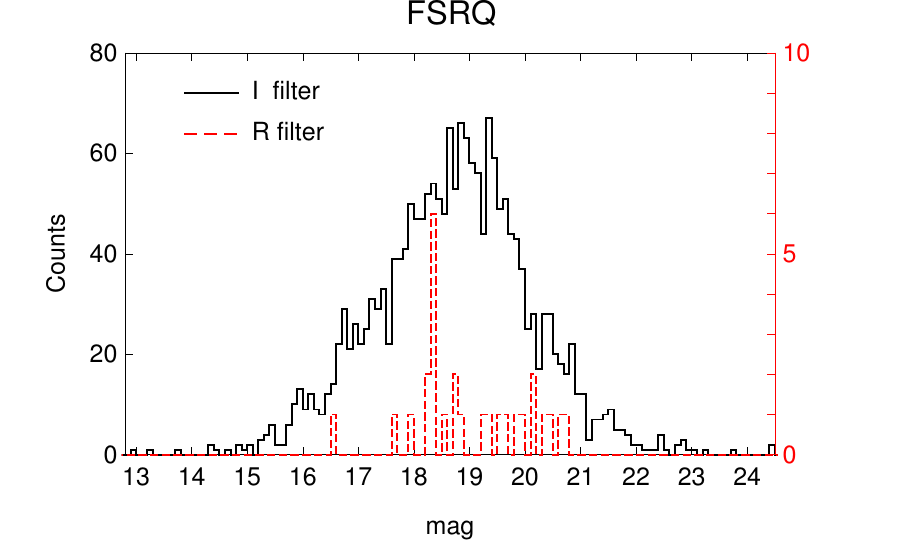}
\includegraphics[angle=0,scale=1.0]{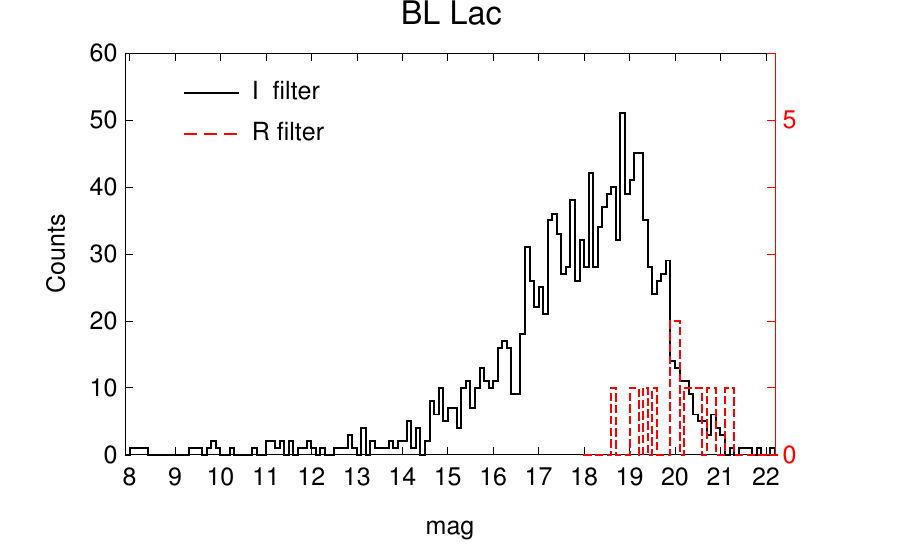}
\caption{Distributions of R mag and apparent I magnitude of the FSRQ (upper panel) and BL Lac (bottom panel) blazars from the Roma-BZCAT (solid black lines) and blazar candidates identified in this work (dashed red lines). } \label{MagComparison}
\end{figure}

The distributions of radio flux spectral densities of both blazar type candidates are presented in Figure~\ref{fluxComparison}. In case of the OGLE blazar candidates only the 0.843\,GHz flux density measurements were used. Figure~\ref{MagComparison} shows the distributions of optical apparent magnitude in R filter of the Roma-BZCAT sources and in I filter of the OGLE blazar candidates (note the very similar effective wavelengths for both filters) 

A Kolmogorov-Smirnov test \citep{fran51,tarn15} was performed to verify the null hypothesis, $H_0$, whether the samples examined in this work have the same distributions as the corresponding datasets from Roma-BZCAT. Bear in mind, however, that the K-S test was used here to compare a small sample of newly identified blazar candidates, consisting of 12 up to 27 sources depending on the compared parameters, with a relatively large Roma-BZCAT sample. In the case of the redshift distribution of FSRQs, $H_0$ is not rejected at the $\alpha=0.01$ significance level. For the radio flux distribution of BL Lacs, the $H_0$ is also not rejected, unlike in the case of radio flux distribution of FSRQs, for which  the $H_0$ is rejected with the $p$-value$ = 6 \times 10^{-11}$. For the apparent magnitude distributions in $I$ and $R$ filters, $H_0$ hypothesis is not rejected for FSRQs, but is rejected for BL Lacs.

All in all, a comparison between the list of blazar candidates selected here and the Roma-BZCAT, indicates that even though the redshift and the apparent optical magnitude distributions of FSRQs in both samples are similar, our FSRQ candidates are characterized by lower radio fluxes, on average. In the case of BL Lacs, on the other hand, while the radio flux distributions seem comparable, the objects from our list appear on average dimmer in optical.

\subsection{Mid-IR colours}

The mid-IR colour criteria provide a useful tool enabling theidentification of AGN among other objects, such as stars or normal galaxies. In particular,  \citet{ster05} and later \citet{mass11} and \citet{dabr12} proposed diagnostic tools to separate blazars based on their IR colours. 

\begin{figure}[!th]
\includegraphics[angle=0,scale=.35]{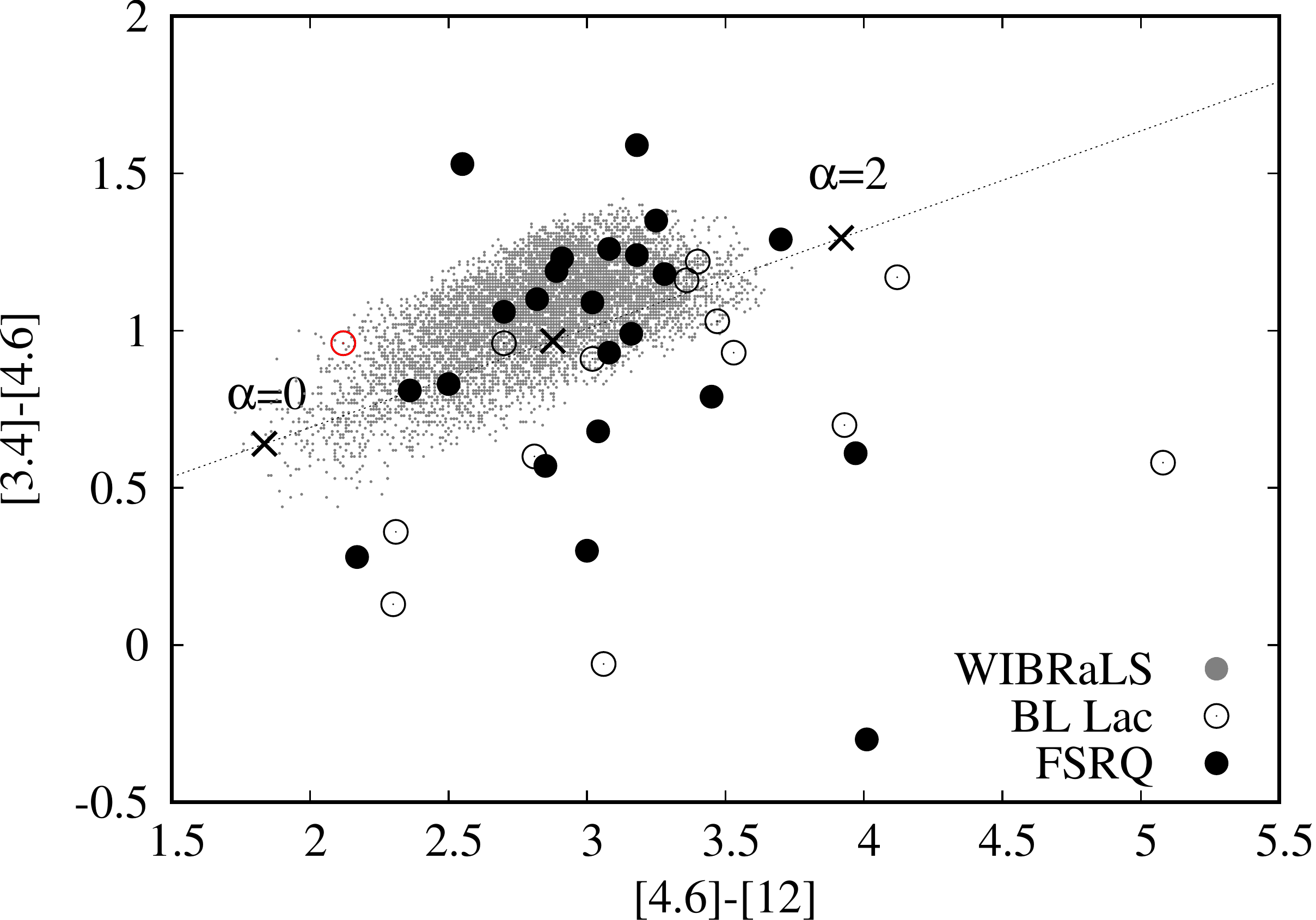}
\includegraphics[angle=0,scale=.35]{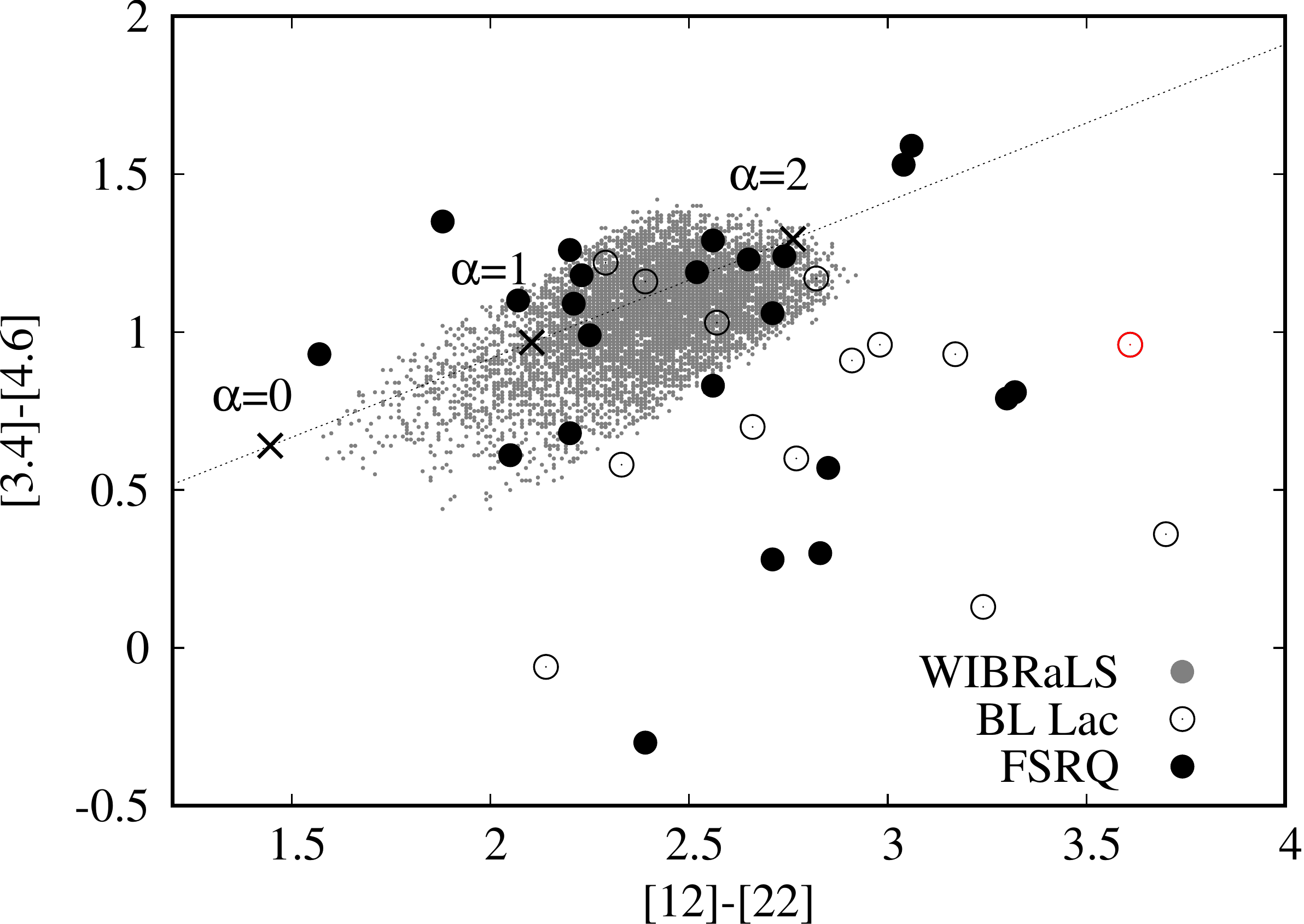}
\caption{The WISE [3.4]-[4.6]-[12] (top panel) $\mu$m and [3.4]-[4.6]-[12]-[22] $\mu$m (bottom panel) colour-colour diagrams of OGLE blazar candidates. Open and filled black circles denote BL Lacs and FSRQs, respectively, while red open circles mark the BL Lac candidate J0545-6846; small grey dots stand for the $\gamma$-ray blazar candidates from WIBRaLS \citep{dabr14}. Black dashed lines correspond to the case of a single power-law emission continuum within the entire range of wavelength considered (i.e., 3.4--12\,$\mu$m in the upper panel, and 3.4--22\,$\mu$m in the lower panel), with the particular values of mid-IR spectral indices $\alpha = 0$, 1, and 2 denoted by \ding{53} symbols.} \label{IR_WISE} 
\end{figure}

{We analyzed our sample of blazar candidates using the data collected by WISE and the Spitzer Space Telescope (the IRAC instrument), generating four mid-IR colour-colour planes, i.e. [3.4]-[4.6]-[12] and [3.4]-[4.6]-[12]-[22] $\mu$m based on the WISE observations (Figure~\ref{IR_WISE}), and [3.6]-[4.5]-[8.0] $\mu$m and [3.6]-[4.5]-[5.8]-[8.0] $\mu$m with the IRAC data (Figure~\ref{IR_IRAC}). In the figures, black filled circles denote the mid-IR colours of the FSRQ blazar candidates, and black open circles indicate the mid-IR colours of the BL Lac candidates; black dashed lines correspond to the case of a single power-law emission continuum within the entire range of wavelength considered (e.g., 3.4--12\,$\mu$m in the upper panel of Figure~\ref{IR_WISE}, or 3.4--22\,$\mu$m in the lower panel of Figure~\ref{IR_WISE}), with the particular values of mid-IR spectral indices $\alpha = 0$, 1, and 2 denoted by \ding{53} symbols; finally, red open circles mark the BL Lac candidate J0545-6846, which coincides with the $\gamma$-ray transient detected by the \textit{Fermi}-LAT (see Section~\ref{HE}).

Figure~\ref{IR_WISE} displays in addition a comparison between our blazar candidates and 7,855 $\gamma$-ray blazar candidates (FSRQs, BL Lacs, and mixed-type objects, all denoted by small grey dots) selected based on the \textit{AllWISE Data Release} products \citep{cutr13} and listed in the \textit{WISE blazar-like radio-loud sources} catalogue \citep[WIBRaLS,][]{dabr14}. Some of our blazar candidates occupy the same area in the plot as the WIBRaLS objects, signalling a single power-law character of their mid-IR emission continua, with mid-IR spectral indices roughly within a broad range of 0.5--2.0 (see Table~\ref{blazarlist}), consistently with the blazar classification. About half of our sample, however, is scattered much below the black dashed lines in the diagrams, indicating that their mid-IR emission continua are curved, with flatter slopes at longer wavelengths. This could be a signature of either a hot dust emission from host galaxies dominating the mid-IR output of the systems, or a non-negligible contamination of WISE fluxes by nearby foreground/background sources.

\begin{figure}[!th]
\includegraphics[angle=0,scale=.35]{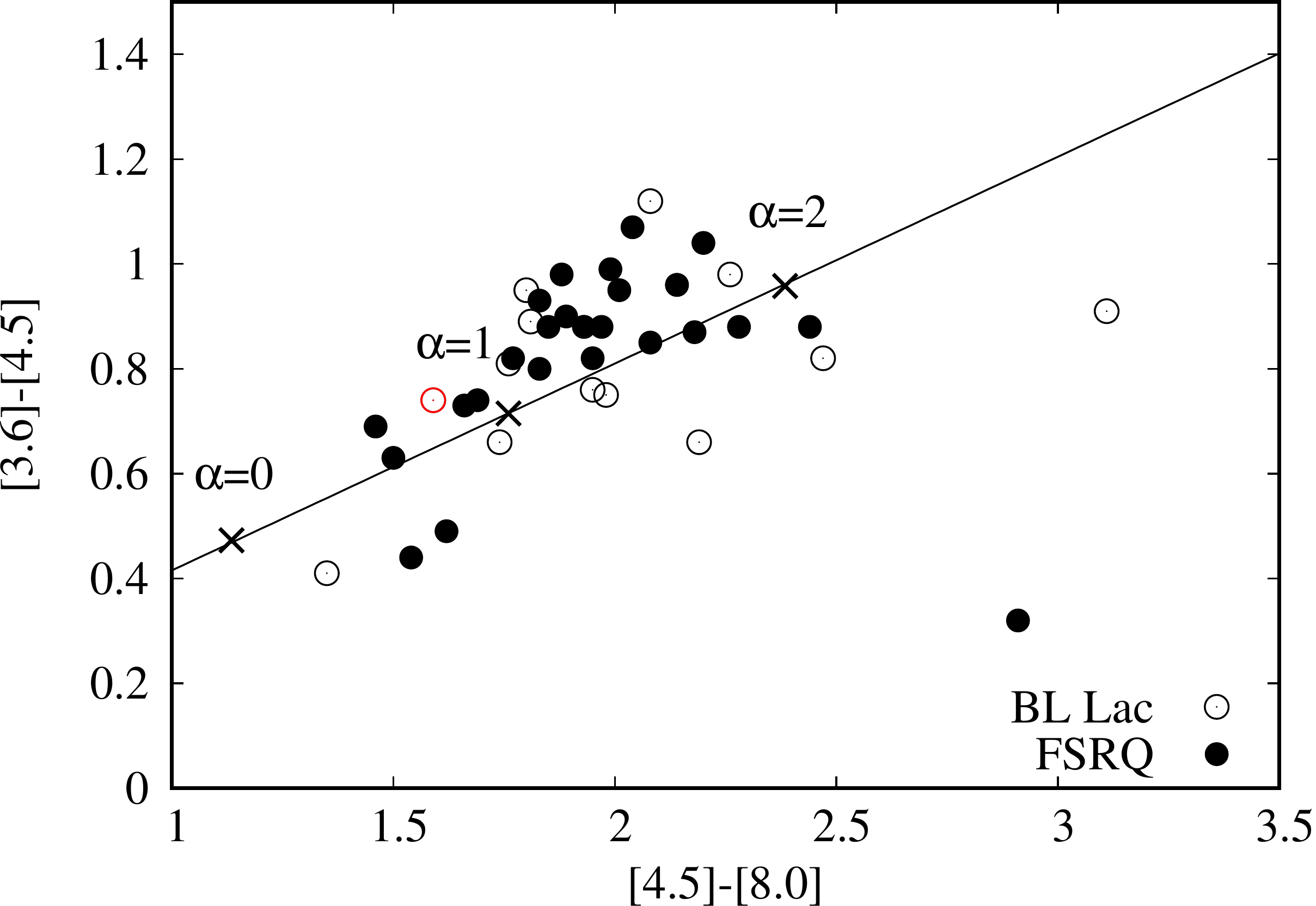}
\includegraphics[angle=0,scale=.35]{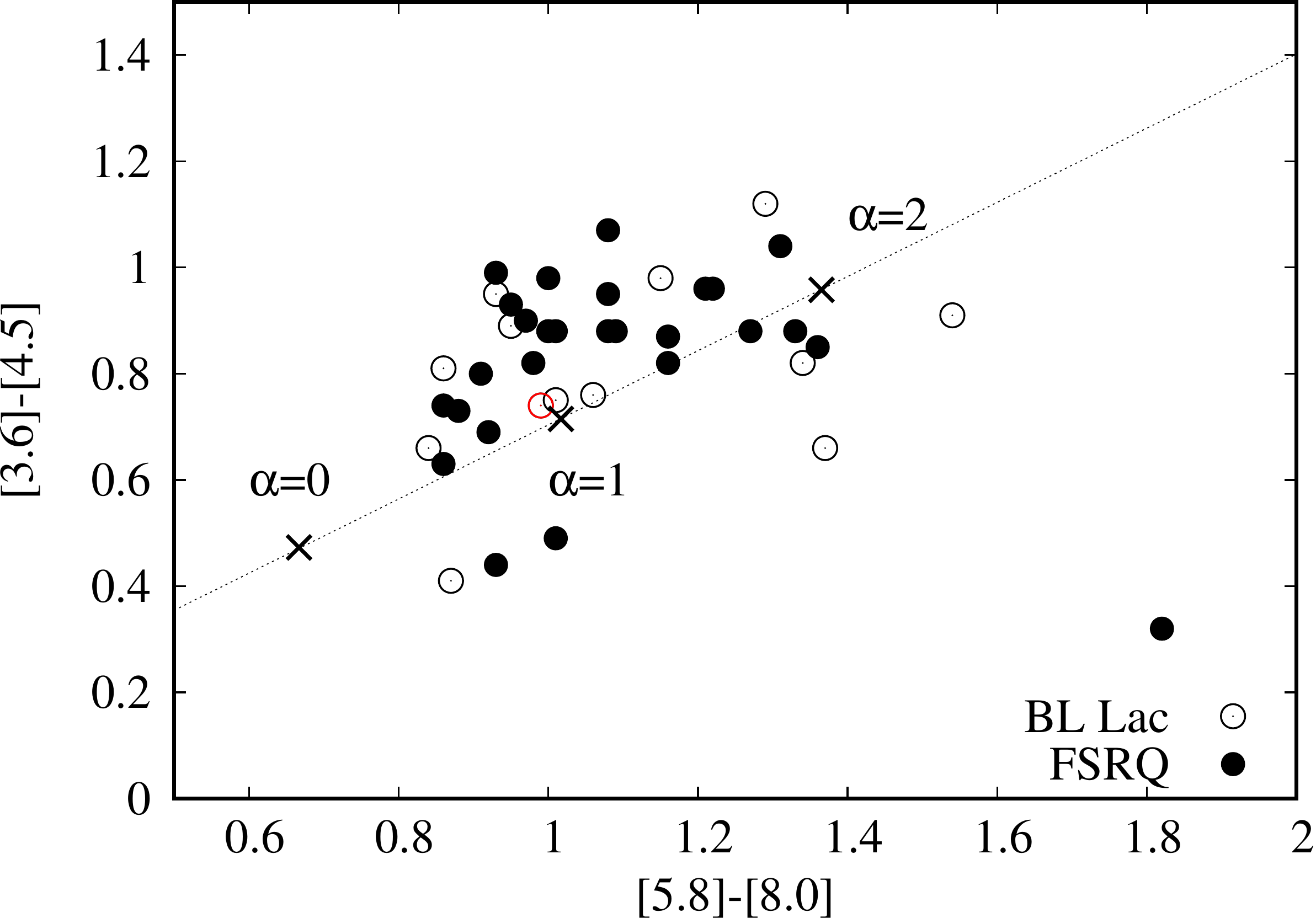}
\caption{The IRAC [3.6]-[4.5]-[8.0] $\mu$m (top panel) and [3.6]-[4.5]-[5.8]-[8.0] $\mu$m (bottom panel) colour-colour diagrams of OGLE blazar candidates. Open and filled black circles denote BL Lacs and FSRQs, respectively, while red open circles mark the BL Lac candidate J0545-6846. Black dashed lines correspond to the case of a single power-law emission continuum within the entire range of wavelength considered (3.6--8.0\,$\mu$m), with the particular values of mid-IR spectral indices $\alpha = 0$, 1, and 2 denoted by \ding{53} symbols.} \label{IR_IRAC}
\end{figure}

Subsequently, we investigated the IRAC colours of our blazar candidates using the [3.6]-[4.5]-[8.0] $\mu$m and [3.6]-[4.5]-[5.8]-[8.0] $\mu$m colour-colour diagrams, motivated by the supreme angular resolution of the IRAC instrument when compared to WISE ($\sim 2\arcsec$ versus $\sim 6\arcsec$). As shown in the resulting Figure~\ref{IR_IRAC}, the IRAC colours of our blazar candidates are now more consistent with a single power-law character of their mid-IR emission continua, as expected for blazar sources, suggesting that a large scatter in the WISE colour-colour diagrams is mostly due to the contamination effect. Interestingly, we note that our sample is homogenous in that we do not see any clear separation between the FSRQ or the BL Lac blazar candidates. This may suggest that the majority of our BL Lac candidates  are of the LBL type.
\vspace{0.5cm}

\subsection{Radio polarization}

Since the radio-optical emission of blazars is synchrotron in origin, a significant amount of polarization at radio and optical frequencies is expected. We therefore gathered all the available archival data for the linear radio degree PD$_{\rm{r}}$ and polarization angle (PA$_{\rm{r}}$) from the AT20G catalogue and polarized flux density maps at 4.8 and 8.6 GHz for the LMC\footnote{\url{http://www.atnf.csiro.au/research/lmc\_ctm/index.html}} and SMC\footnote{\url{http://www.atnf.csiro.au/research/smc\_ctm/index.html}} to investigate polarization properties for the selected blazar candidates in Tables \ref{blazarFluxes} and \ref{blazarlist}. As a result, we extracted nine objects (see Table~\ref{Polar}), out of which six are FSRQ and three are BL Lac blazar candidates; all of them are strongly polarized sources with PD$_{\rm{r}}$ $>$ 3\% at all the analyzed radio frequencies. These should be considered as ``secure'' blazar candidates.

The 4.8 and 5 GHz data were obtained with the same telescope but presumably on different dates. The fact that for the two objects, i.e., J0512-6732 and J0111-7302, observed at both 4.8 and 5 GHz, there are very different PD$_{\rm{r}}$ in these close radio bands, suggests polarization variability, which is a further indication that they are blazar-like objects.

\begin{deluxetable*}{lccccccc}
\tabletypesize{\footnotesize}
\tablecolumns{8}
\tablewidth{0pt}
\tablecaption{The degree of linear polarization and polarization angle of FSRQ and BL Lac blazar type candidates. \label{Polar}}
\tablehead{
\multicolumn{1}{c}{Object} & \multicolumn{5}{c}{PD$_{\rm{r}}$} & \multicolumn{2}{c}{PA$_{\rm{r}}$} \\
 & \colhead{4.8 GHz} & \colhead{5 GHz} & \colhead{8 GHz} & \colhead{8.6 GHz} & \colhead{20 GHz} & \colhead{4.8 GHz} &  \colhead{8.6 GHz} \\ 
  & [\%] & [\%] & [\%] & [\%] & [\%] & [$^{\circ}$] & [$^{\circ}$] \\ 
\multicolumn{1}{c}{\tiny{(1)}} & \tiny{(2)} & \tiny{(3)} & \tiny{(4)} & \tiny{(5)} & \tiny{(6)} & \tiny{(7)} & \tiny{(8)} 
}
\startdata 
\cutinhead{FSRQs}  
J0114$-$7320 & 9.5$\pm$0.5 & & & 7.0 & & 70.7$\pm$14.7 & $-$12.9$\pm$14.3 \\
J0120$-$7334 & 5.0$\pm$0.3 & & & & & $-$52.2$\pm$0.7 &  \\
J0442$-$6818 & & 11.7 & 10.0 &  & 8.1 & &  \\
J0512$-$6732 & 3.3$\pm$0.2 & 12.7 & 10.7 &  & 13.6 & 13.6$\pm$2.1 &  \\
J0551$-$6916 & 7.3$\pm$0.4 & & & & & 8.3$\pm$38.2 &  \\
J0551$-$6843 & 9.1$\pm$0.5 & & & & & $-$3.9$\pm$40.4 &  \\
\cutinhead{BL Lacs}
J0111$-$7302 & 4.1$\pm$0.2 & 8.3 & 8.3 &  & 9.7 & 4.7$\pm$34.1 &  \\
J0501$-$6653 & 10.6$\pm$0.5 & & & 22.7$\pm$1.1 &  & $-$2.8$\pm$25.6 & $-$11.1$\pm$46.1\\
J0518$-$6755$^{\rm{a}}$ & 12.6$\pm$0.6 & & & & & $-$16.6$\pm$11.0 & 
\enddata
\tablecomments{$^{\rm{a}}$Dubious; Columns: (1) source designation, (2)-(6) linear polarization degree (PD$_{\rm{r}}$) at 4.8, 5, 8, 8.6, and 20 GHz, (7)-(8) linear polarization angle (PA$_{\rm{r}}$) at 4.8 GHz and 8.6 GHz.}
\end{deluxetable*}

\subsection{X-ray and $\gamma$-ray counterparts}
\label{HE}

We cross-matched our blazar candidates with the \textit{Fermi} 2FGL catalogue \citep{nola12} and the ROSAT All-Sky Survey Catalogue \citep[RASS-BSC;][]{voge99} to search for their counterparts in $\gamma$- and X-ray regime. We did not find any match between the positions of our blazar candidates and the \textit{Fermi} and RASS-BSC catalogues.

Recently, \citet{acke16a} reported the  detection of four point-like sources in the area of the LMC. One of them is identified based on its characteristic pulsed $\gamma$-ray emission as pulsar PSR J0540$-$6919, while possible associations of the three remaining sources are still pending. None of these sources coincides with our blazar candidates.

\begin{figure}[!b]
\includegraphics[angle=0,scale=.3]{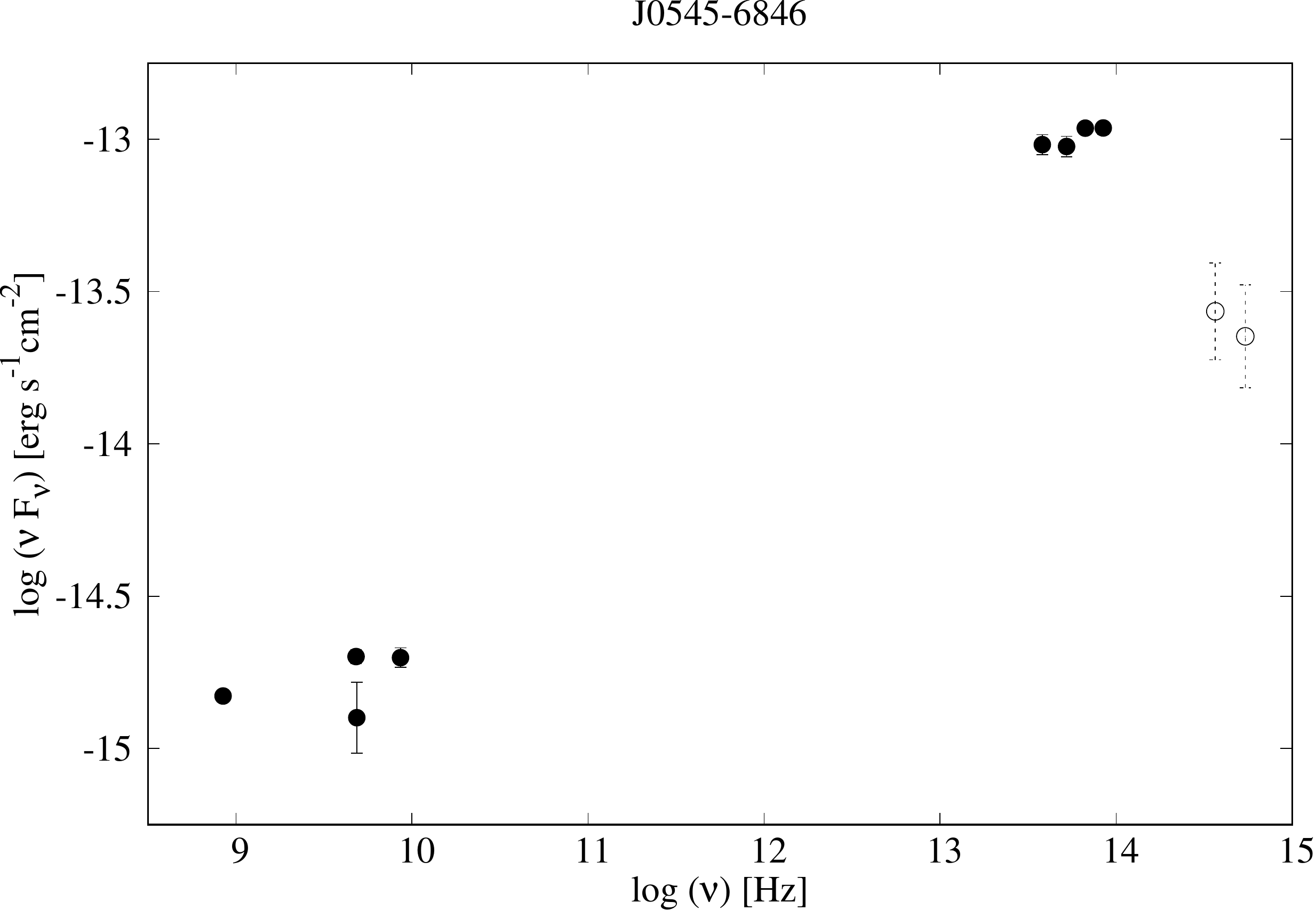}
\caption{SED of the J0545$-$6846 BL Lac candidate, which coincides with the $\gamma$-ray transient detected by the \textit{Fermi}-LAT; open circles denote optical data from the FS list, black filled circles are for the archival radio and mid-IR data.}\label{J0545-6846}
\end{figure}

Interestingly, \citet{abdo10b} noticed a flaring activity from the direction of LMC during the 4th month of the \textit{Fermi}-LAT observations (at the turn of July and August 2008). The location of this event was estimated by fitting a point source with the 2DG model and HII gas map as templates for the LMC emission to the datasets. Using the 2DG model, the source was found at RA = 05$^{\rm{h}}$46\fm2 and DEC= $-$69\dg01\arcmin with a detection significance of 4.5$\sigma$ and 36\arcmin\, containment radius, while with the HII map, at the position of RA = 05$^{\rm{h}}$ 46\fm4 and DEC= $-$69\dg01\arcmin, with 4.6$\sigma$ significance and 29\arcmin\, radius. In Figure~\ref{clouds} (top panel), we show both possible locations of the flaring activity (green circle denotes the 2DG model and black circle -- H II gas map) which, moreover, coincide with positions of four our blazar candidates, namely J0545$-$6846, J0551$-$6843, J0551$-$6916, and J0552$-$6850. The BL Lac candidate J0545$-$6846, for which the broad-band SED is shown in Figure~\ref{J0545-6846}, is consistent with both estimated locations, and is characterized by a particularly large radio flux and high radio-loudness in our sample. We however note here a rather large integrated flux $>1$ GeV of the LAT transient, $\simeq 9.6^{+2.8}_{-6.1}\times10^{-12}$\,erg\,s$^{-1}$\,cm$^{-2}$; if related to the flaring activity of J0545$-$6846, this would imply an unusually large value of the Compton dominance parameter as for a BL Lac candidate.

The \textit{Fermi} group noticed also the second $\gamma$-ray event, which took place a few years later, in April 2015. It is not clear if the two flares are produced by the same object, but they originated from about the same direction. The source, however, did not show up in a full six-year data analysis.

\subsection{Identification procedure: a summary} \label{results}

Figure~\ref{BlazarMaps} shows the radio contours overlayed on the grey scale optical images for the blazar candidates indentified in the OGLE data for the LMC and the SMC. We used the Digitized Sky Survey (DSS) R band optical images as grey scales and the radio data at 8.6, 4.85 or 0.845 GHz as contours. A black cross marks the optical position from the MQS catalogue or the FS list for each blazar candidate. All the sources which were found in the PMN catalogue only, as well as other sources which are not associated with the strongest central radio emission indicated by the radio contours, are considered as dubious and marked with the ``a'' symbol  in Tables \ref{blazarFluxes} and \ref{blazarlist}. The main results of our further analysis of the selected blazar candidates are:
\begin{enumerate}
\item As a result of the cross-matching procedure, we selected a sample of 44 objects, including 27 FSRQ and 17 BL Lac blazar candidates. This significantly increased the sample of blazar candidates behind LMC and SMC.
\item Each selected object is optically faint with the I band magnitude between 16 and 21. 
\item All the FSRQ candidates are distant sources with redshifts from 0.286 up to 3.320. The redshift distribution for BL Lacs is unknown due to the lack of lines in their optical spectra. 
\item The radio-loudness parameter $R$ varies from 12 to 4,450 for the FSRQ blazar candidates and from 171 to 7,020 for the BL Lac candidates, implying that all the sources included in the sample are radio-loud indeed.
\item The estimated values of the radio spectral index $\alpha_r$ vary from $-$0.57 to 1.37. Since we used non-simultaneous archival radio data in the analysis, the $\alpha_r$ may indicate radio flux variability for the sampled sources, as expected for blazars, rather than steep radio spectra. 
\item The infrared colours indicate that the infrared continua of the selected sources are non-thermal in origin, as expected for blazars. The estimated values of mid-IR spectral indices for the majority of sources are between 0.5 and 2.0.
\item The PD$_{\rm{r}}$ and PA$_{\rm{r}}$ were found in the AT20G catalogue or were measured by us from the archival radio maps of the 4.8 and 8.6 GHz surveys of the LMC and SMC. All the blazar candidates listed in the Table~\ref{Polar} are strongly polarized sources with the average polarization of PD$_{\rm{r},4.8}$ $\sim$ 6.8\% at 4.8 GHz.
\item Although we did not find any clear associations with high-energy $\gamma$-ray emitters, we note a repeating flaring activity detected with \textit{Fermi}-LAT from the direction of the J0545$-$6846 BL Lac candidate.
\end{enumerate}

\section{Discussion: relevance of the analysis} \label{discussion}

\subsection{Physics of blazar sources}

 The majority of blazars of the FSRQ and LBL types are identified in the radio surveys \citep[e.g.,][]{stic91,perl96,perl98,cacc02}, while HBLs are mostly discovered via the X-ray surveys \citep[e.g.,][but see \citealt{plot10} for the optically selected BL Lac sample using the SDSS data]{purs13,rect00,land01}. This naturally leads to various selection biases, as any flux-limited survey will detect sources which are the brightest in a given frequency band. As an illustration of the problems arising in this context one can note that the luminosity evolution of the radio-selected BL Lacs was previously claimed to be different from that of the X-ray-selected BL Lacs, the finding which was however not confirmed in the follow-up studies using larger samples \citep[e.g.,][and references therein]{pado07}. Our blazar candidates are selected uniquely from the deep, flux-limited optical survey, based on the \emph{time-domain} analysis, and as such could in principle be of relevance for in-depth population studies of blazar sources.

 We also emphasize here the optical variability of blazars on different timescales, which was and is a subject of an intense research, albeit typically limited to rather small and sparse samples of sources \citep[e.g.,][]{webb88, ghos00}. Although the variable optical emission of blazars originates predominantly in nuclear relativistic jets, it may also contain a pronounced contribution from accretion disks, which are modulating the jet activity, as seen directly in a few luminous radio galaxies and quasars \citep{lohf13,bhatta18}. Interestingly, all the newly identified blazar candidates in our sample have high photometric accuracy, high cadence optical light curves from the long-term ($\sim$ 2 decades) photometric monitoring from OGLE (OGLE-III and OGLE-IV phases). 
\vspace{0.5cm} 

\subsection{MC magnetic field study}

 Since magnetic field may play an important role in the cosmological evolution of galaxies, it is mandatory to disclose in detail its origin, structure and strength in nearby galaxies, including the nearest dwarf galaxies, i.e., the SMC and LMC. The commonly used method for investigating the galactic magnetic field morphology, is the Faraday rotation measure (RM) study of background polarized sources (as implemented by, e.g., \citealt{burn66,kawa69}, or more recently \citealt{bren05}). In this way, the magnetic field of the MCs system, i.e., the SMC, LMC, and Magellanic Bridge (MB), was intensively investigated in a last few decades, first using the optical \citep{schm70,math70} and then radio data \citep[e.g.,][]{hayn86,klei93,gaen05, mao12,kacz17}. Our sample of blazar candidates should be therefore relevant in this context as well, as it identifies potentially \emph{variable} polarized emitters behind the MCs system, which should be treated with caution when included in the RM studies.
  
\subsection{$\gamma$-ray study}

One of the key projects of the newly emerging very high energy $\gamma$-ray observatory Cherenkov Telescope Array (CTA), is to conduct a deep LMC survey, expanding over previous results provided by the currently operating $\gamma$-ray observatories, namely \textit{Fermi}-LAT and H.E.S.S. \citep[see Chapter~7 in][]{acha17}. This project will allow to study in detail the diffuse $\gamma$-ray emission of the LMC, including processes of cosmic ray acceleration and propagation within the interstellar medium, but also enabling the dark matter research. Clearly, a proper identification of all the potential background $\gamma$-ray emitters behind the MC system, i.e., blazars, is indispensable for a correct analysis and interpretation of the $\gamma$-ray data on the LMC in general, and the planned CTA observations in particular.
 
\acknowledgments
We thank Dr Pierrick Martin and Dr Shane O'Sullivan for fruitful discussions and comments on the analysis. N\.Z work was supported by the Polish National Science Centre (NCN) through the grant DEC-2014/15/N/ST9/05171. AG and MO acknowledge support from the NCN through the grant 2012/04/A/ST9/00083. AG also acknowledges partial support from 2013/09/B/ST9/00026. {\L}S was supported by Polish NSC grant 2016/22/E/ST9/00061. The OGLE project has received funding from the NCN MAESTRO grant no.2014/14/A/ST9/00121 to AU. SK also acknowledges the financial support from the NCN OPUS grant no. 2014/15/B/ST9/00093.


\begin{figure}
\begin{minipage}{0.3\textwidth}
  \centering
\includegraphics[angle=0,scale=.3]{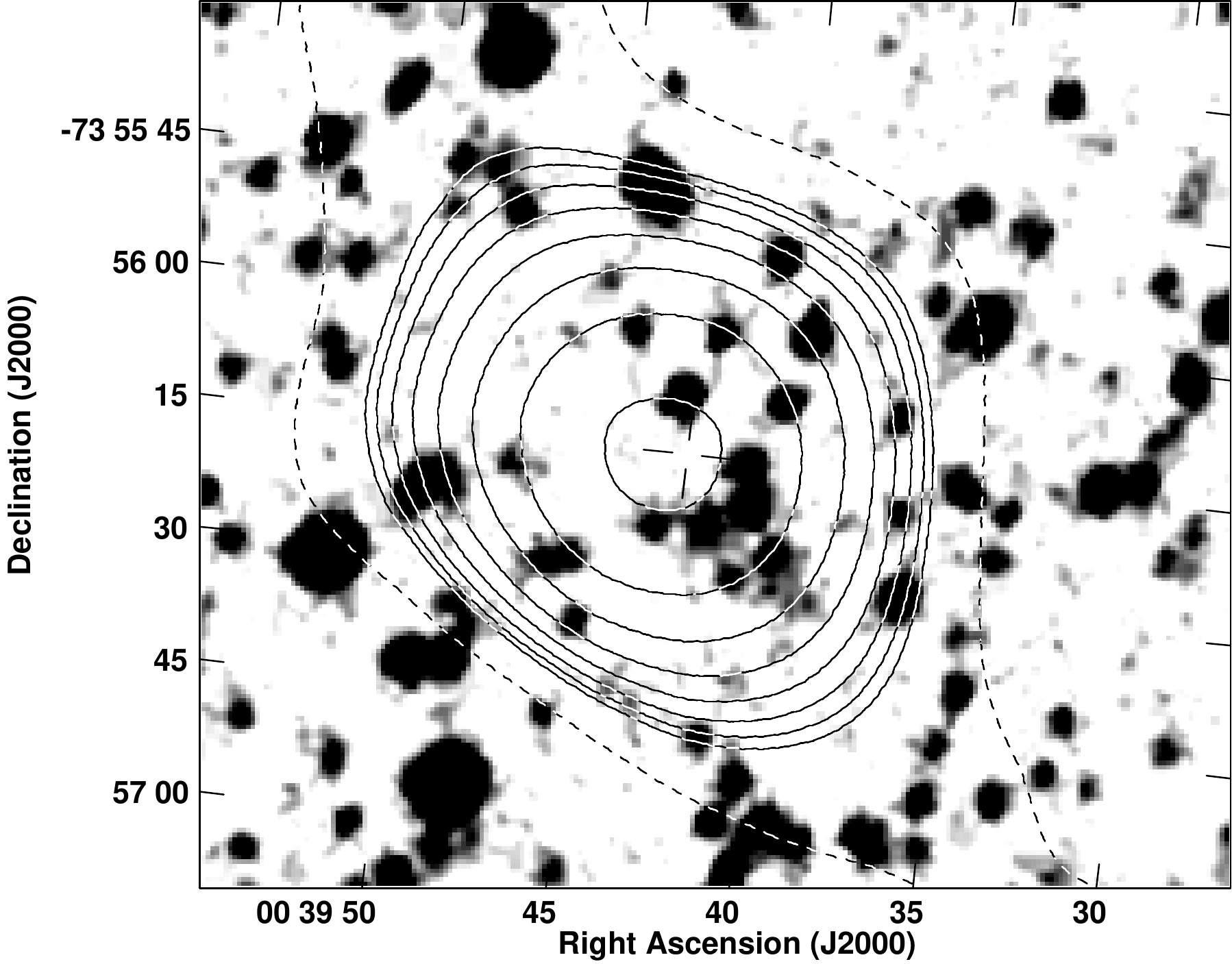}
\includegraphics[angle=0,scale=.3]{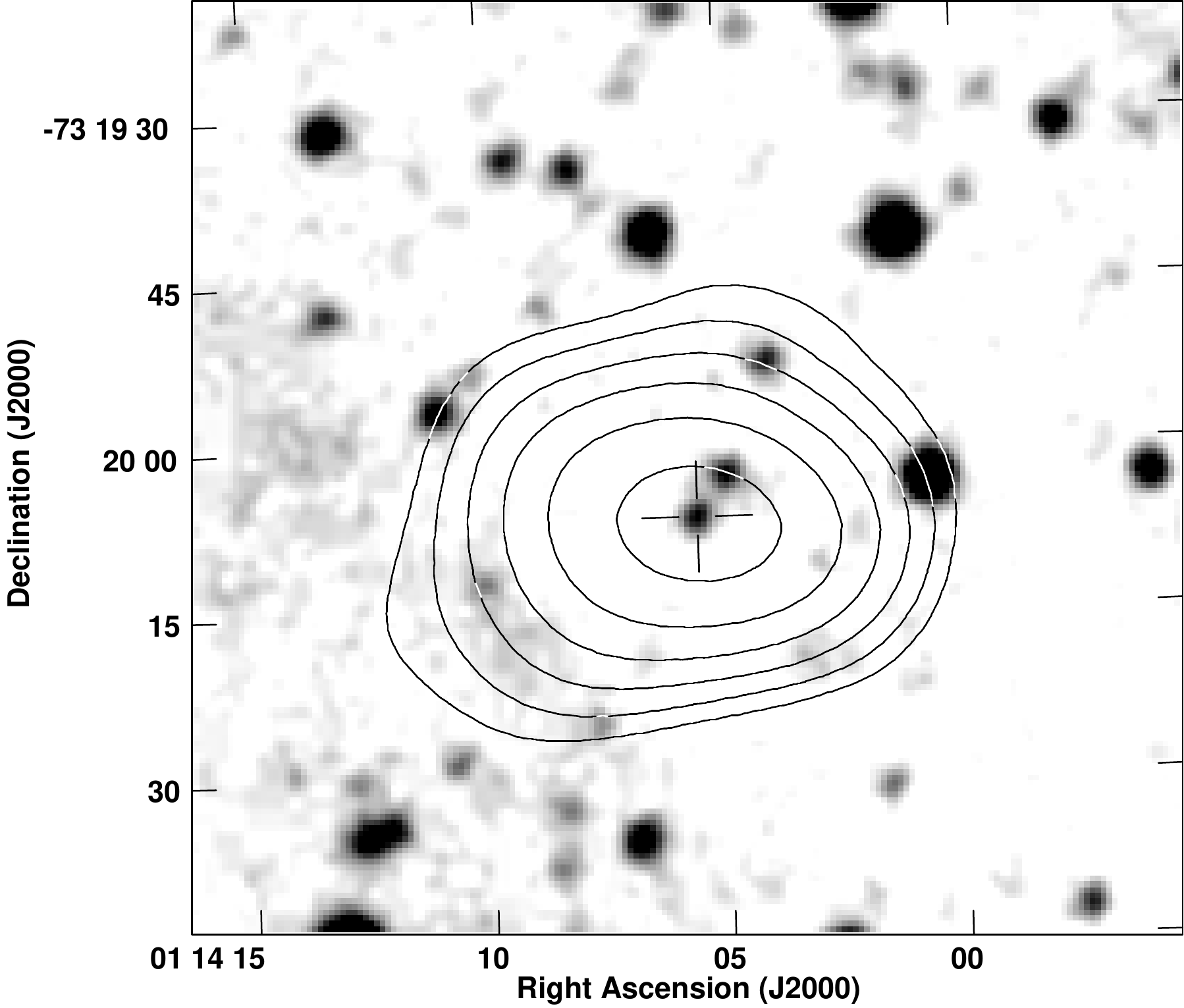}
\includegraphics[angle=0,scale=.3]{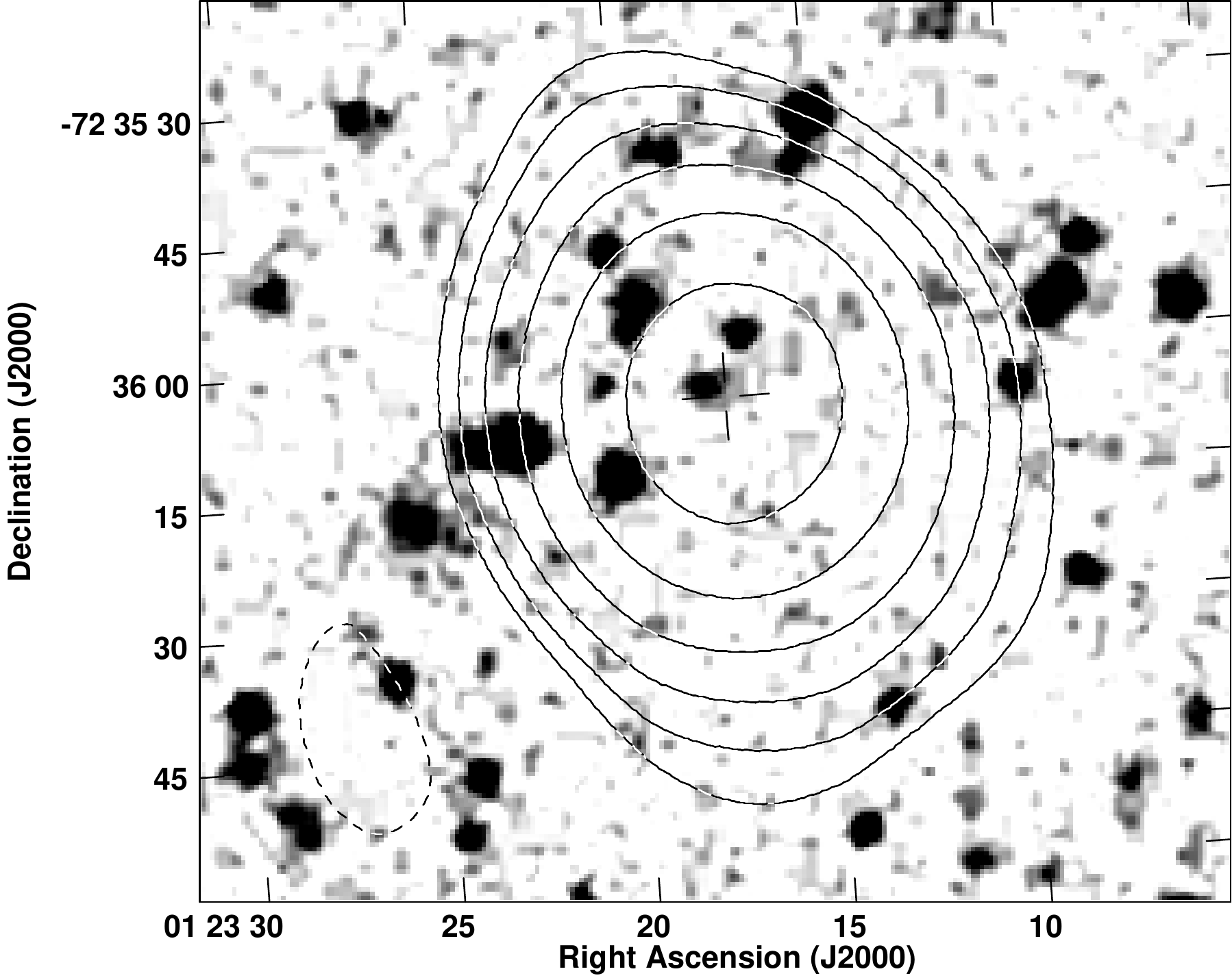}
\includegraphics[angle=0,scale=.3]{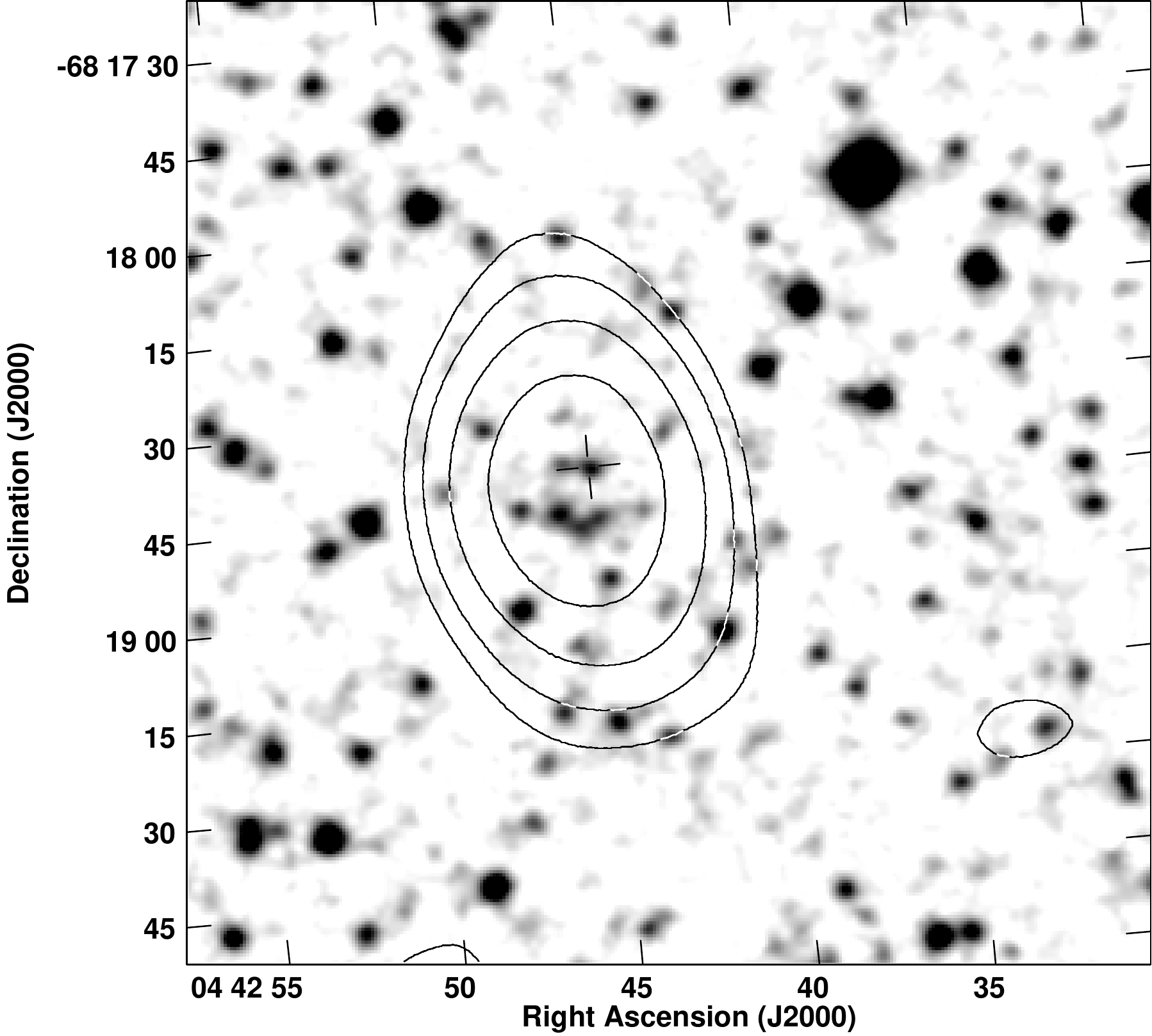}
\end{minipage}%
\begin{minipage}{0.3\textwidth}
  \centering
\includegraphics[angle=0,scale=.3]{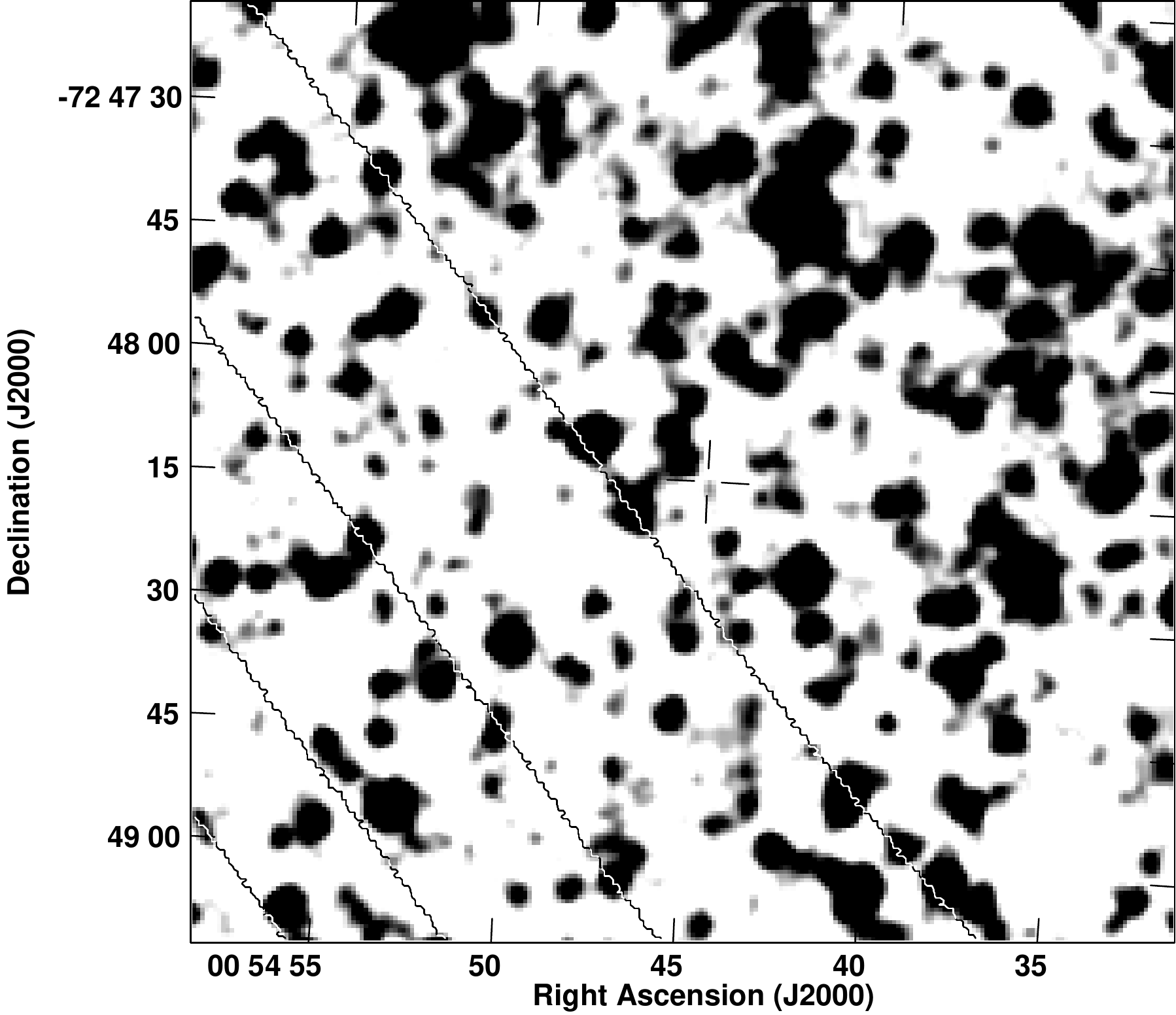}
\includegraphics[angle=0,scale=.3]{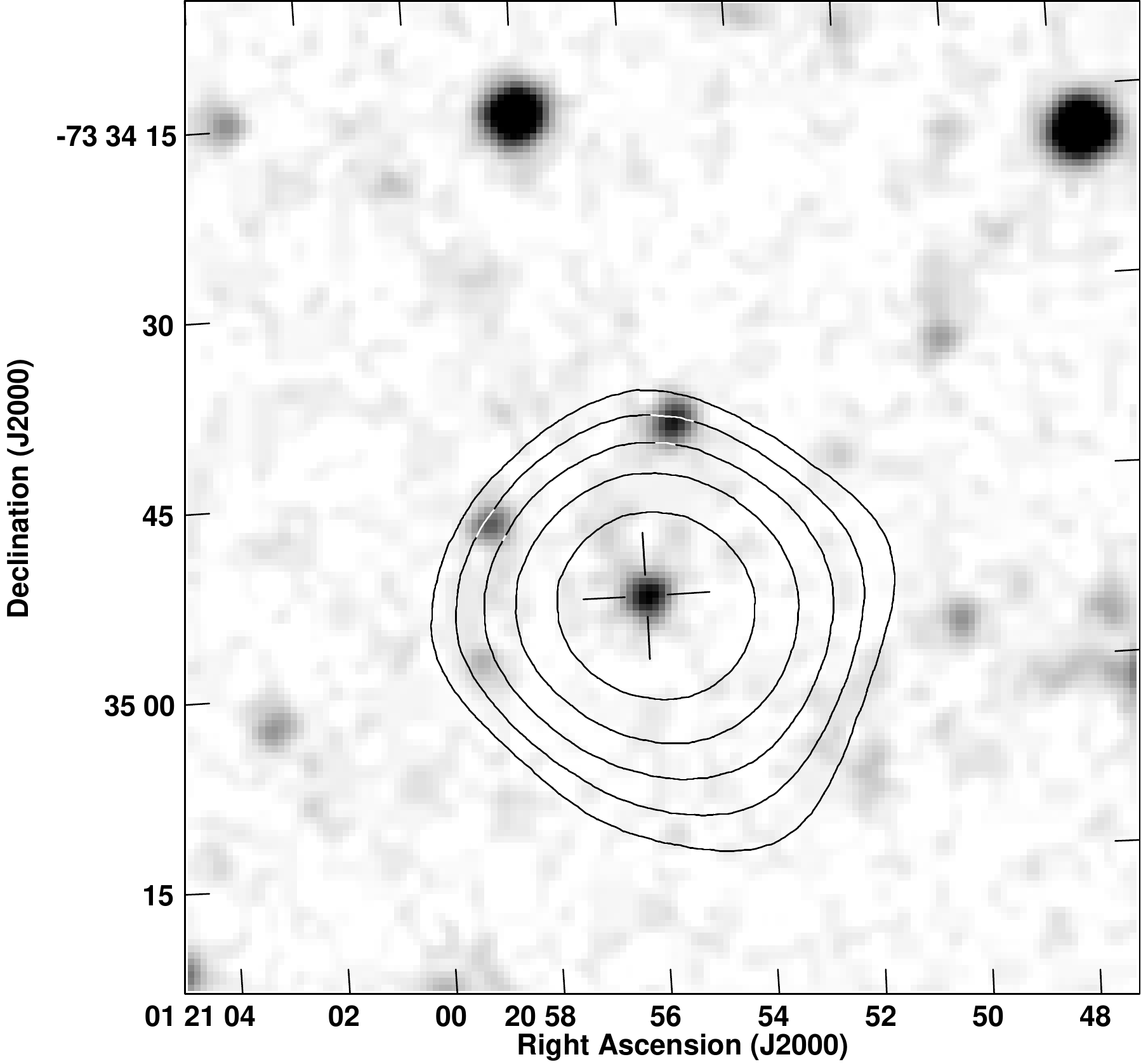}
\includegraphics[angle=0,scale=.3]{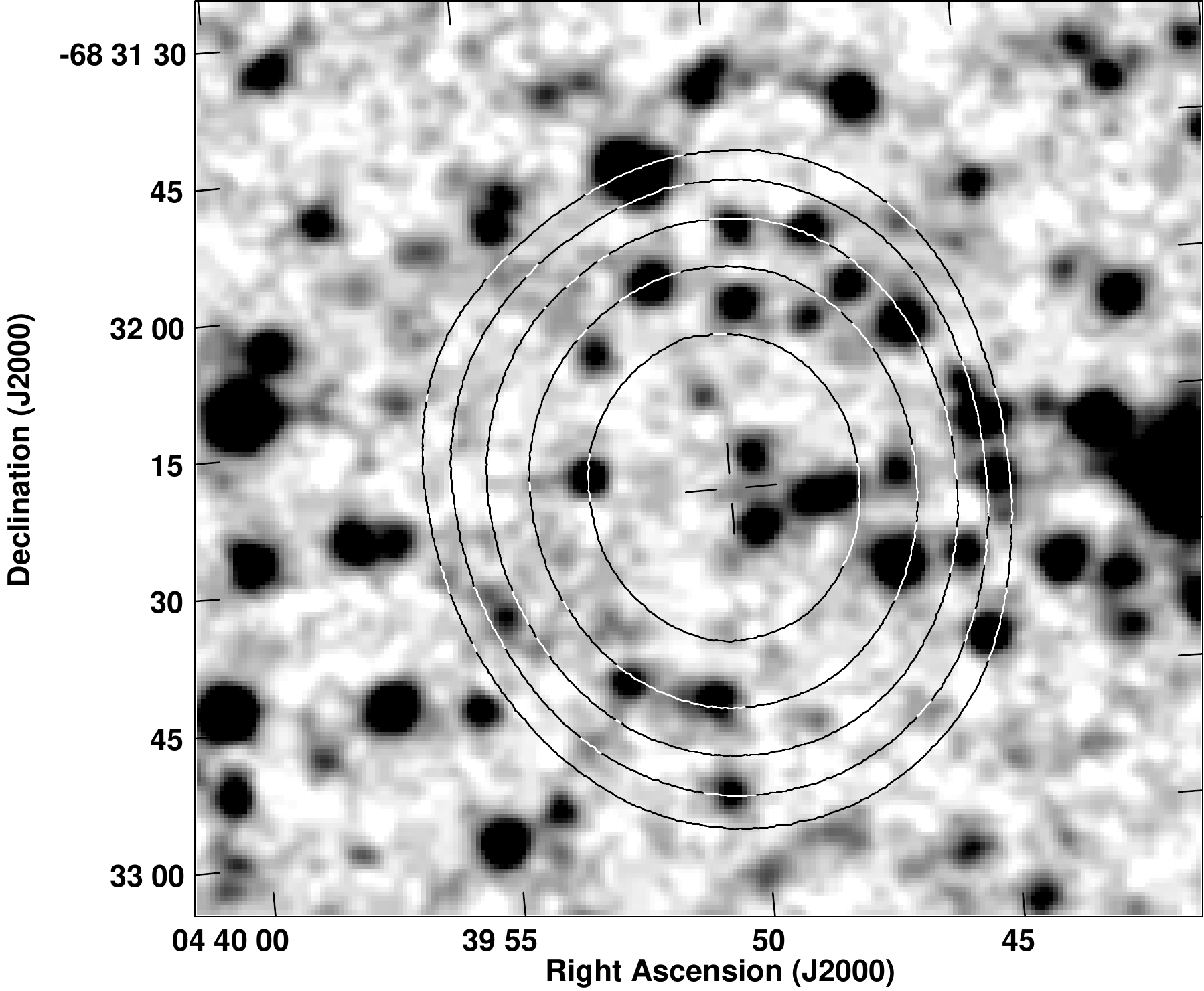}
\includegraphics[angle=0,scale=.3]{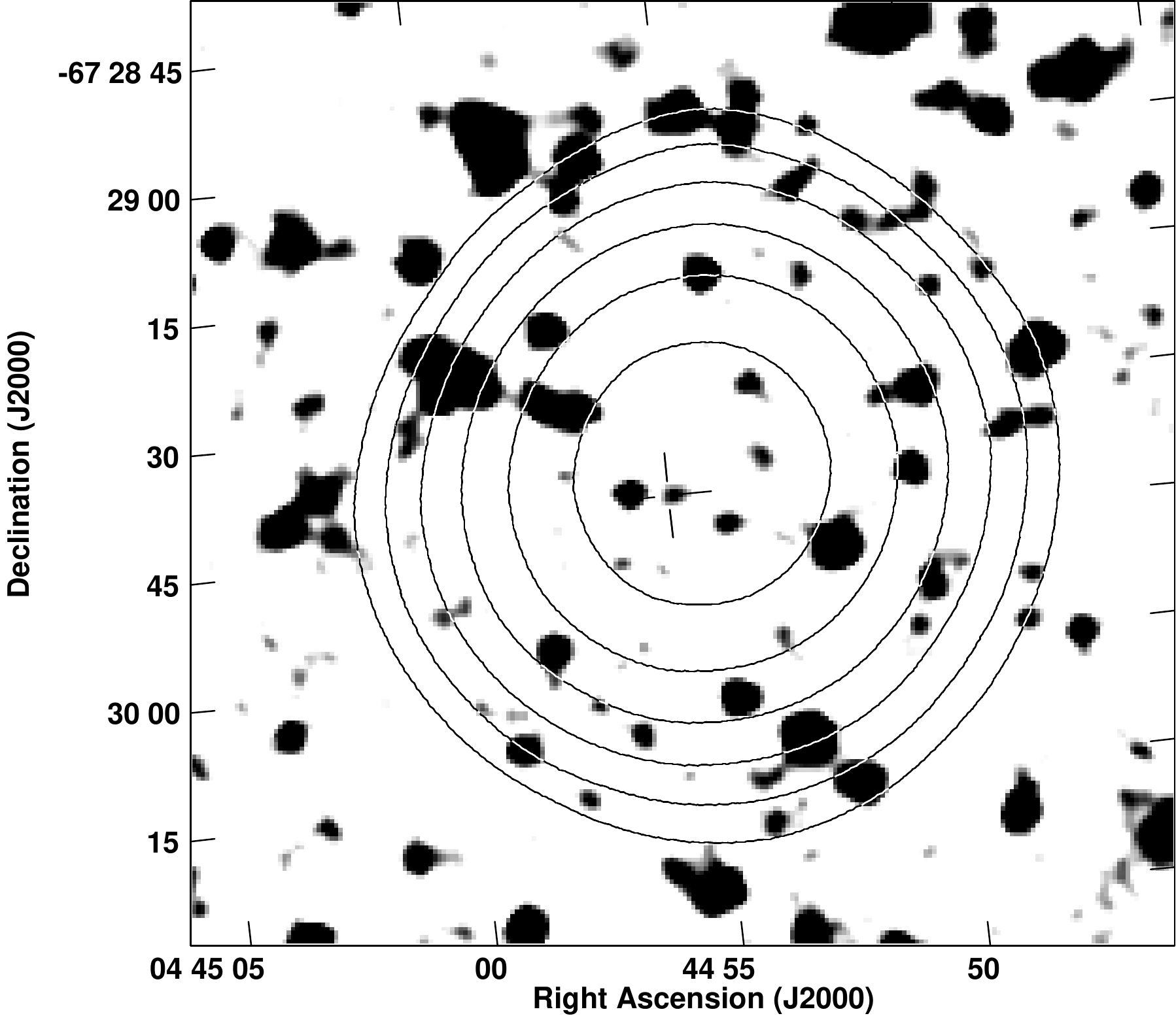}
\end{minipage}%
\begin{minipage}{0.3\textwidth}
  \centering
\includegraphics[angle=0,scale=.3]{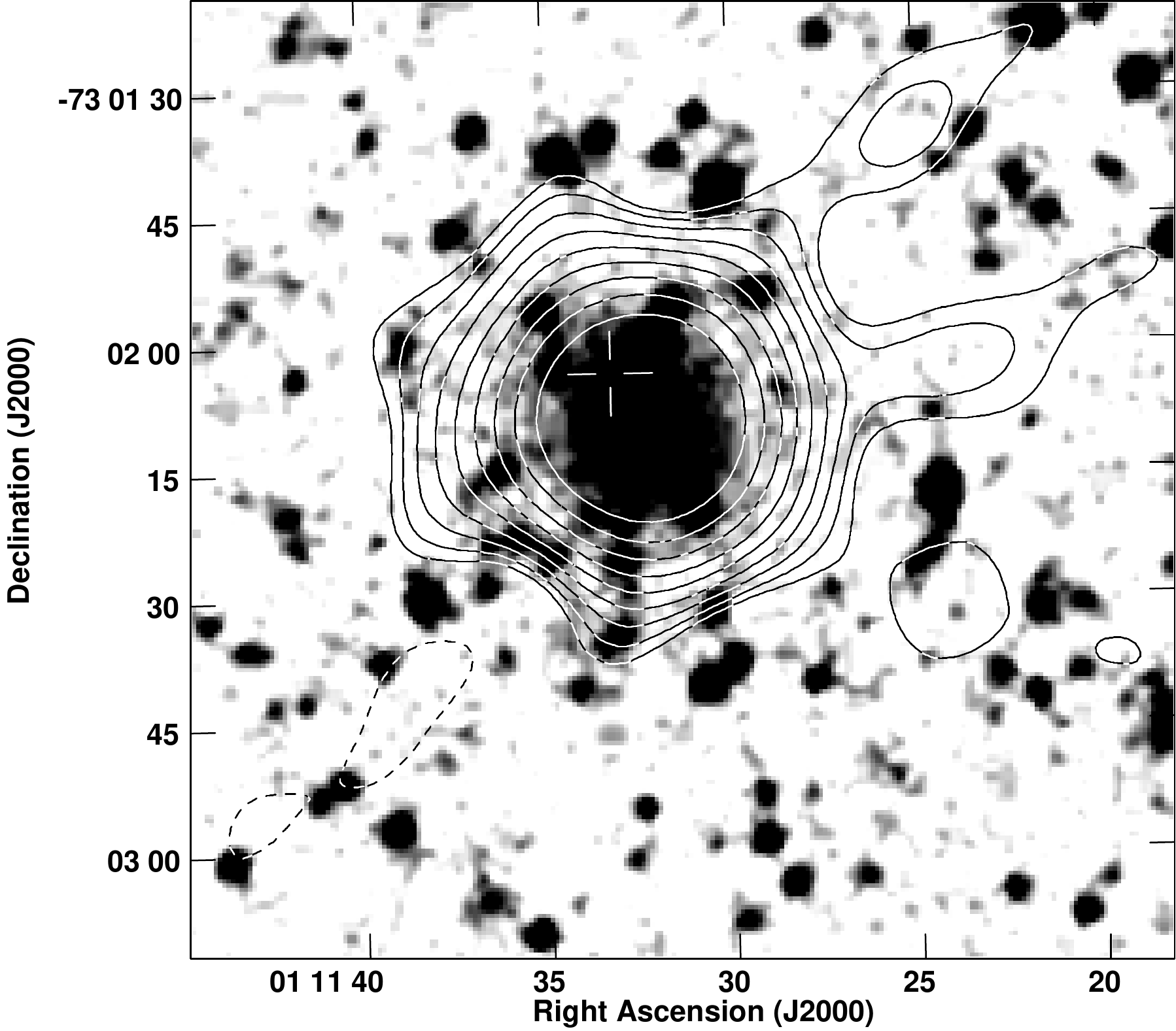}
\includegraphics[angle=0,scale=.3]{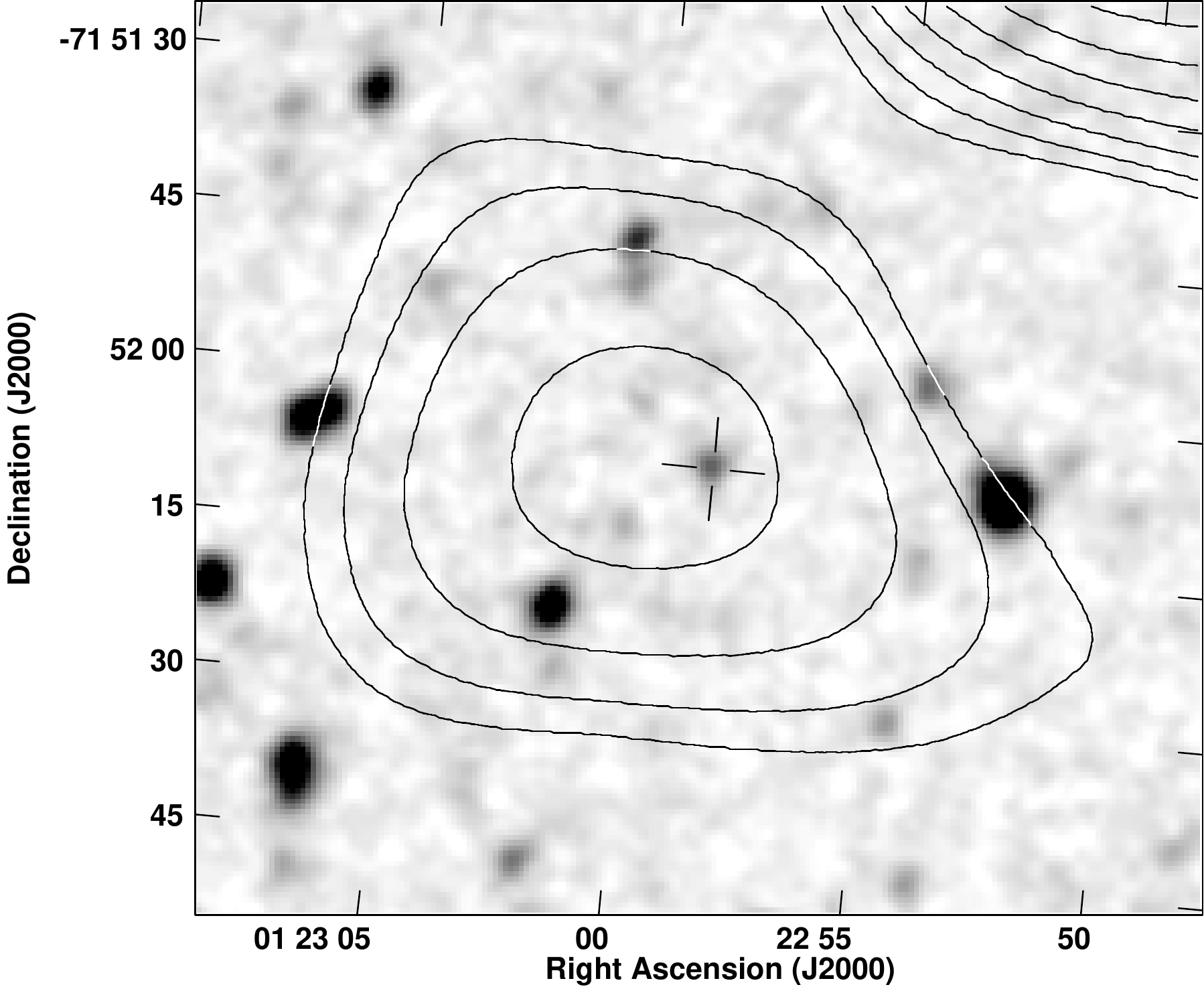}
\includegraphics[angle=0,scale=.3]{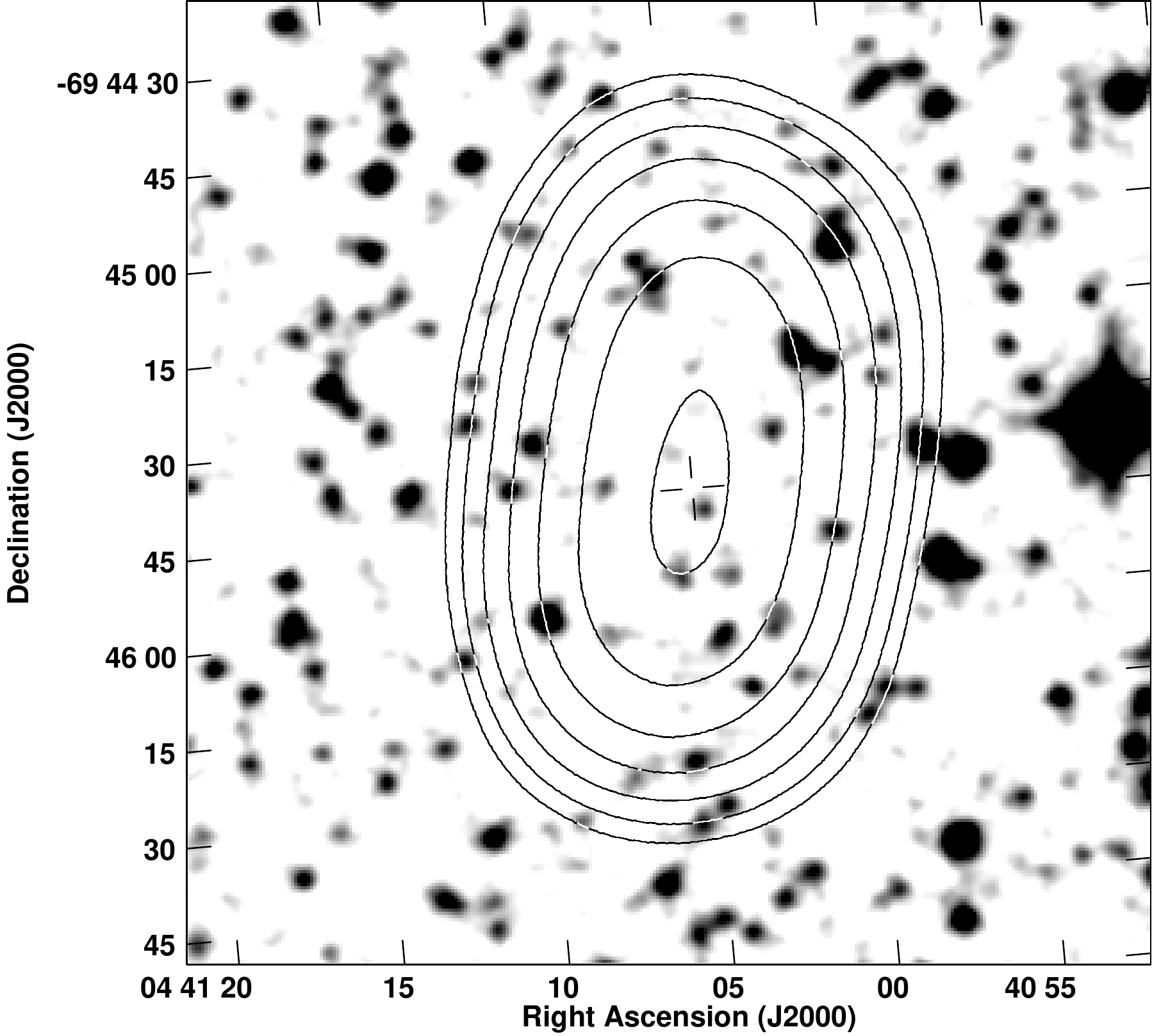}
\includegraphics[angle=0,scale=.3]{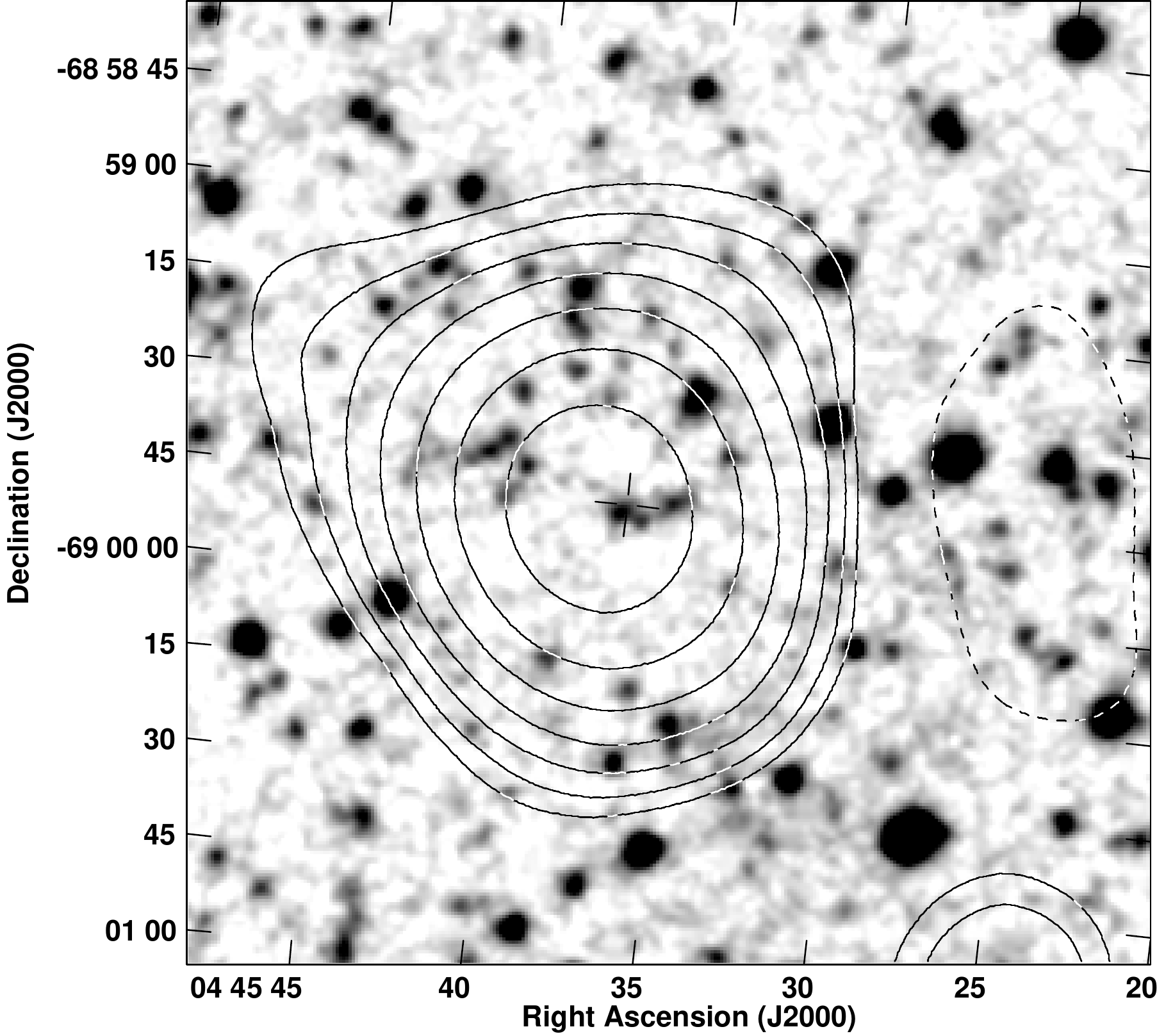}
\end{minipage}%
\caption{Contour maps of all sampled blazar candidates. The contour maps were created with DSS-R band optical images (greys) and 8.6, 4.85 or 0.845 GHz radio data (contours). A black cross marks the blazar candidate optical position from the MQS catalogue or the FS list.}\label{BlazarMaps}
\end{figure}
\setcounter{figure}{9}
\begin{figure}
\begin{minipage}{0.3\textwidth}
  \centering
\includegraphics[angle=0,scale=.3]{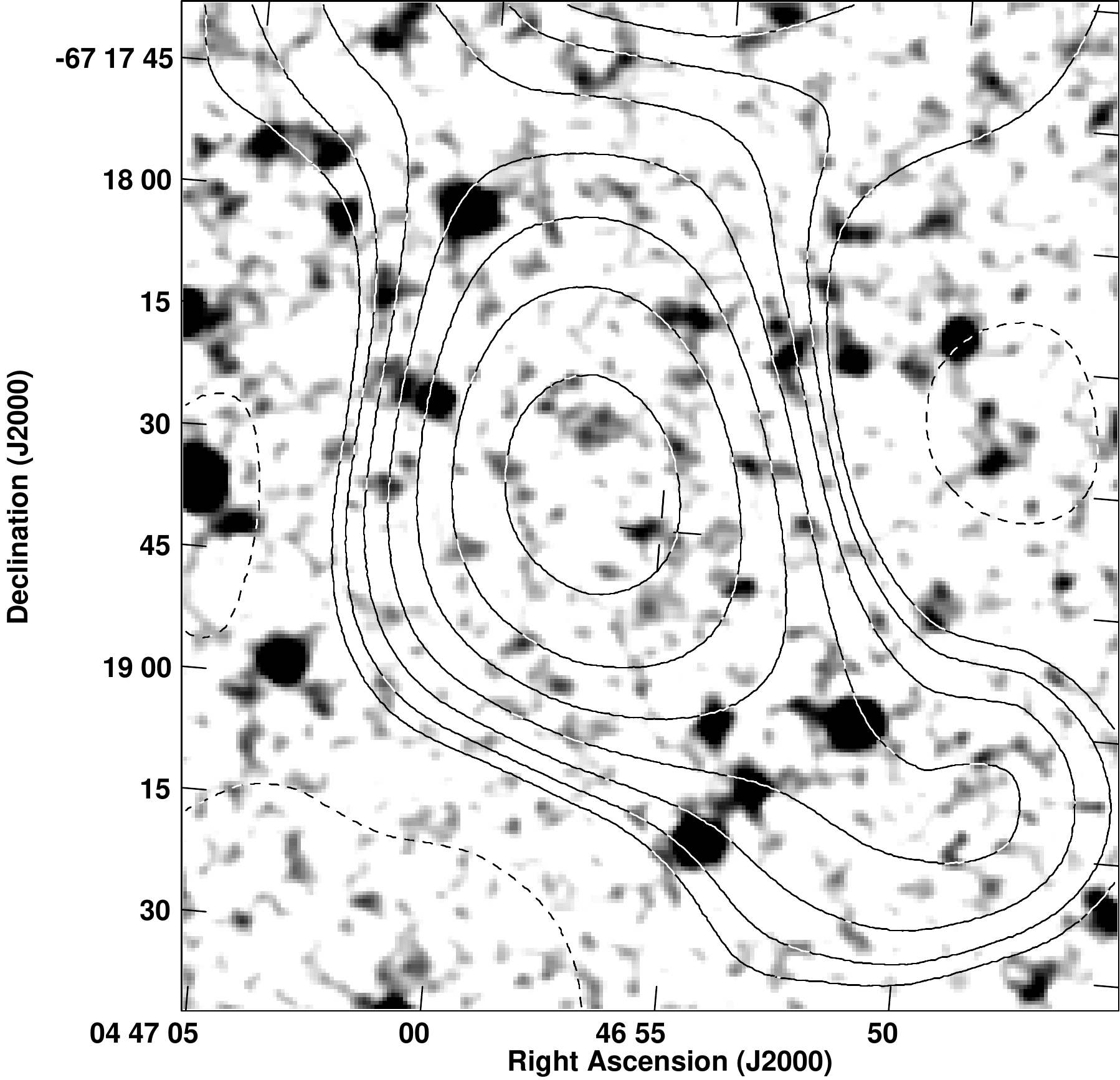}
\includegraphics[angle=0,scale=.25]{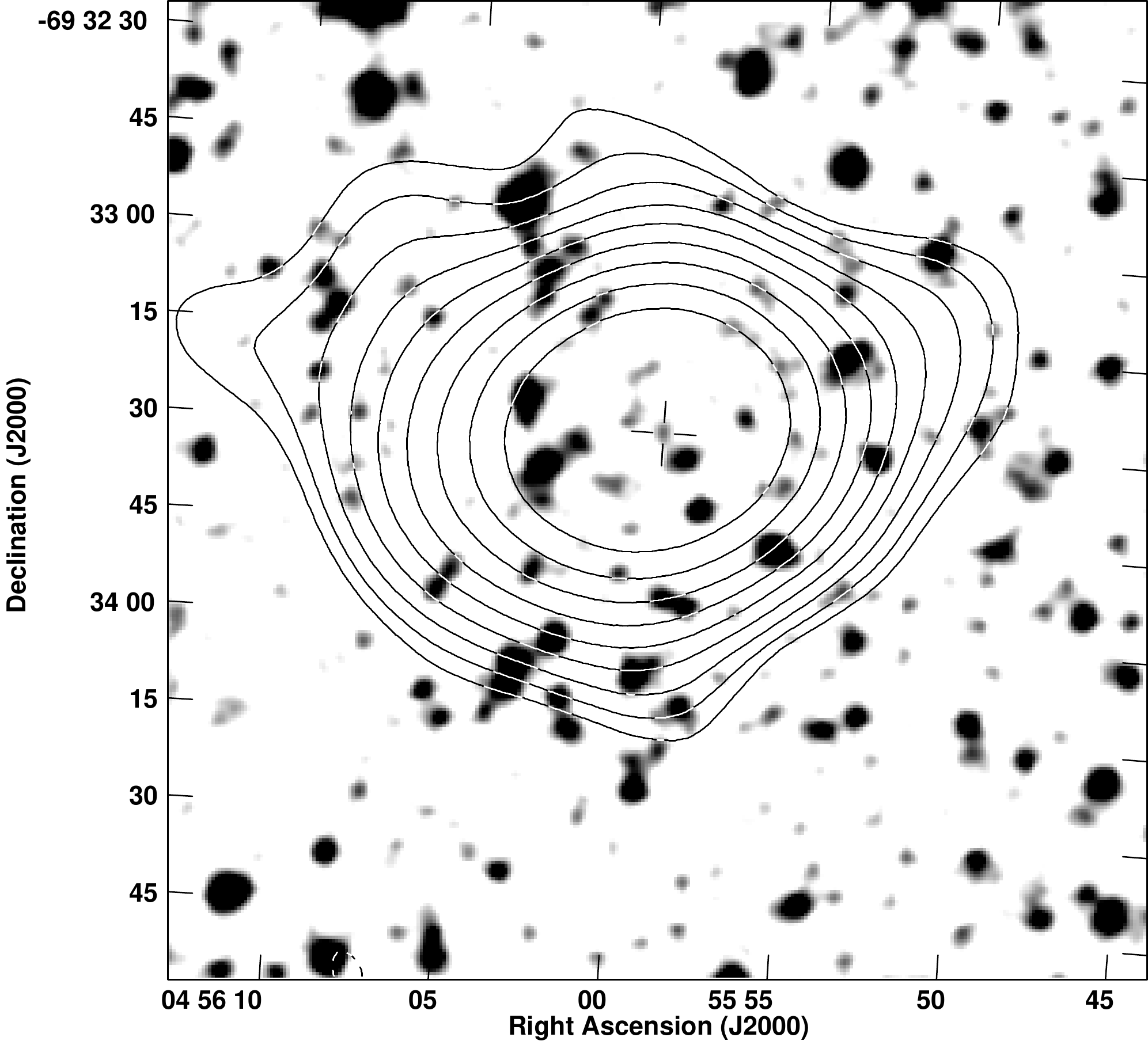}
\includegraphics[angle=0,scale=.25]{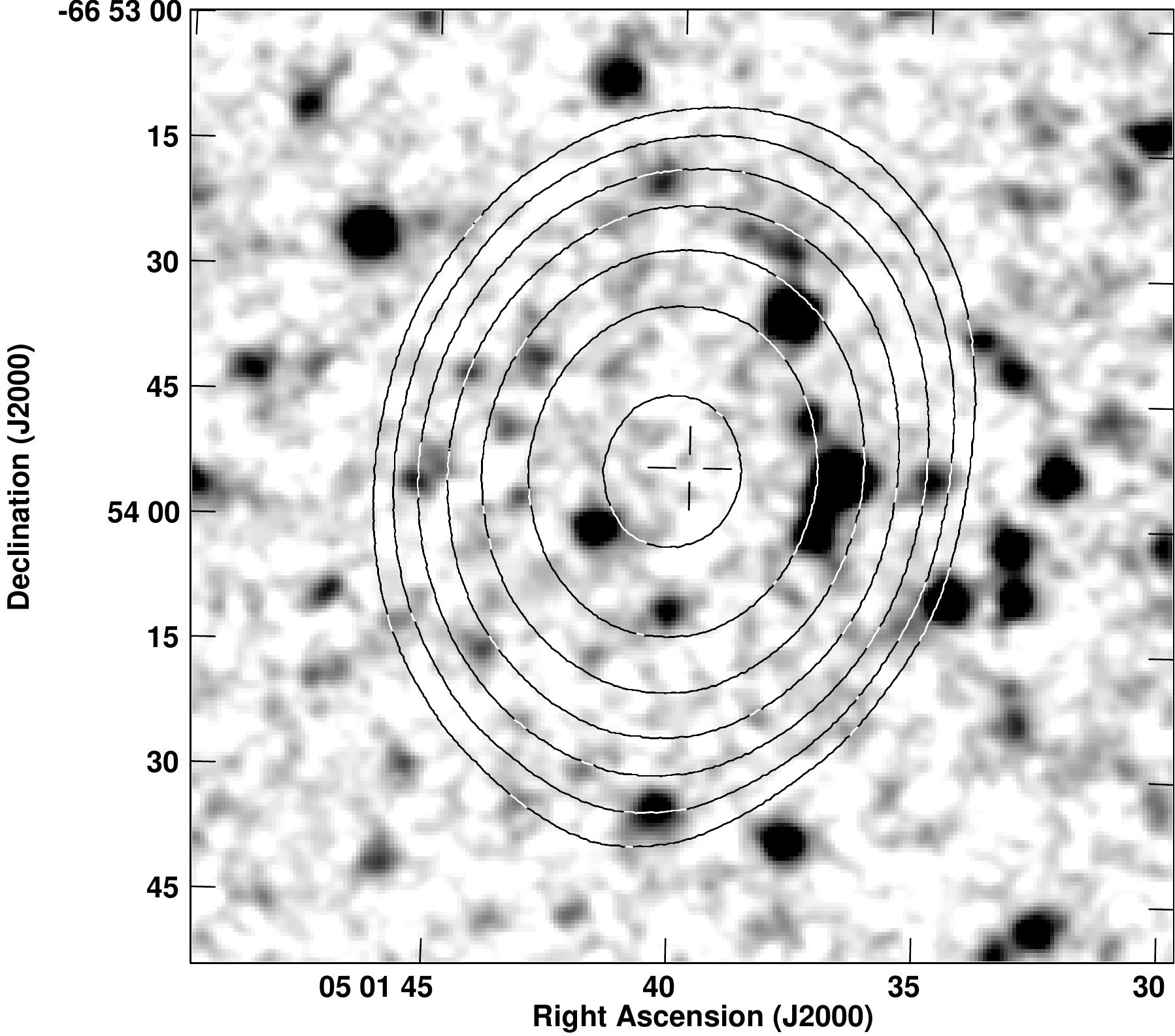}
\includegraphics[angle=0,scale=.3]{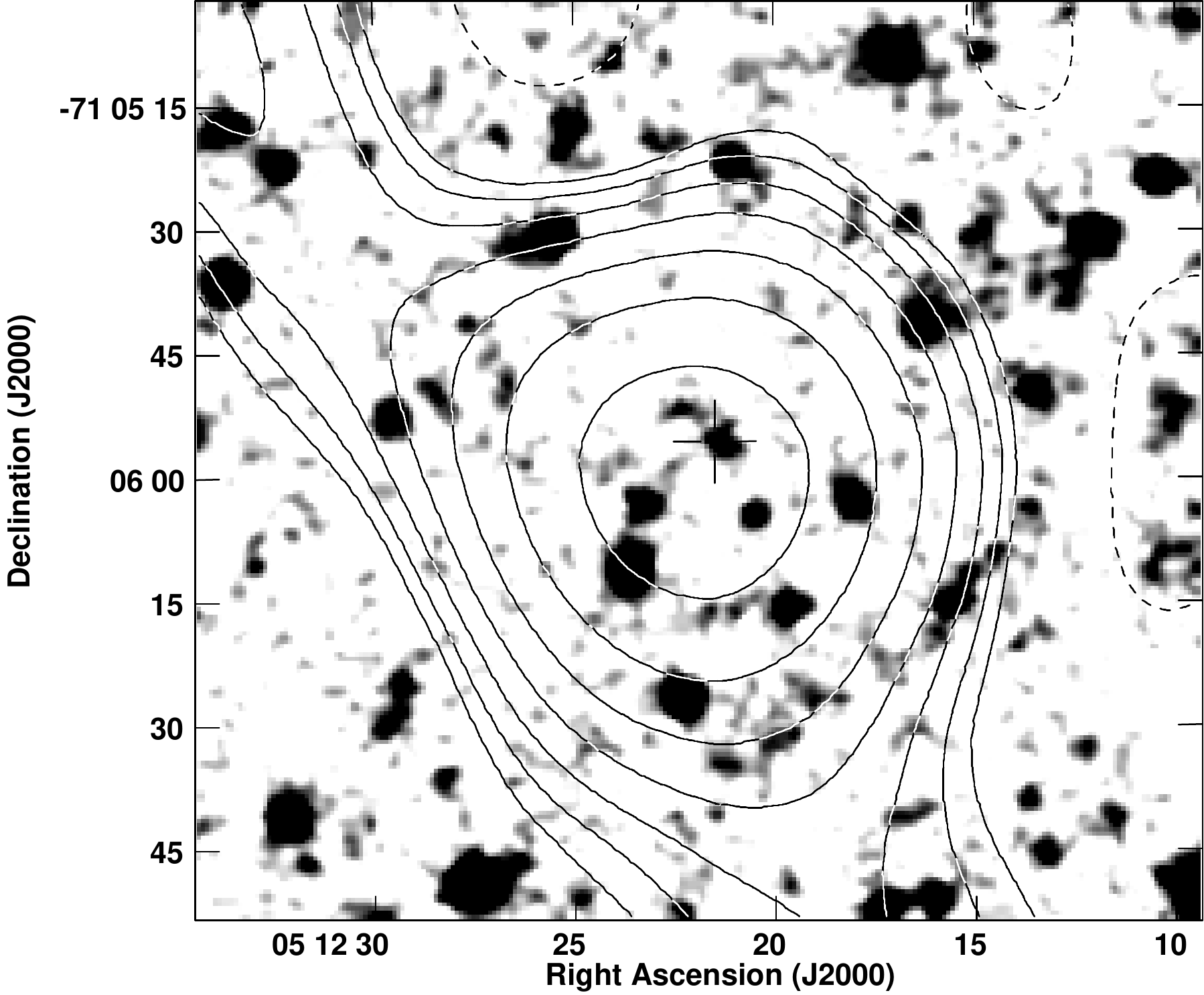}
\end{minipage}%
\begin{minipage}{0.3\textwidth}
  \centering
\includegraphics[angle=0,scale=.3]{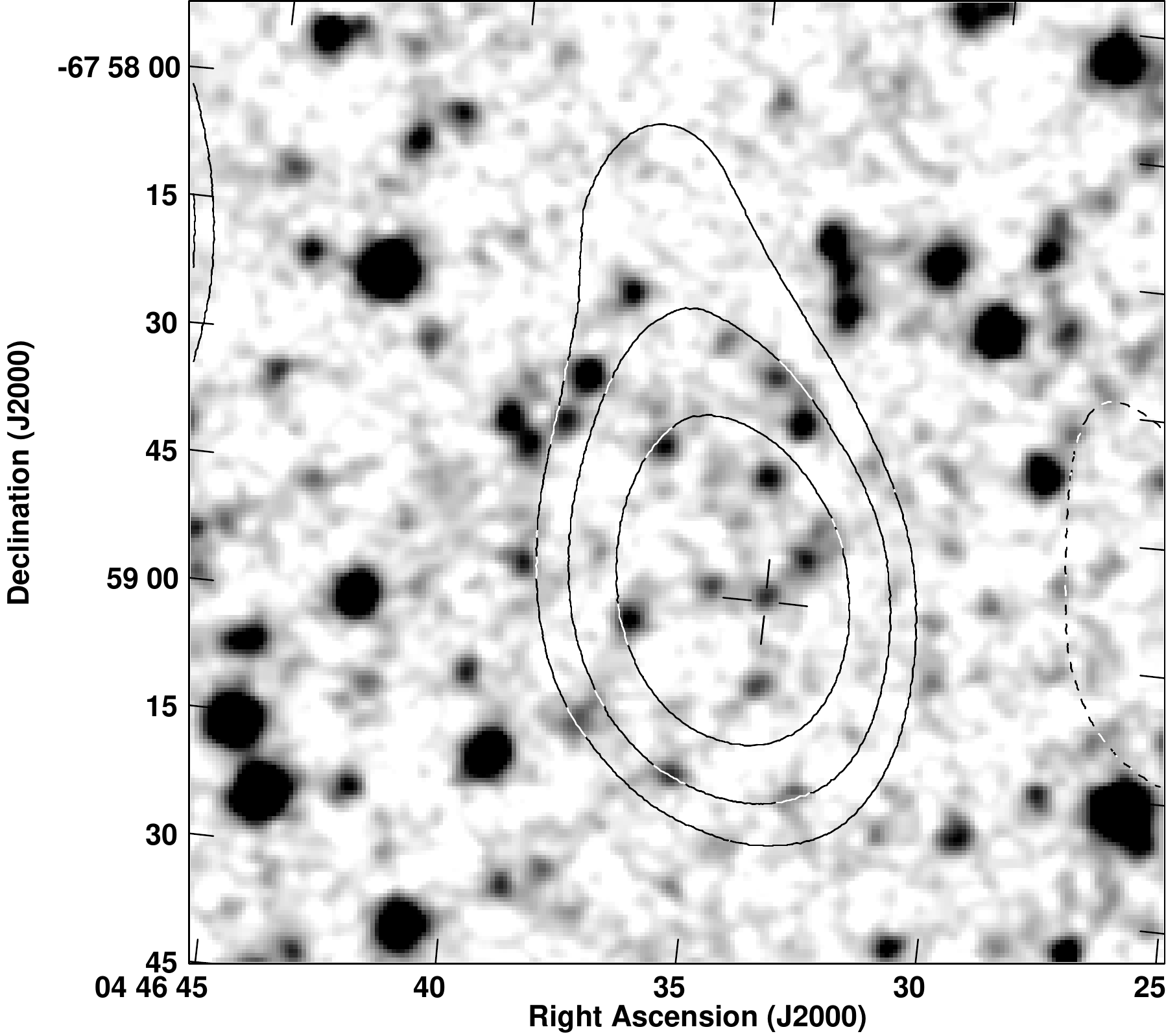}
\includegraphics[angle=0,scale=.25]{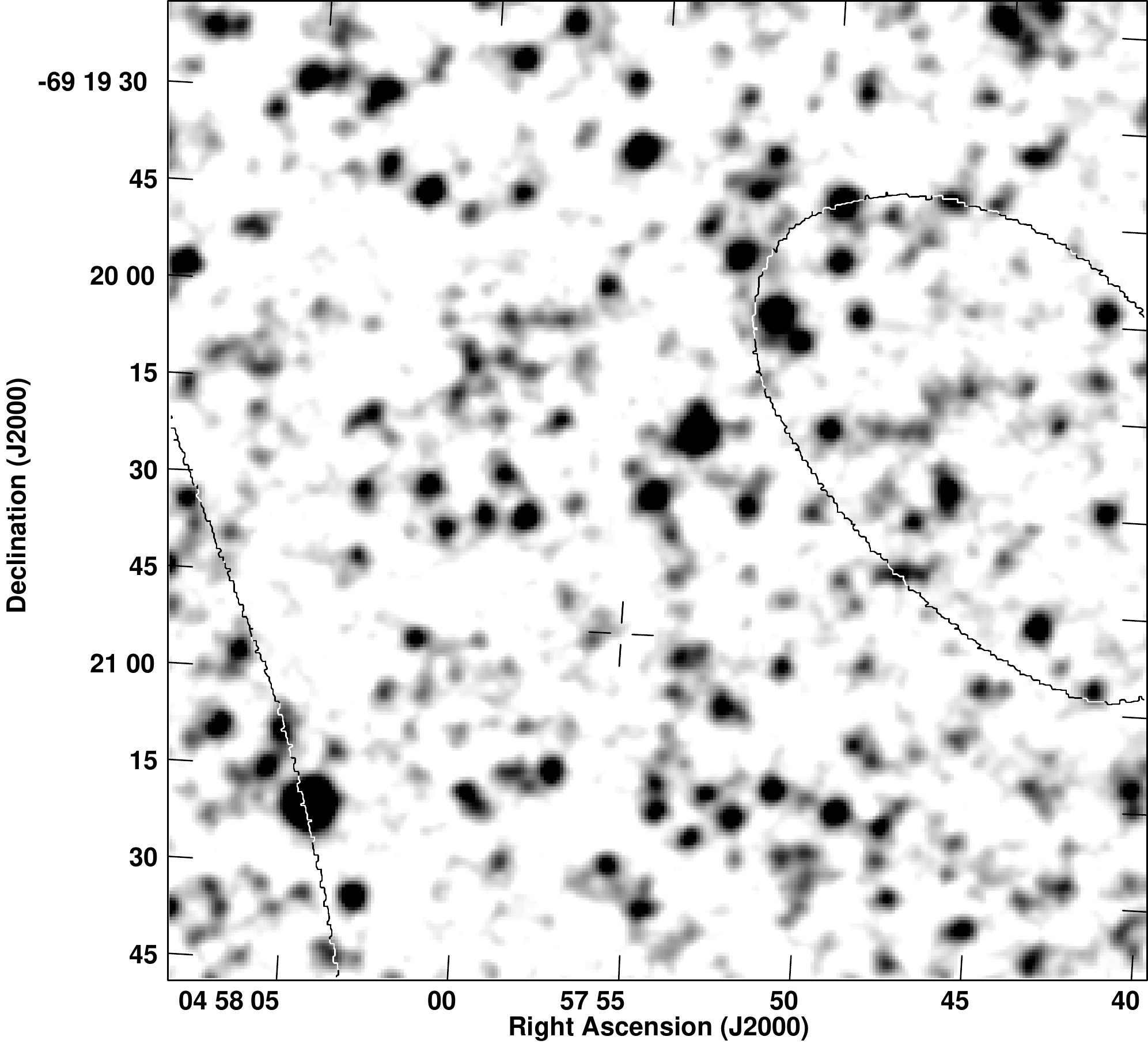}
\includegraphics[angle=0,scale=.25]{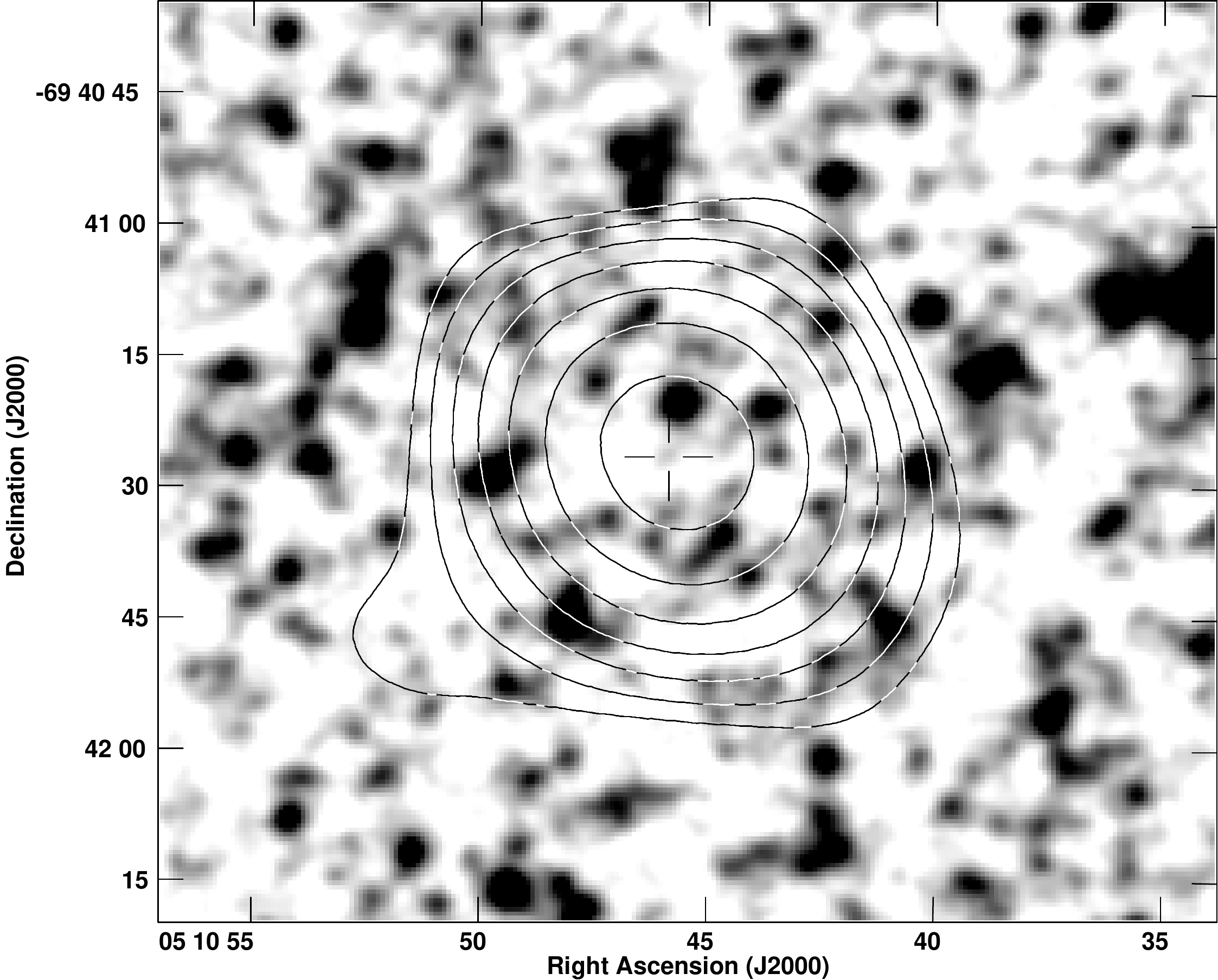}
\includegraphics[angle=0,scale=.25]{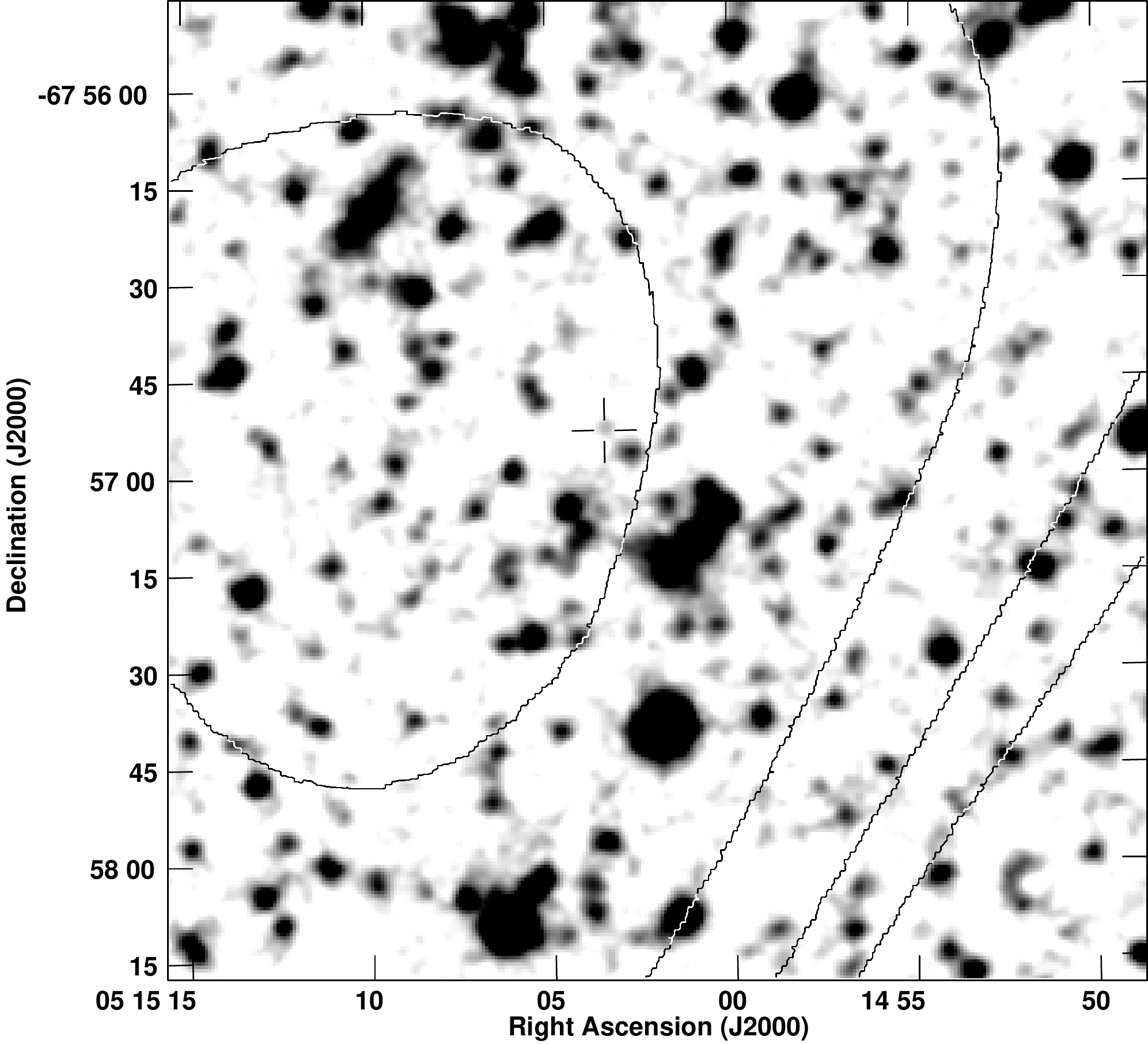}
\end{minipage}%
\begin{minipage}{0.3\textwidth}
  \centering
\includegraphics[angle=0,scale=.35]{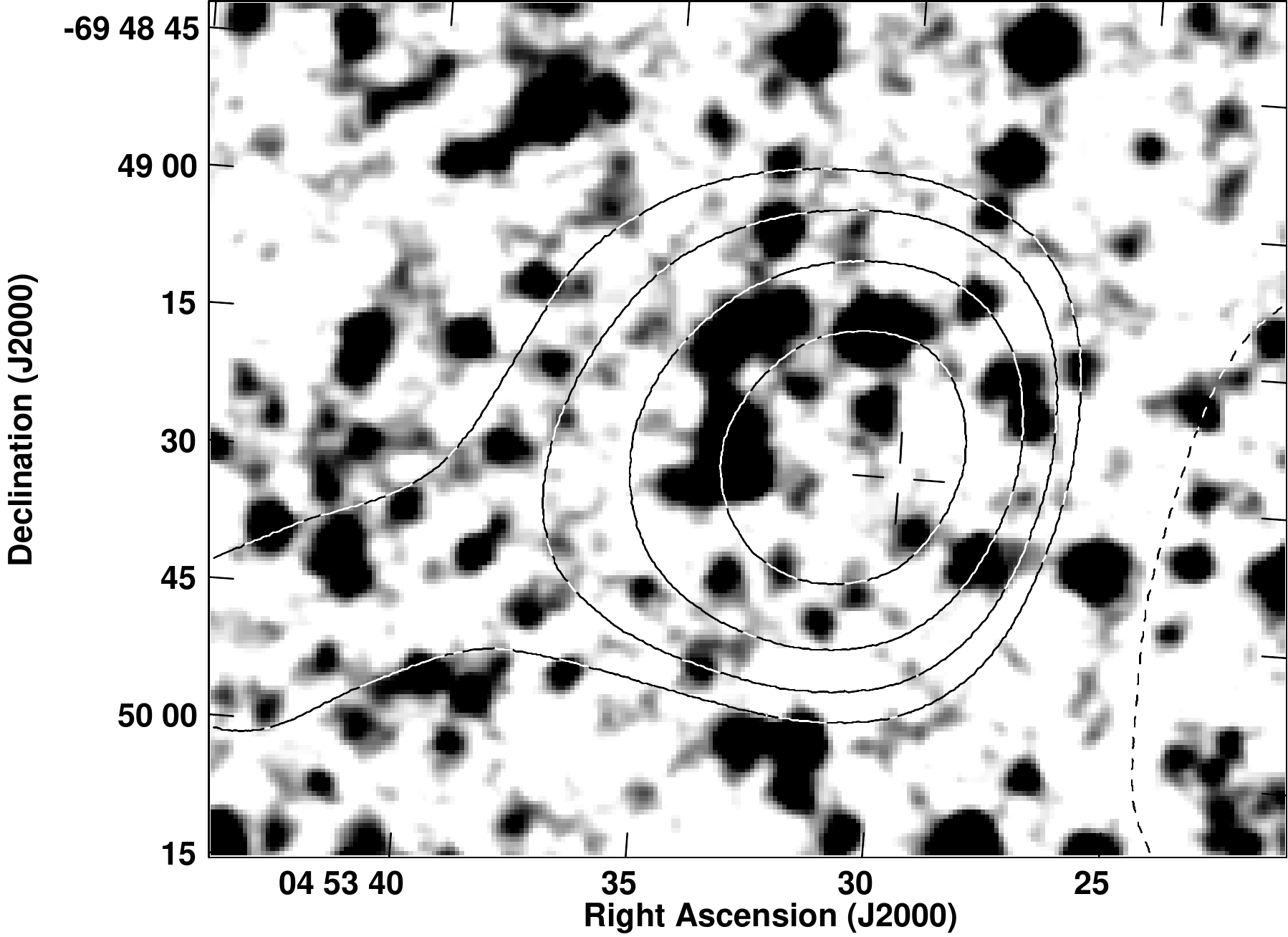}
\includegraphics[angle=0,scale=.35]{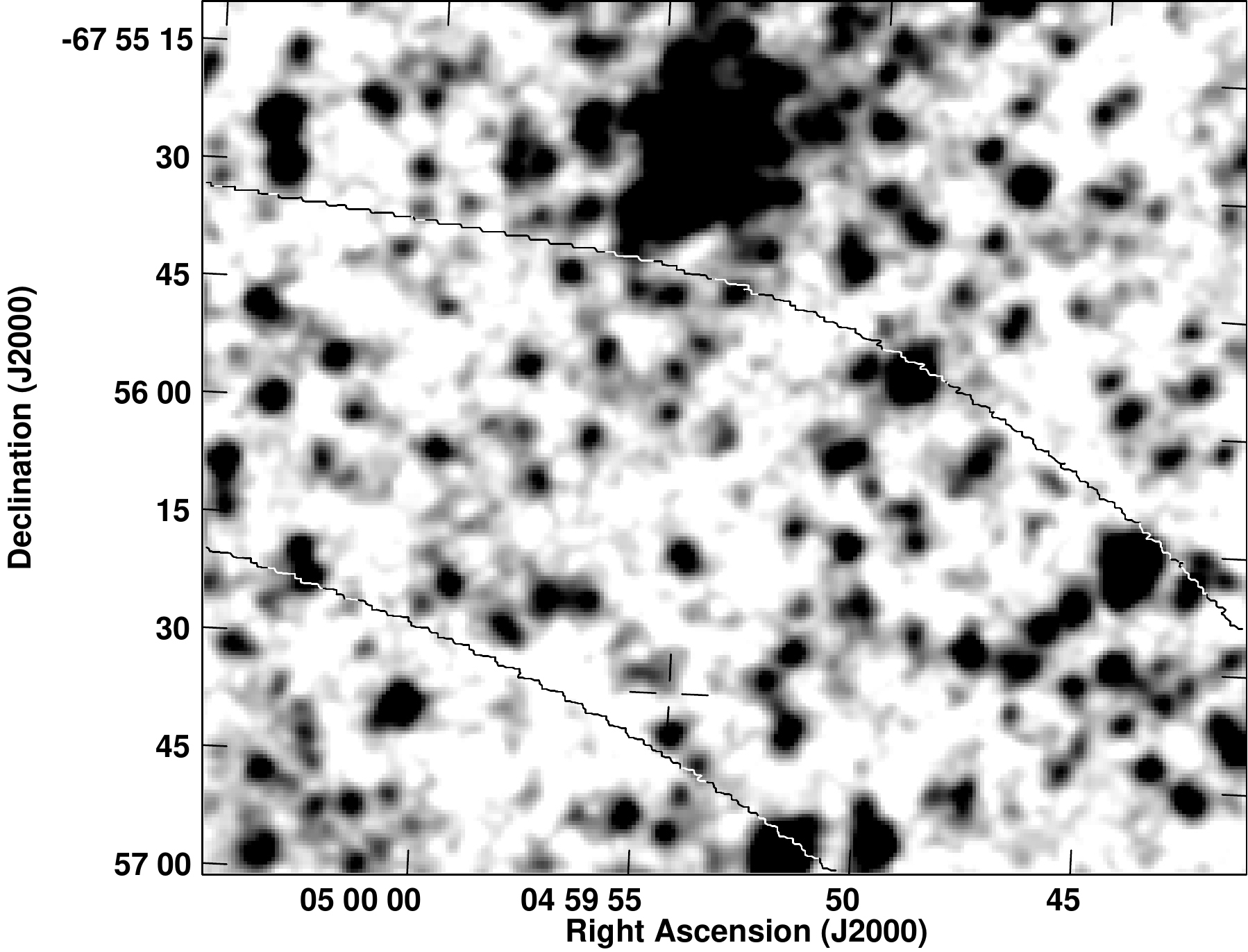}
\includegraphics[angle=0,scale=.35]{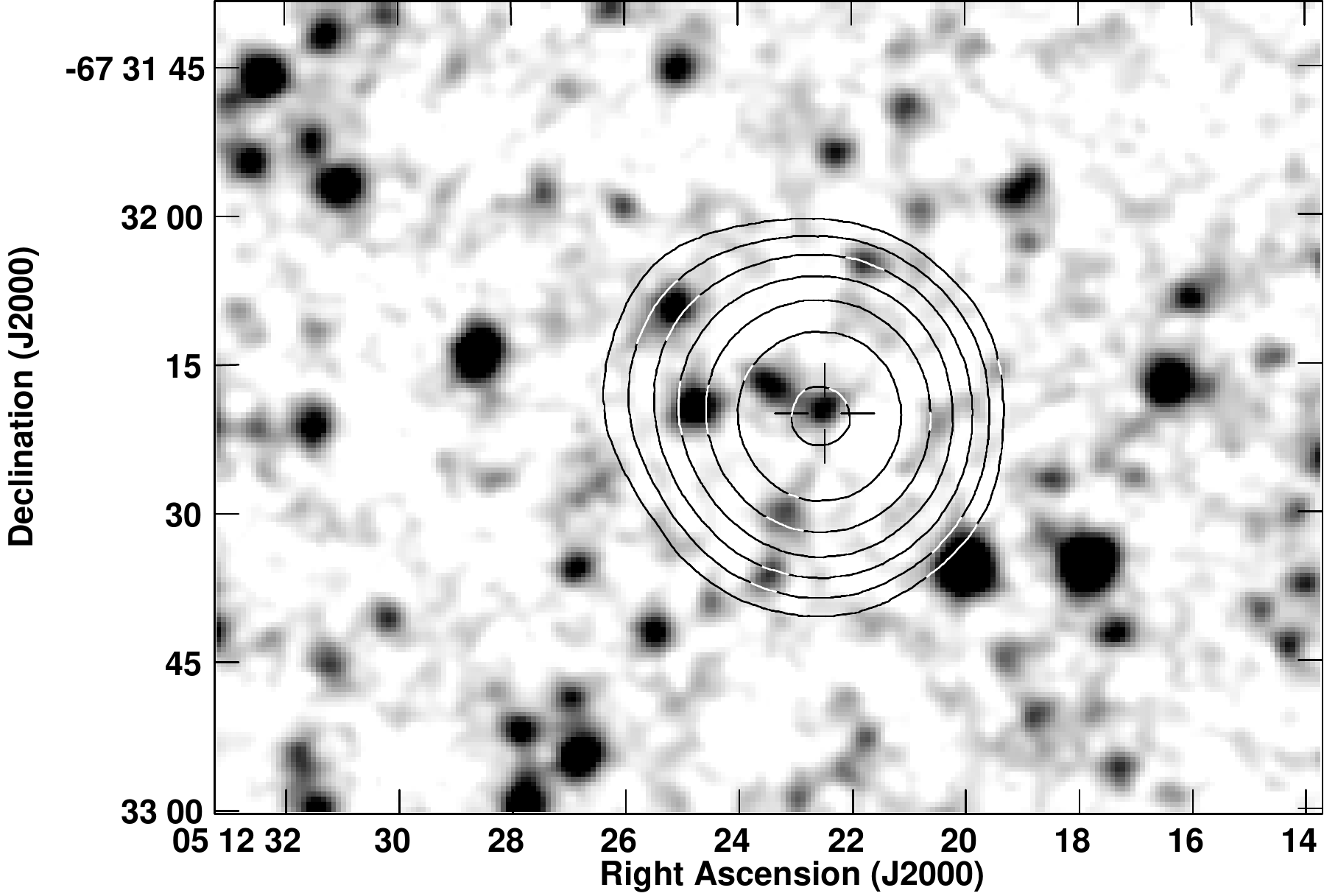}
\includegraphics[angle=0,scale=.25]{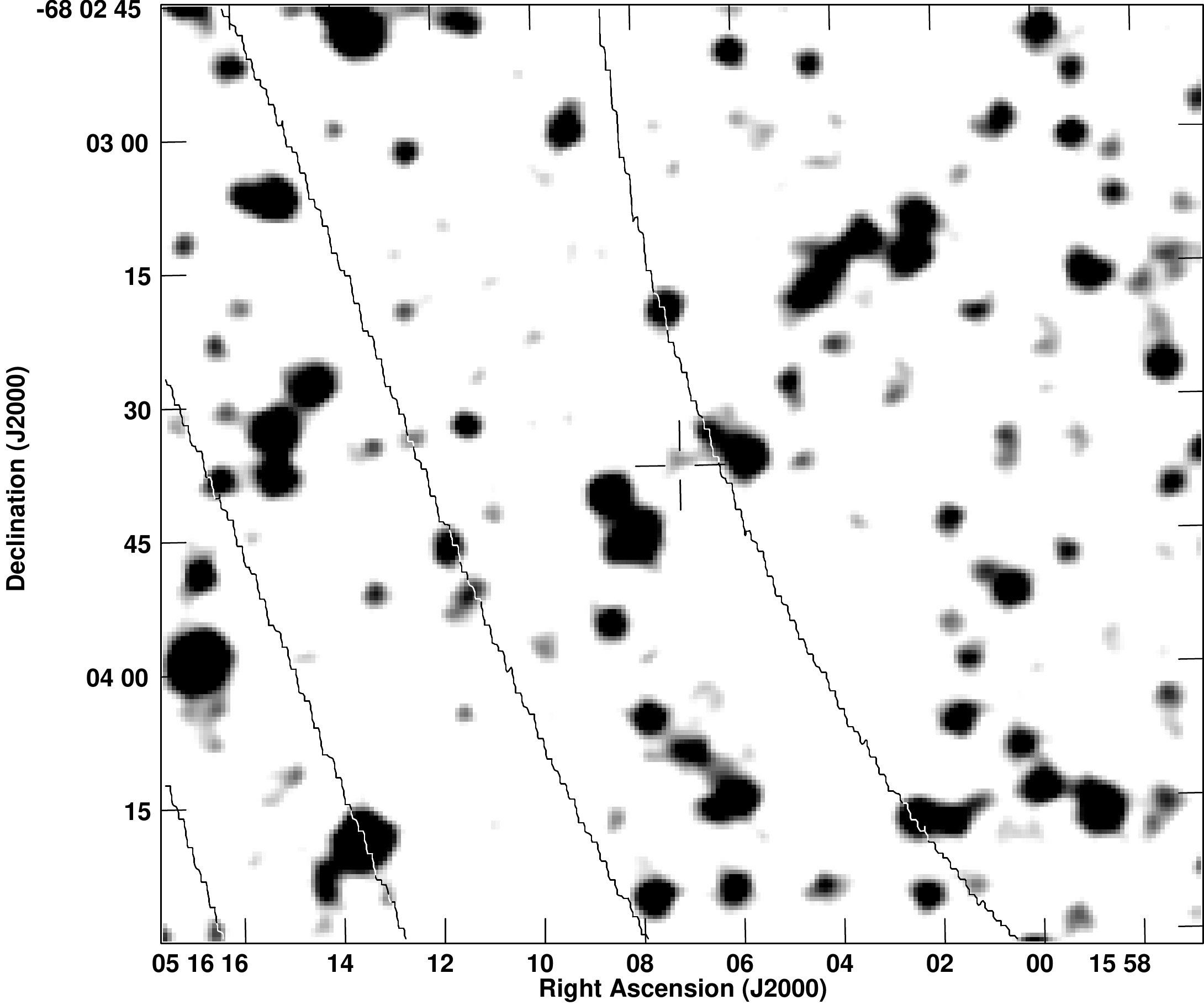}
\end{minipage}%
\caption[]{Continued.}
\end{figure}
\setcounter{figure}{9}
\begin{figure}
\begin{minipage}{0.3\textwidth}
  \centering
\includegraphics[angle=0,scale=.3]{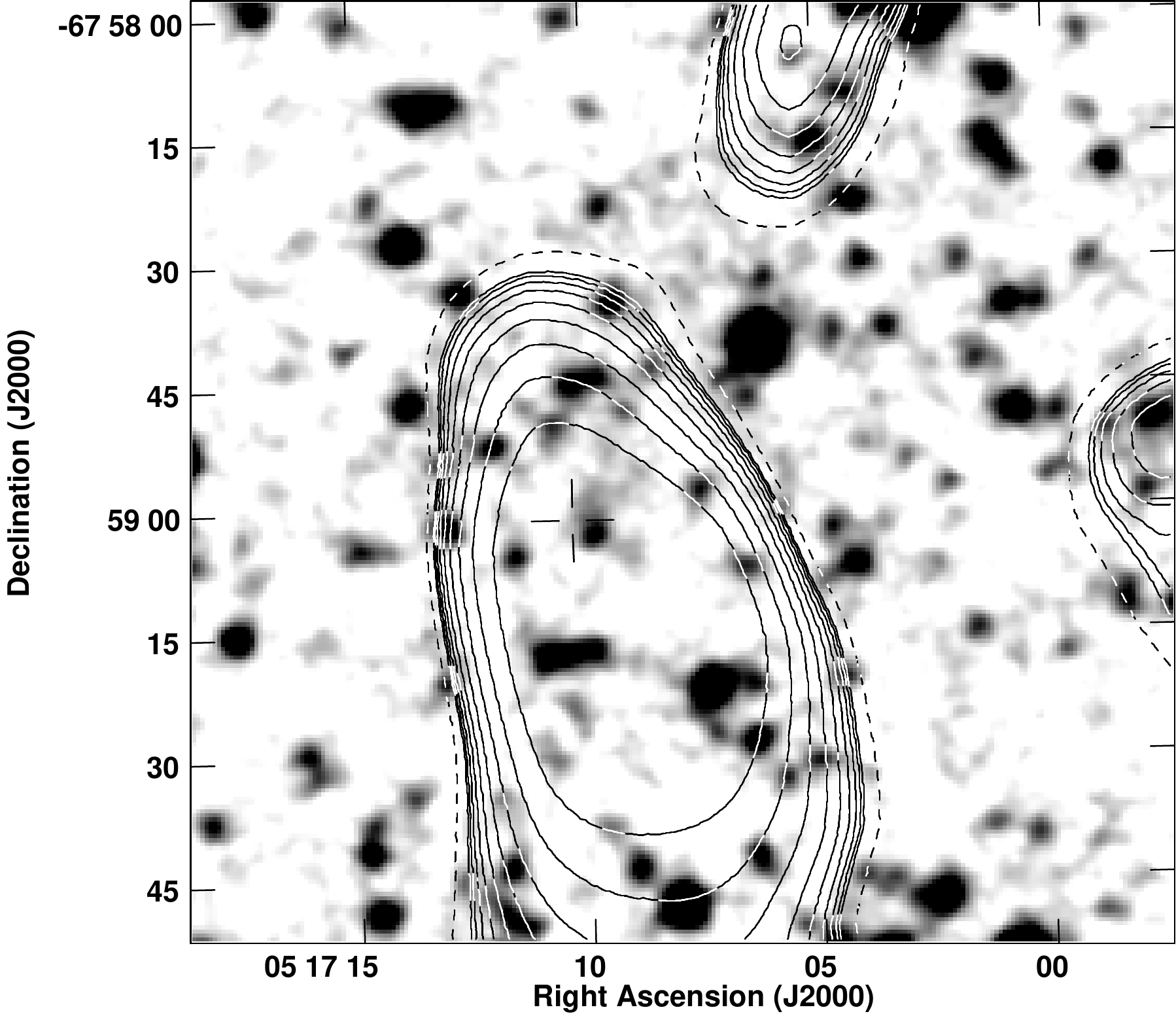}
\includegraphics[angle=0,scale=.28]{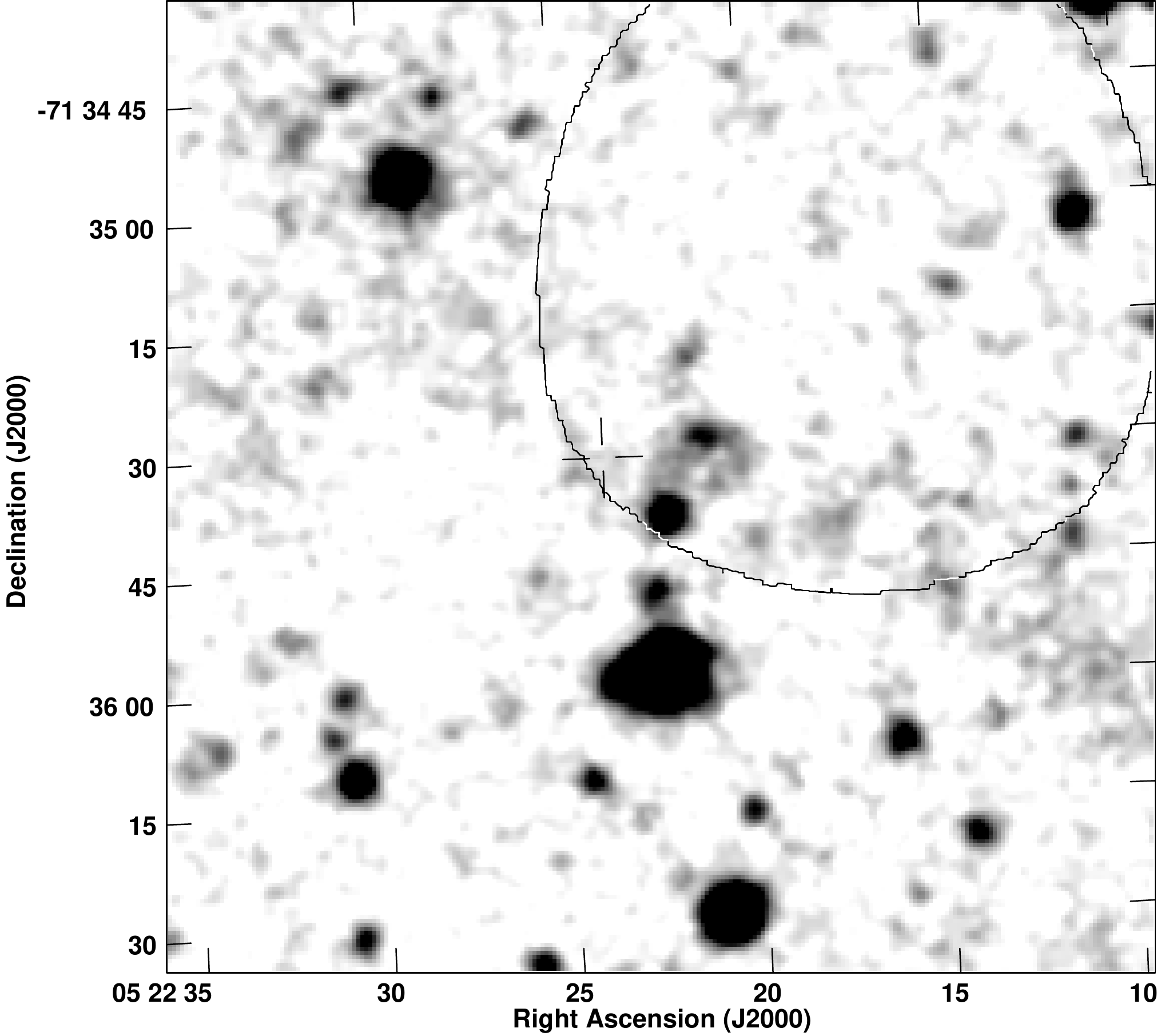}
\includegraphics[angle=0,scale=.3]{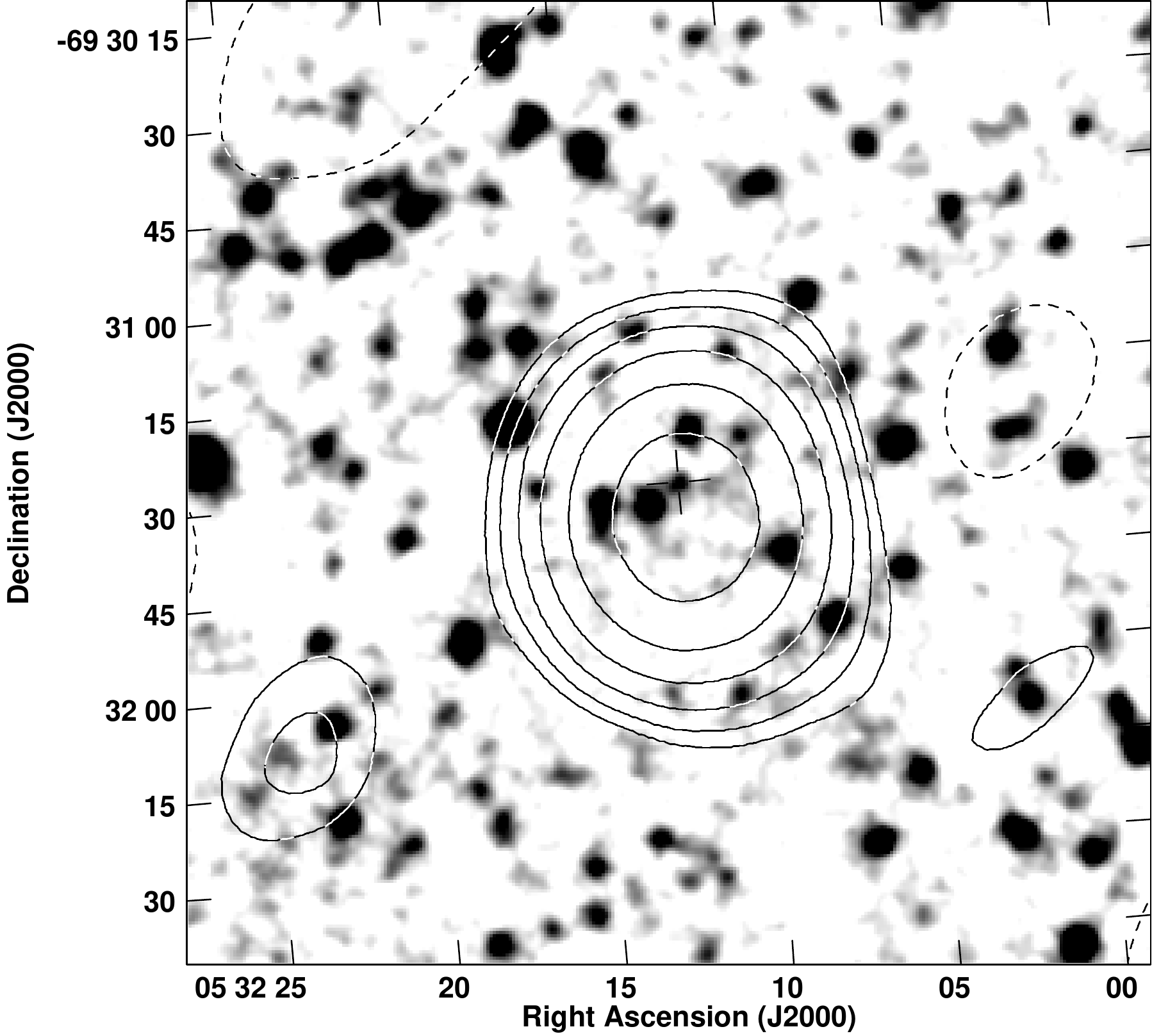}
\includegraphics[angle=0,scale=.3]{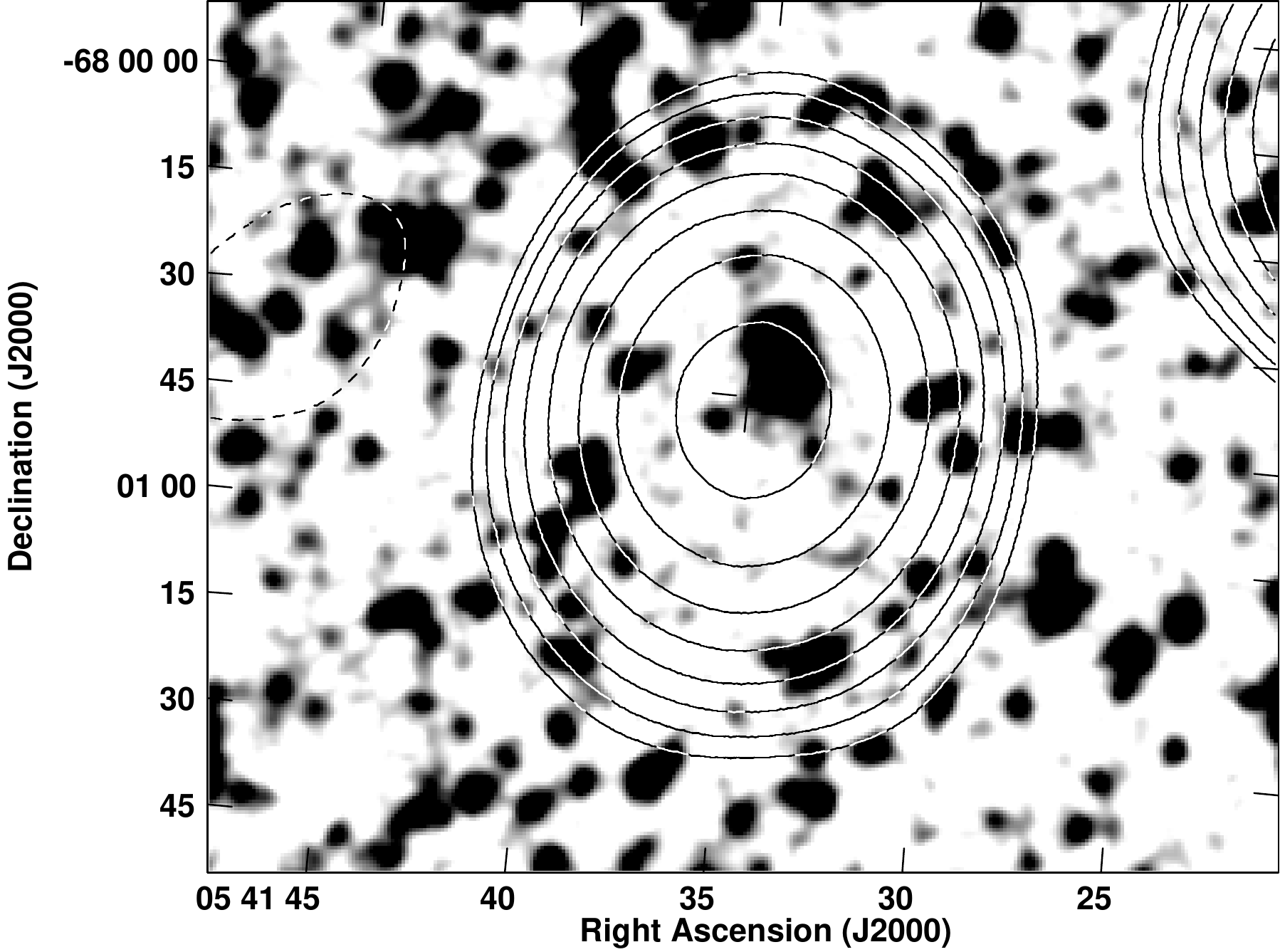}
\end{minipage}%
\begin{minipage}{0.3\textwidth}
  \centering
\includegraphics[angle=0,scale=.25]{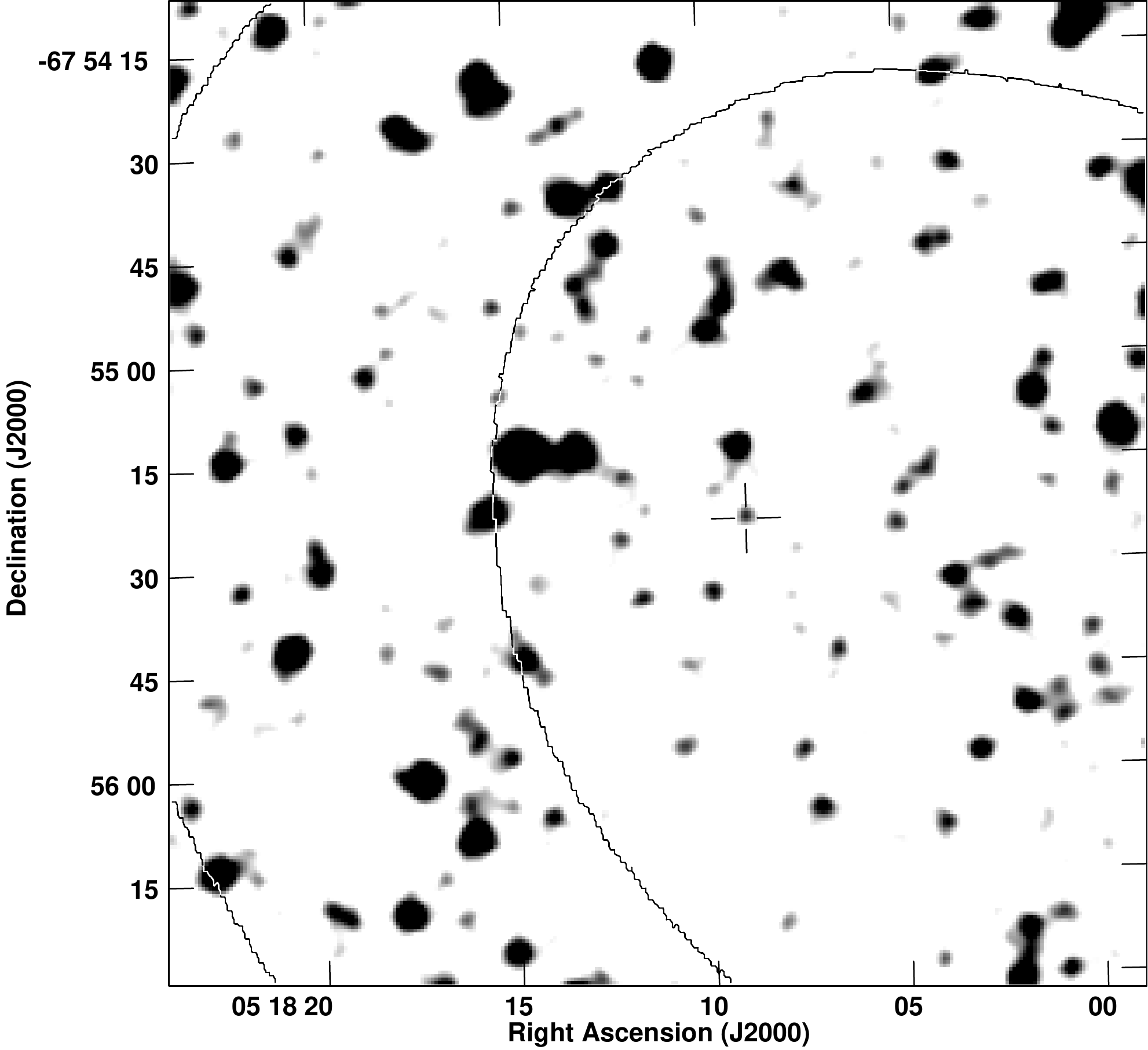}
\includegraphics[angle=0,scale=.3]{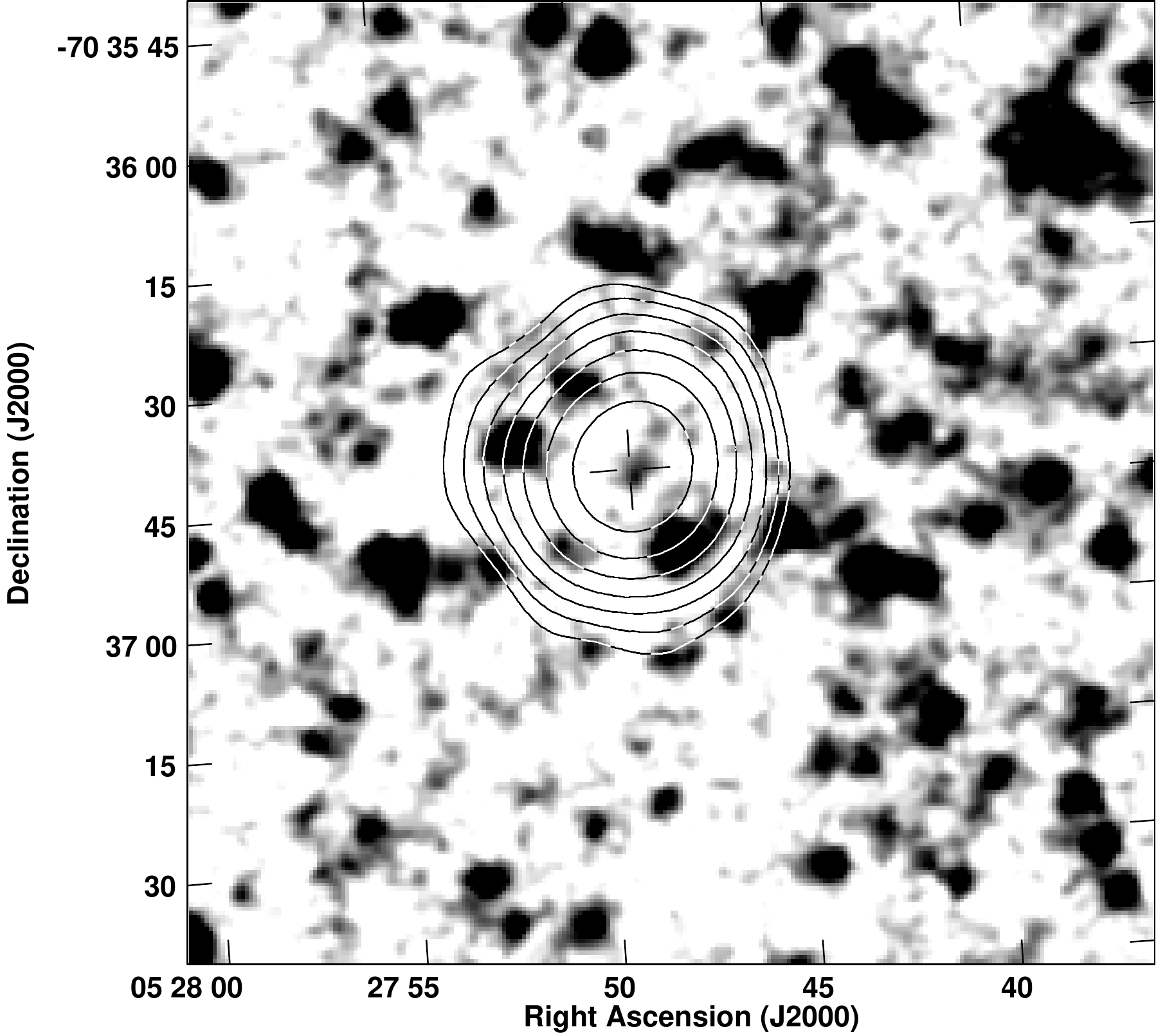}
\includegraphics[angle=0,scale=.3]{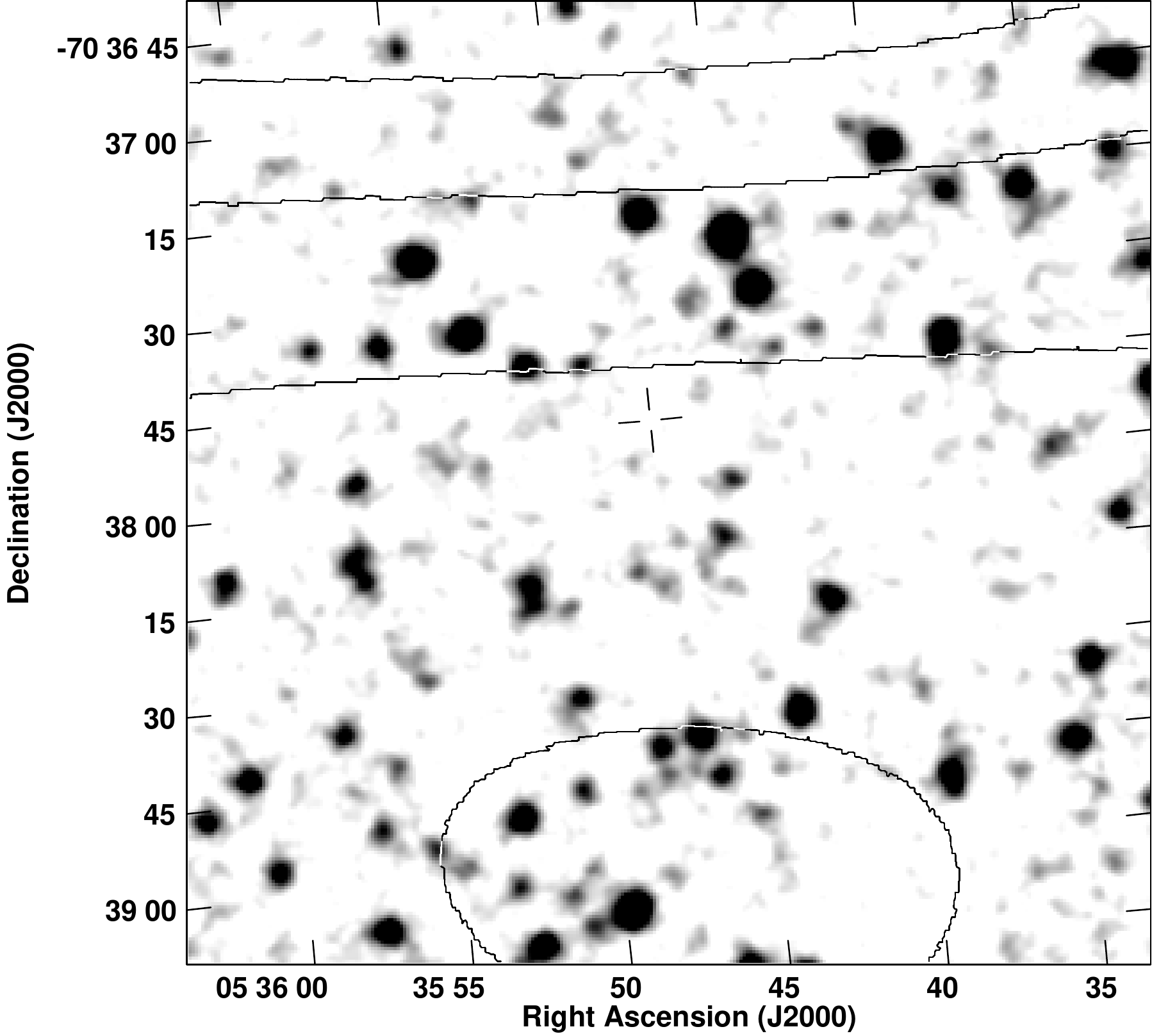}
\includegraphics[angle=0,scale=.3]{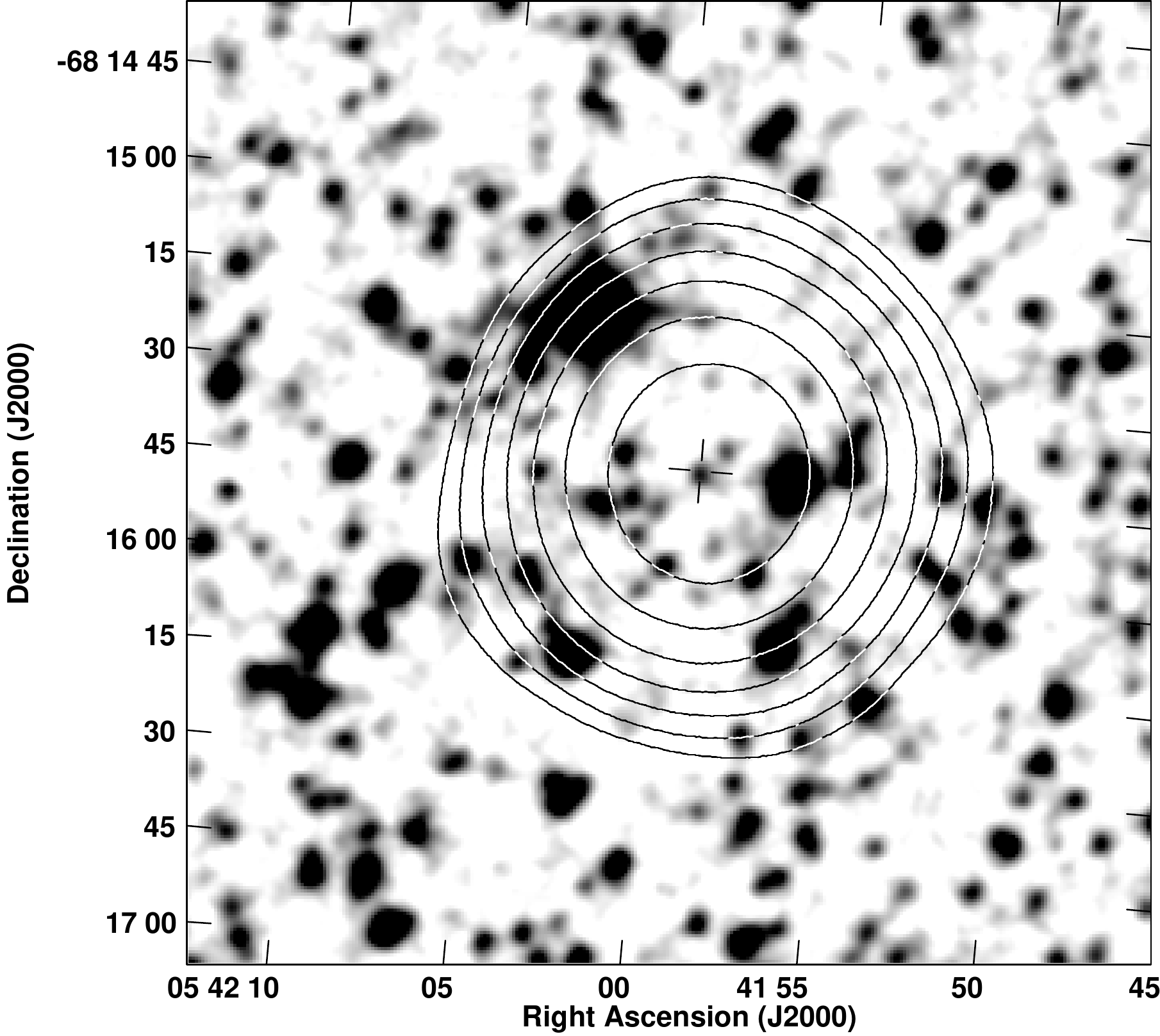}
\end{minipage}%
\begin{minipage}{0.3\textwidth}
  \centering
\includegraphics[angle=0,scale=.25]{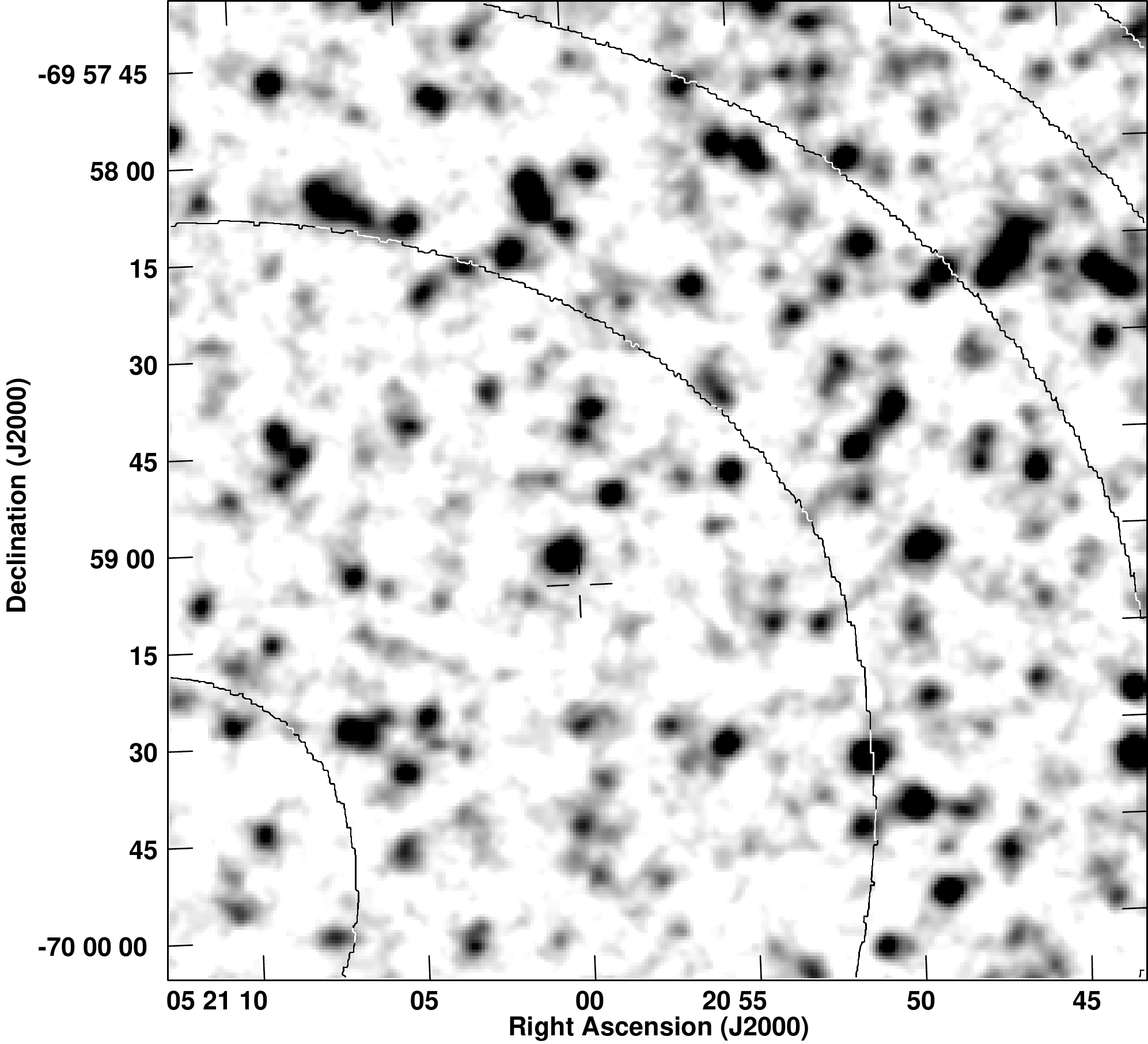}
\includegraphics[angle=0,scale=.3]{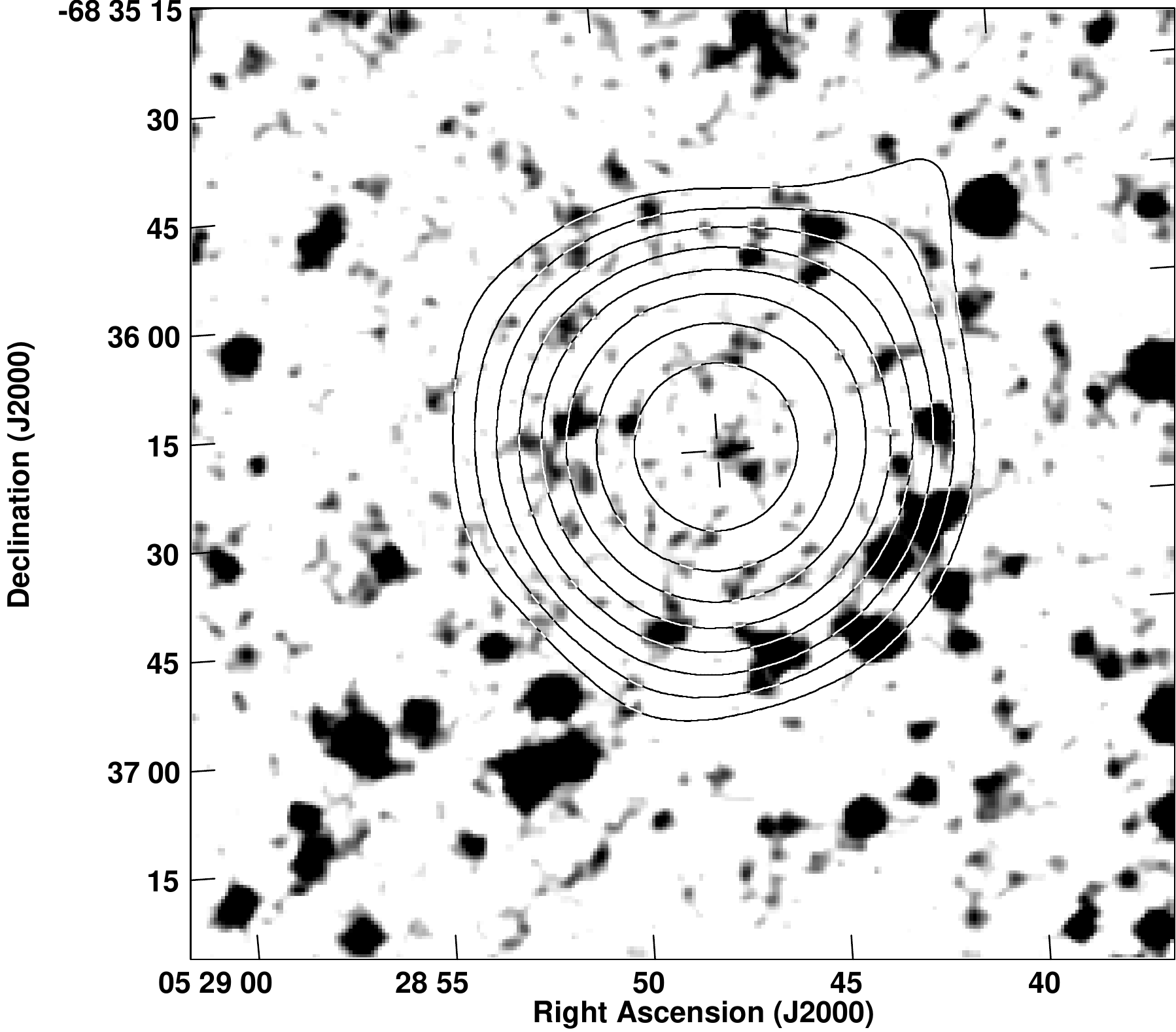}
\includegraphics[angle=0,scale=.3]{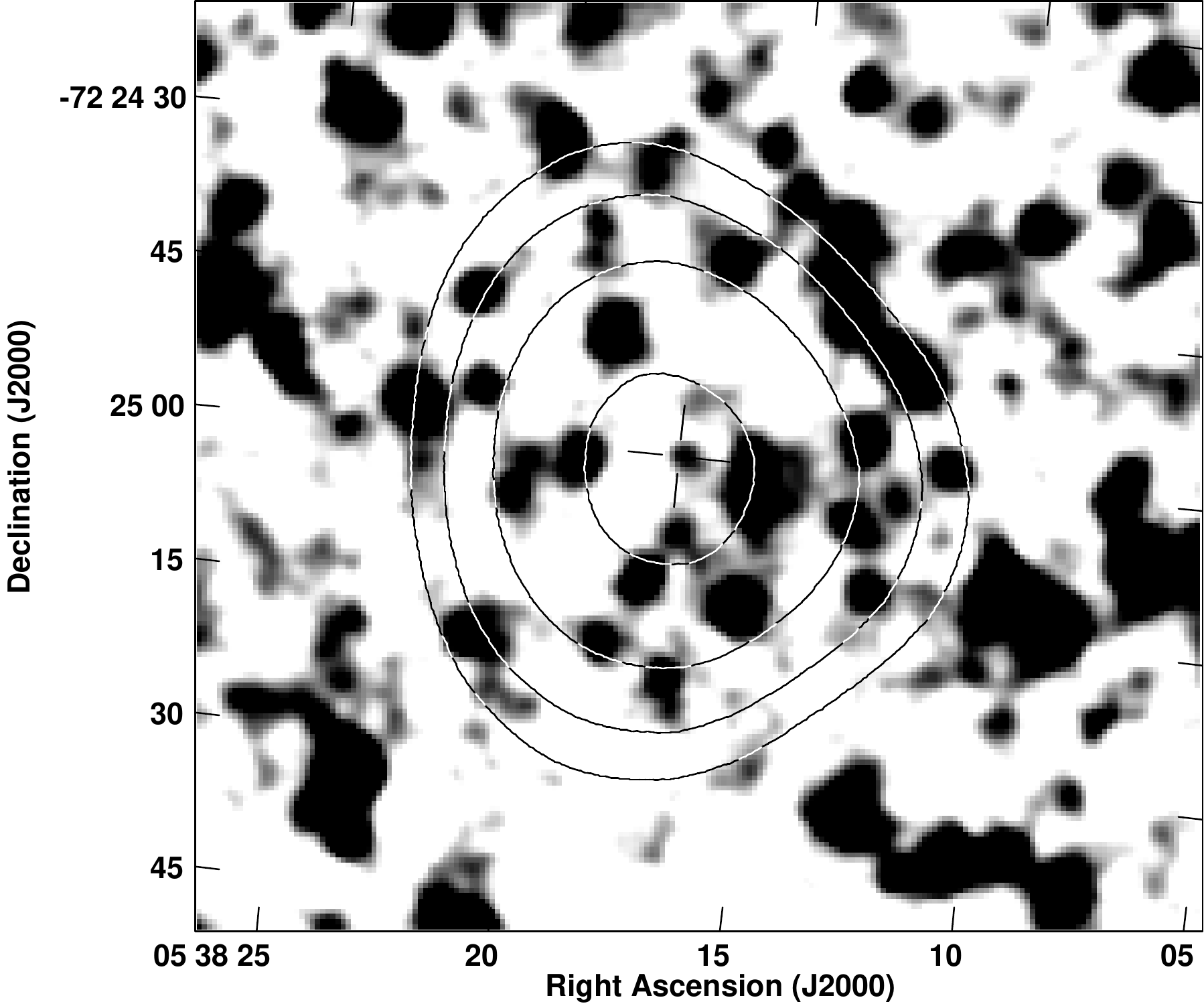}
\includegraphics[angle=0,scale=.3]{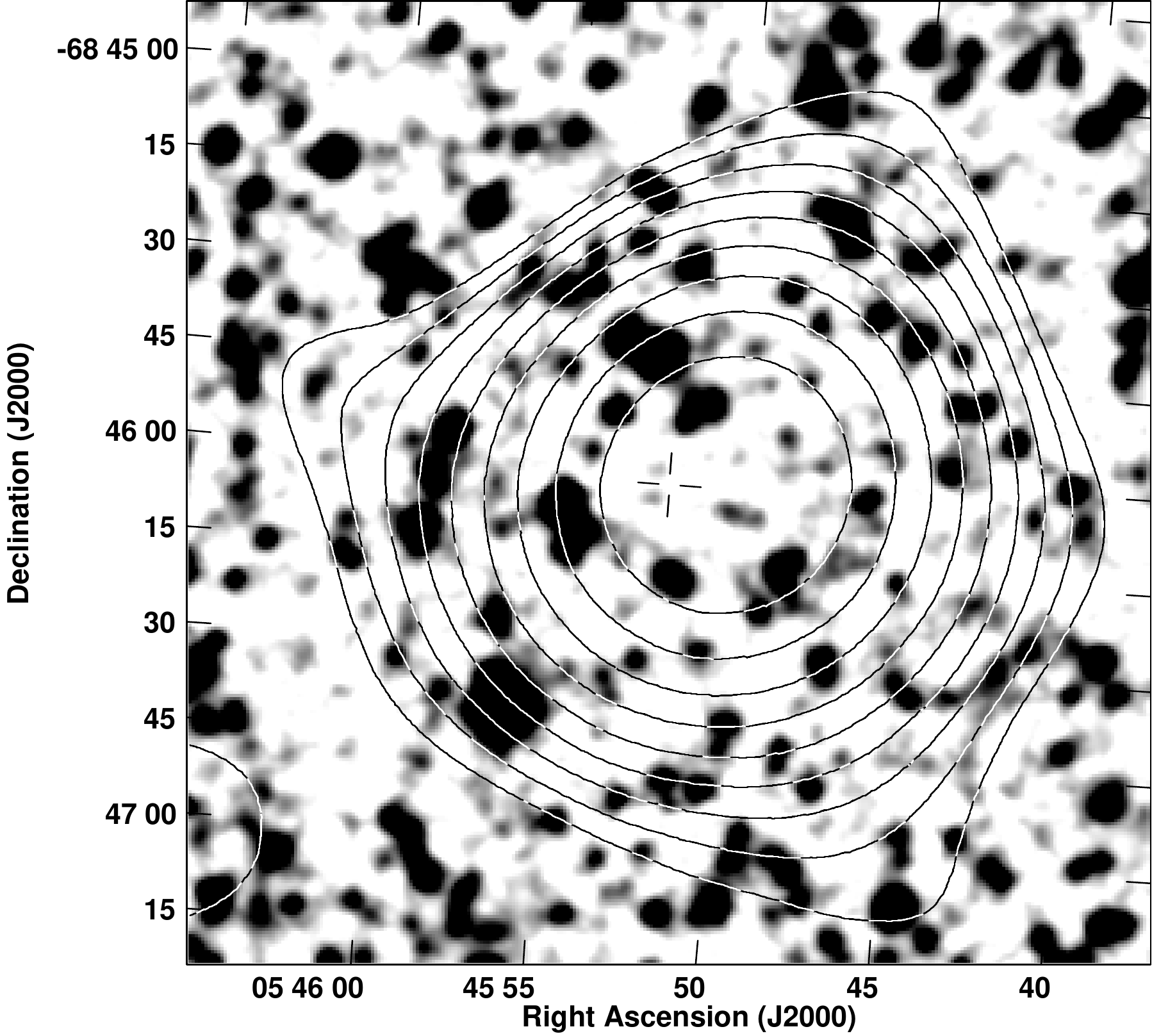}
\end{minipage}%
\caption[]{Continued.}
\end{figure}
\setcounter{figure}{9}
\begin{figure}
\begin{minipage}{0.3\textwidth}
  \centering
\includegraphics[angle=0,scale=.3]{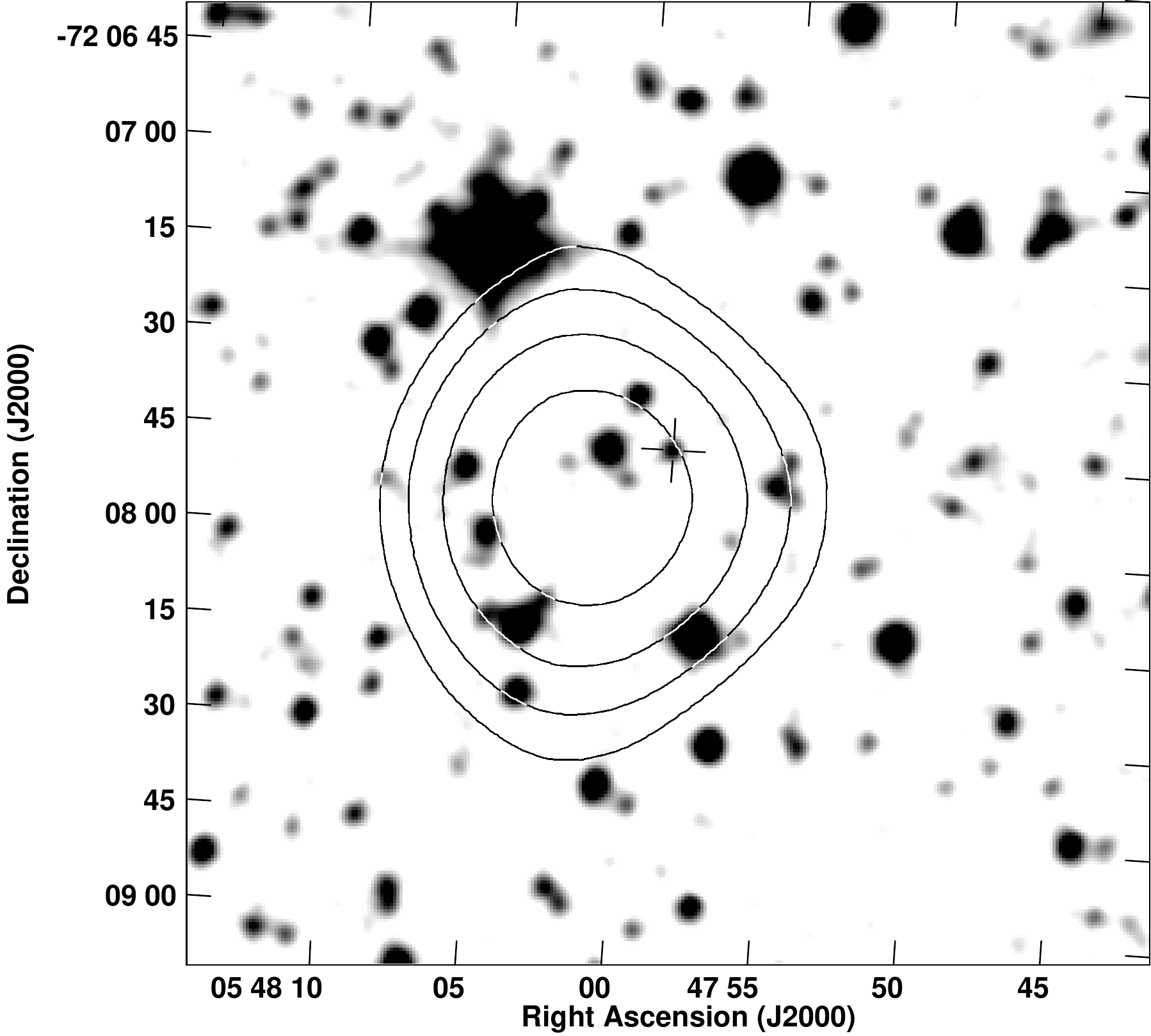}
\includegraphics[angle=0,scale=.25]{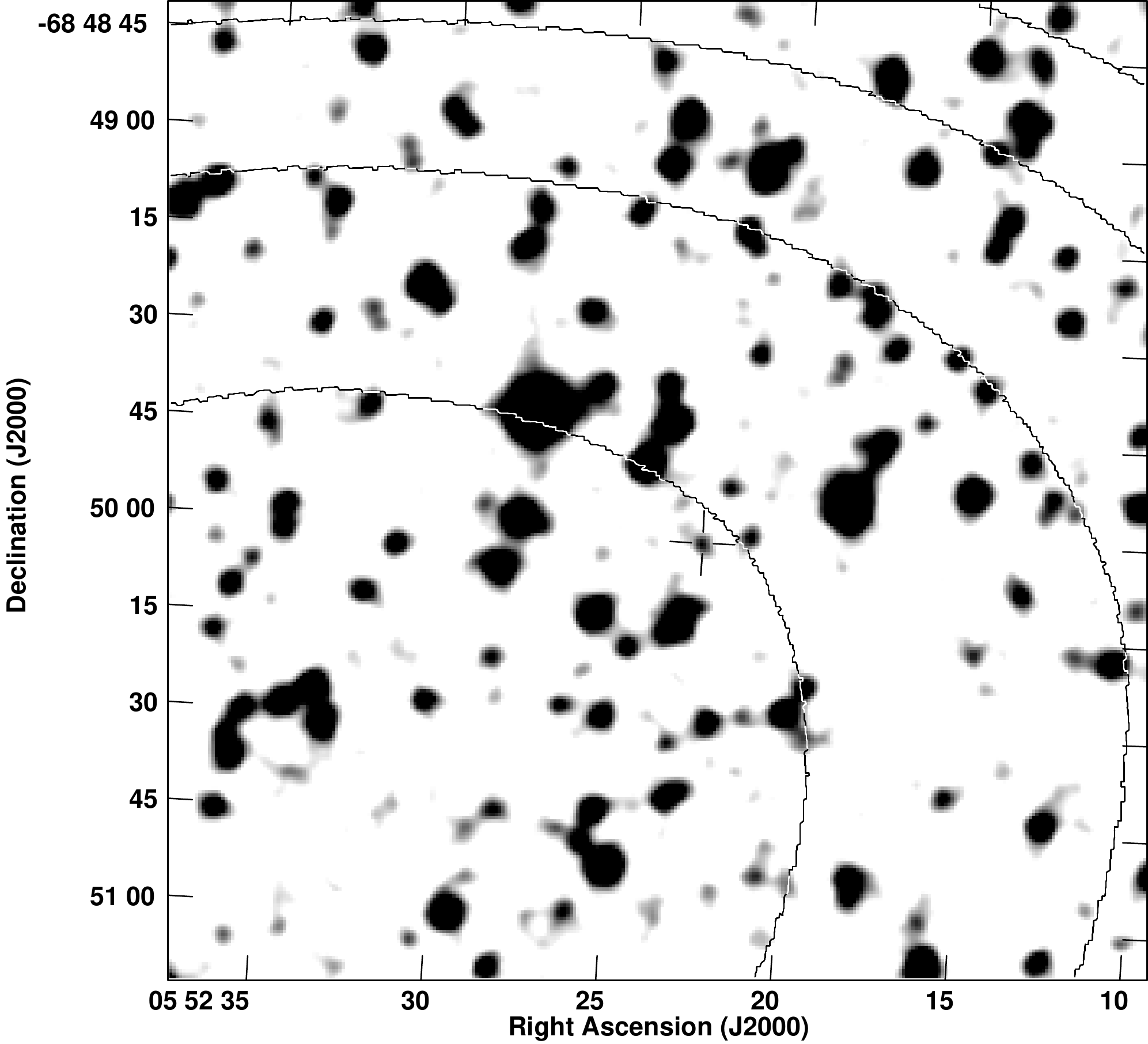}
\includegraphics[angle=0,scale=.3]{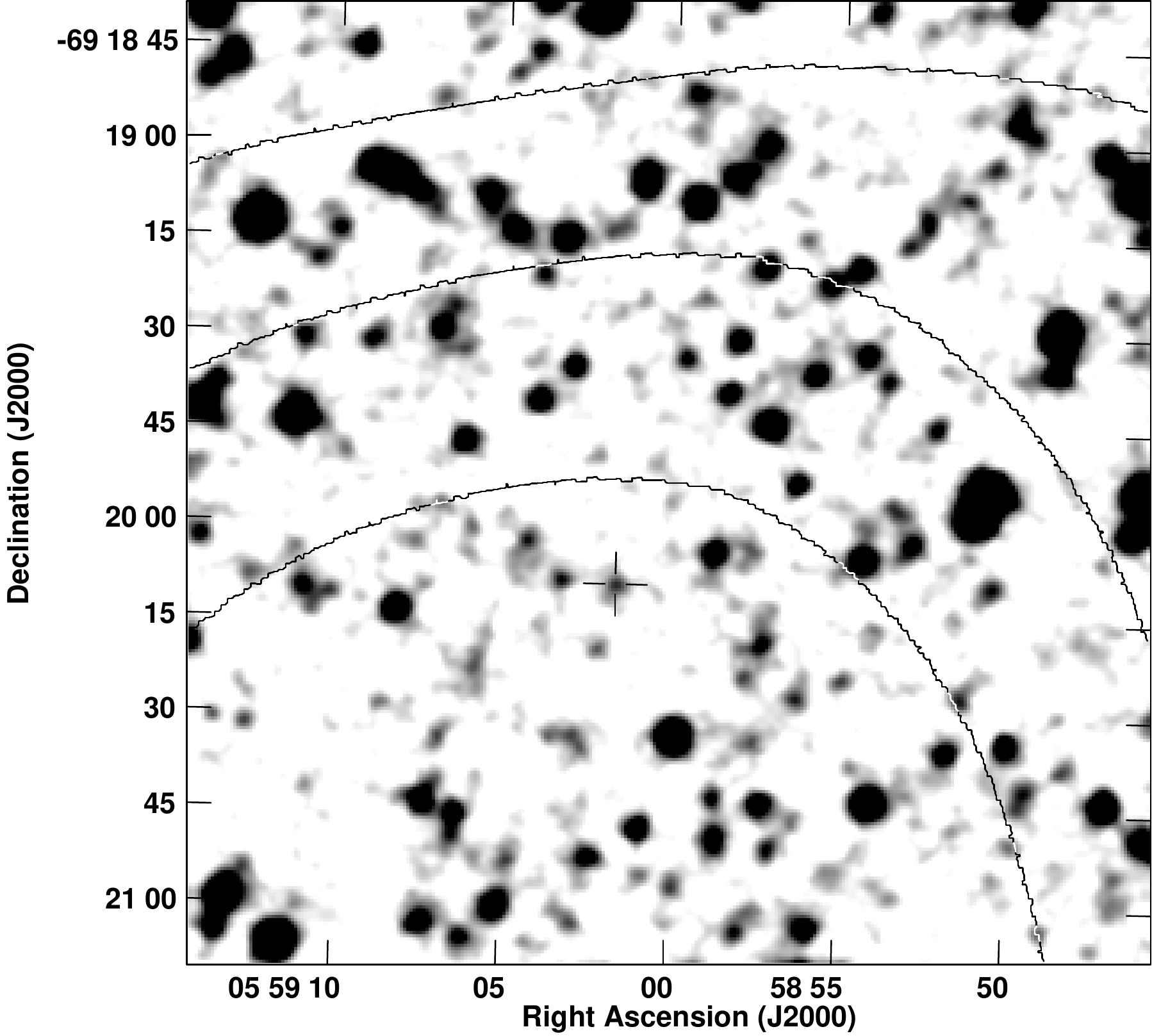}
\end{minipage}%
\begin{minipage}{0.3\textwidth}
  \centering
\includegraphics[angle=0,scale=.3]{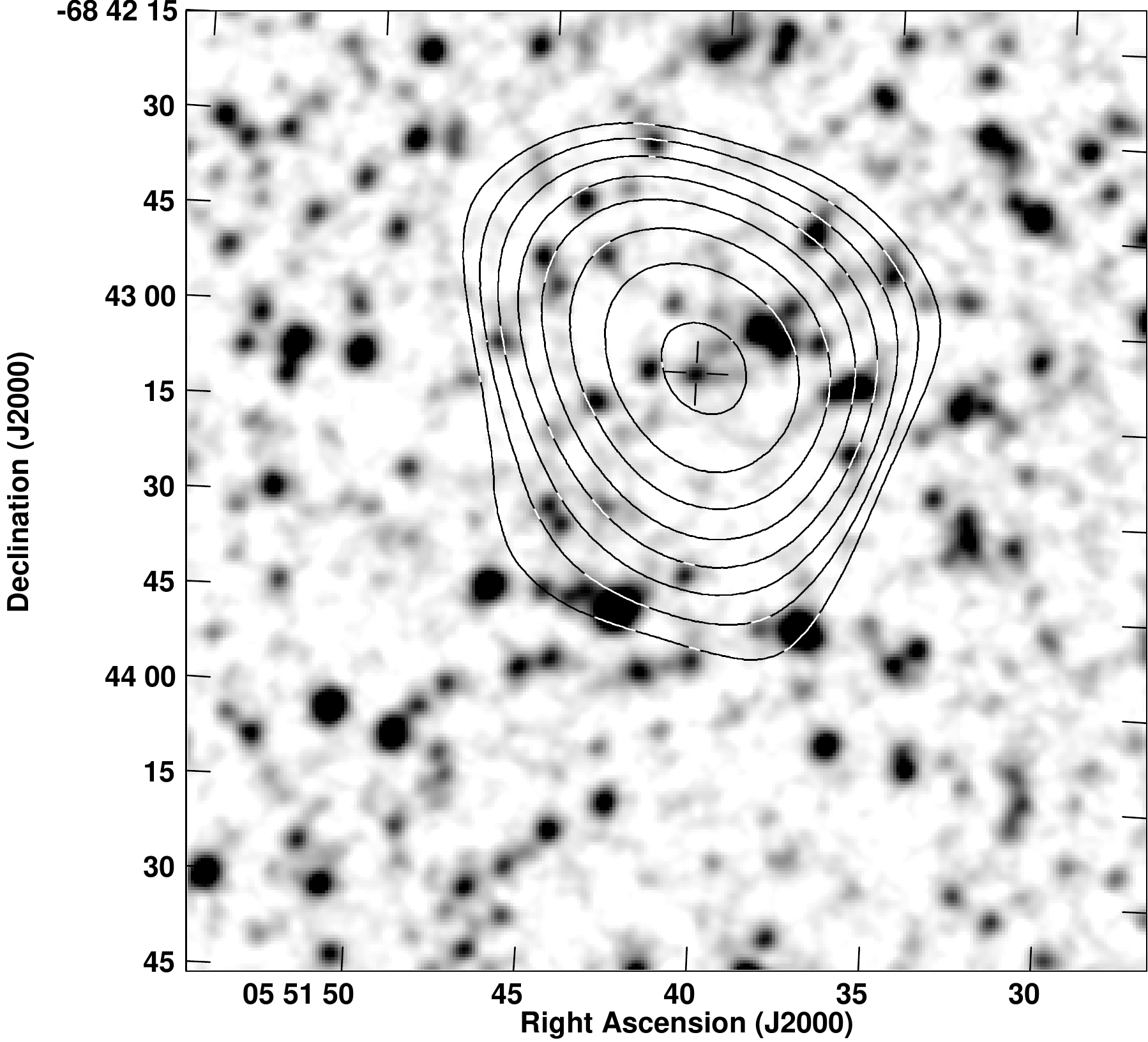}
\includegraphics[angle=0,scale=.3]{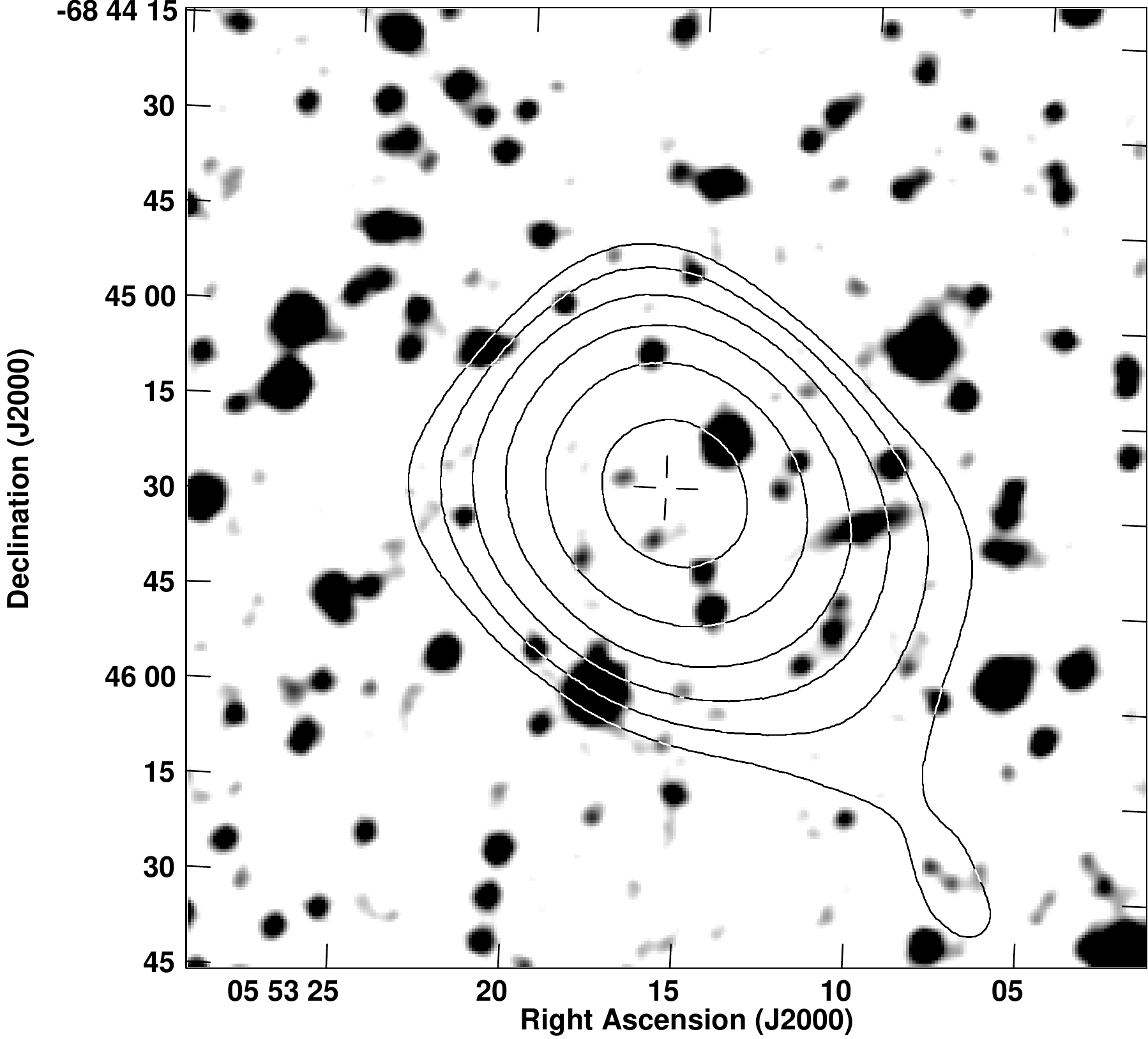}
\includegraphics[angle=0,scale=.3]{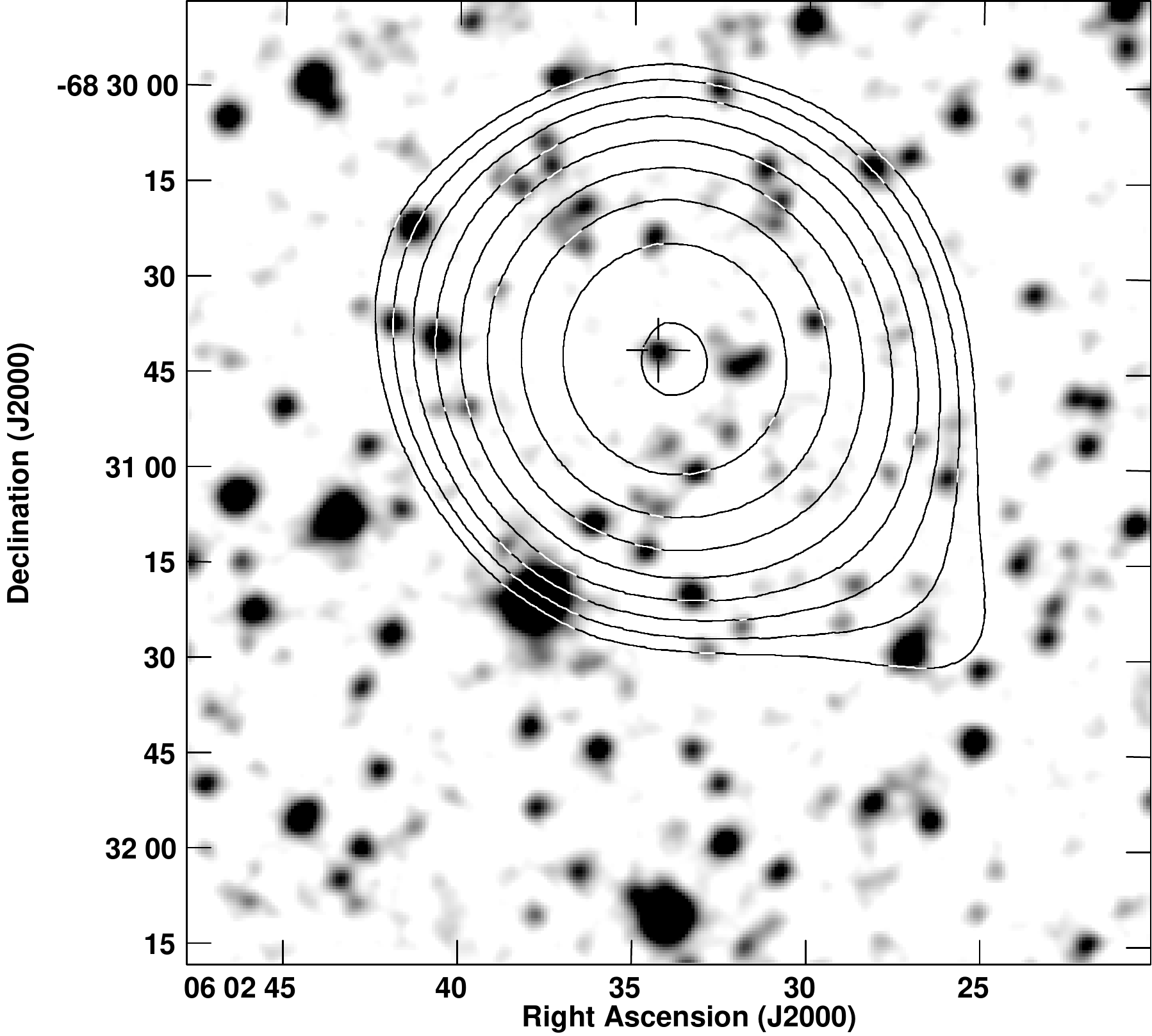}
\end{minipage}%
\begin{minipage}{0.3\textwidth}
  \centering
\includegraphics[angle=0,scale=.25]{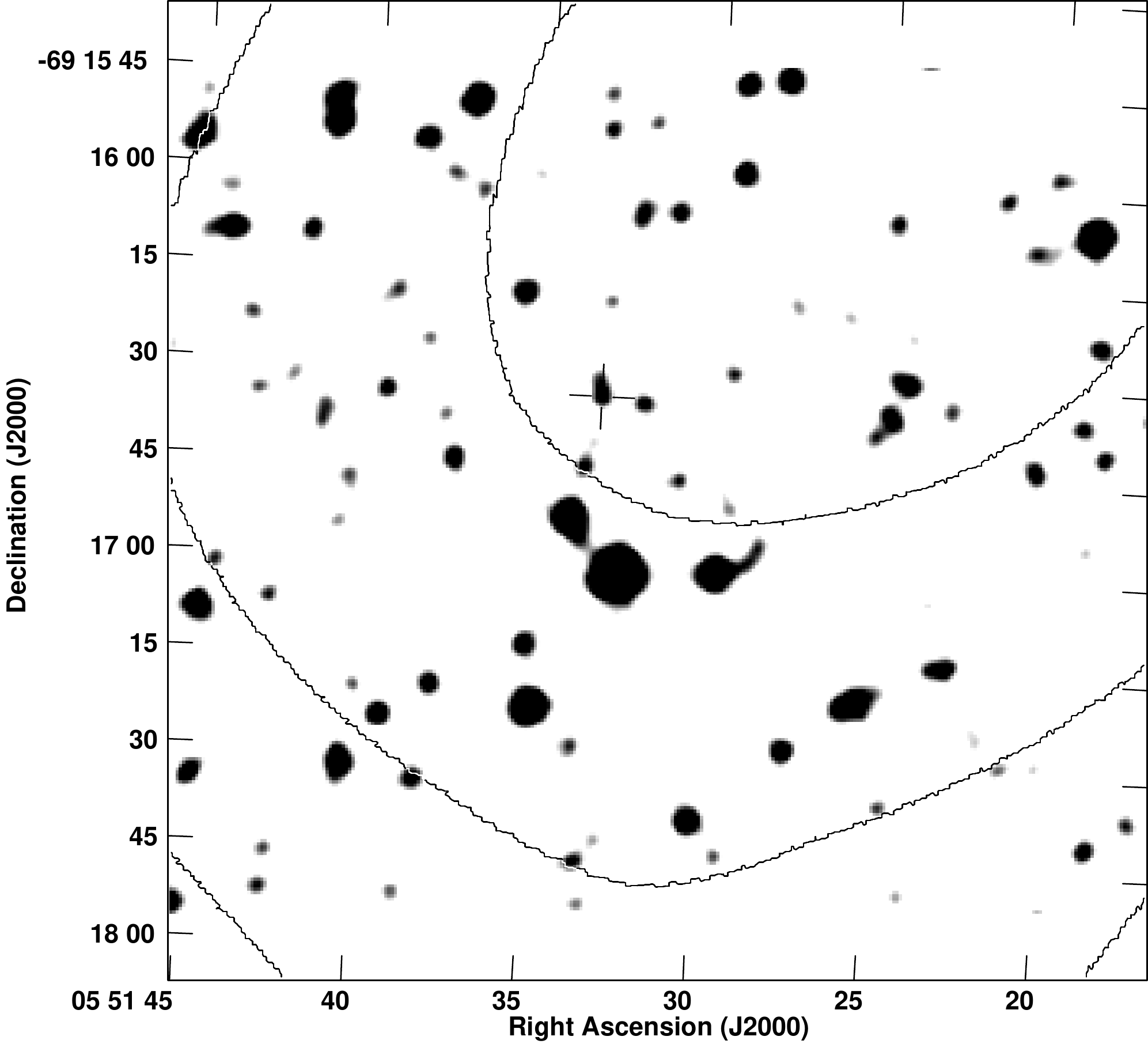}
\includegraphics[angle=0,scale=.3]{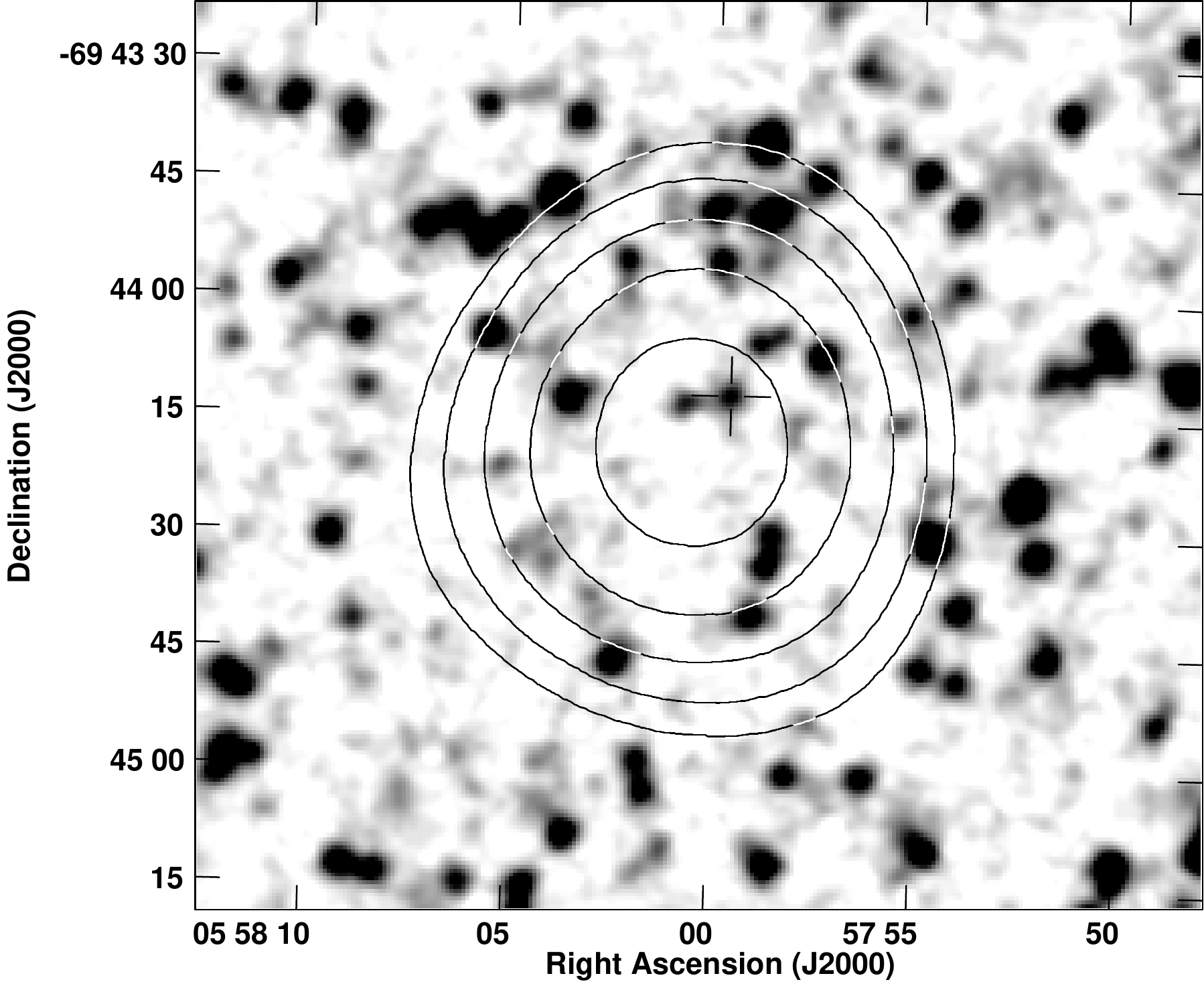}
\includegraphics[angle=0,scale=.3]{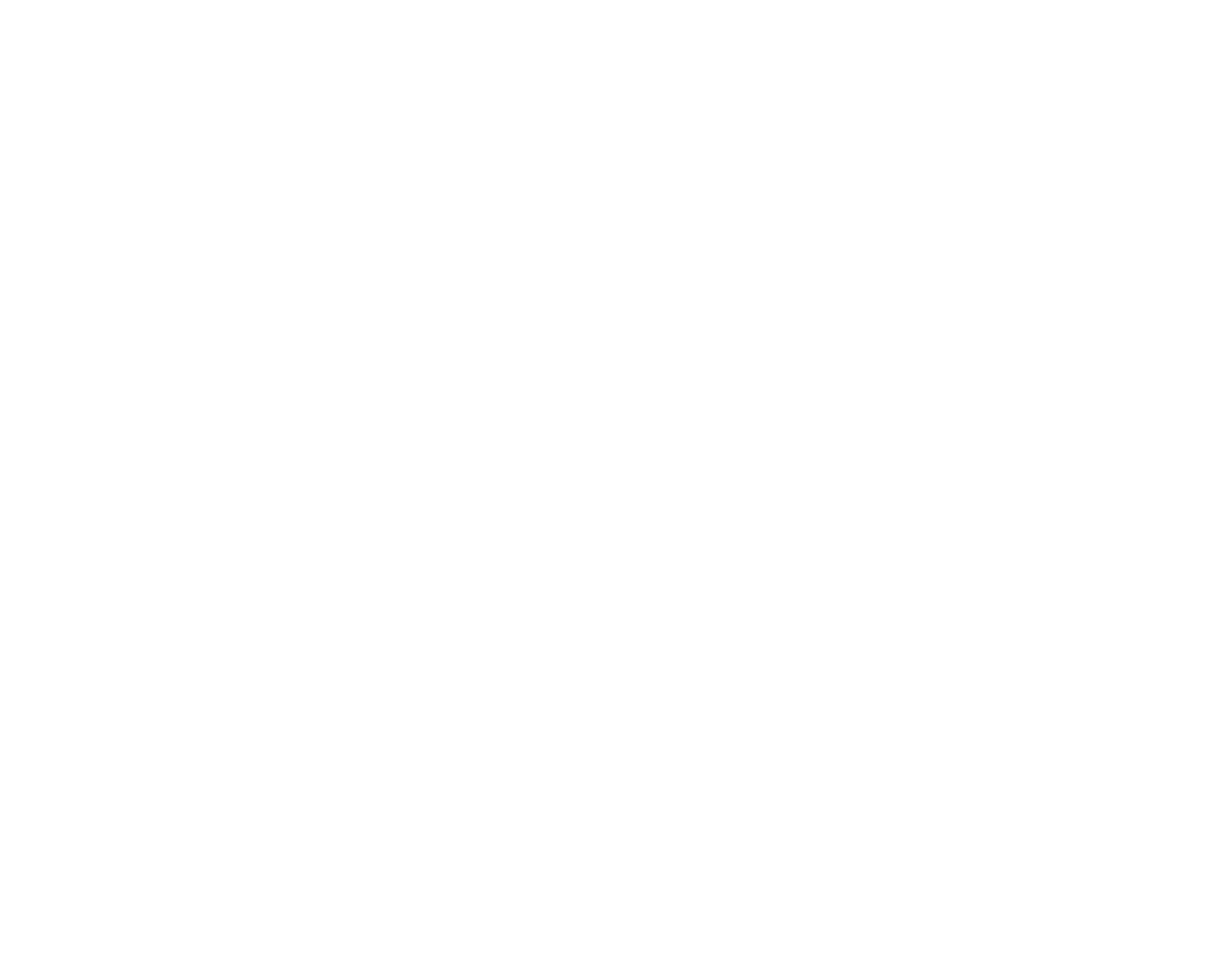}
\end{minipage}%
\caption[]{Continued.}
\end{figure}

\end{document}